\documentclass[12pt]{article}
\pdfoutput=1
\usepackage{jheppub,cancel}
\usepackage[utf8]{inputenc}
\usepackage{mathtools}
\usepackage{enumitem}
\usepackage{caption}
\usepackage{subcaption}
\usepackage{graphicx}
\usepackage{slashed}
\usepackage{booktabs}
\usepackage{mciteplus}
\allowdisplaybreaks

\usepackage{xkeyval}

\usepackage{tikz,fp}
\usetikzlibrary{external,fixedpointarithmetic,decorations.pathmorphing,positioning,calc,fadings,decorations.markings,decorations.pathreplacing,patterns,arrows}

\tikzset{
  label distance=-3pt,
  >=stealth,
  inner sep=1.5pt,
  boson/.style={
    decoration={snake, segment length=2mm, amplitude=0.5mm},
    decorate,
  },
  quark/.style={
    postaction={decorate},
    decoration={markings,mark=at position .5 with {\draw[->] (-0.01,0) -- (0.07,0);}}
  },
  aquark/.style={
    postaction={decorate},
    decoration={markings,mark=at position .5 with {\draw[->] (0.01,0) -- (-0.07,0);}}
  },
  gluonOld/.style={
    line width=0.7pt,
    decorate,
    decoration={coil,aspect=-0.5,amplitude=-2pt,segment length=2.2pt}
  },
  gluon/.style={
    line width=0.5pt,
    decorate,
    decoration={coil,aspect=-0.75,amplitude=-1.5pt,segment length=2pt}
  },
  agluon/.style={
    line width=0.5pt,
    decorate,
    decoration={coil,aspect=0.75,amplitude=1.5pt,segment length=2pt}
  },
  scalar/.style={
    dashed
  },
  dscalar/.style={
    dashed,
    postaction={decorate},
    decoration={markings,mark=at position 0.65 with {\draw[->] (-0.01,0) -- (0.07,0);}}
  },
  adscalar/.style={
    dashed,
    postaction={decorate},
    decoration={markings,mark=at position 0.65 with {\draw[->] (-0.01,0) -- (0.07,0);}}
  },
  line/.style={
  },
  doubleline/.style={
    double
  },
  middleArrow/.style={
    draw=white,
    decoration={markings,mark=at position 0.65 with{\arrow[scale=1,black]{>}}},
    decorate
  },
  dot/.style={
  	postaction={decorate},
    decoration={markings,mark=between positions 0.5 and 0.5 step 1 with {\draw[fill] (0,0) circle [radius=1.5pt];}}  
  },
  dots/.style={
  	postaction={decorate},
    decoration={markings,mark=between positions 0.3 and 0.6 step 0.3 with {\draw[fill] (0,0) circle [radius=1.5pt];}}  
  },
  dotss/.style={
  	postaction={decorate},
    decoration={markings,mark=between positions 0.25 and 0.75 step 0.25 with {\draw[fill] (0,0) circle [radius=1.5pt];}}  
  },
  blob/.style={
  	pattern=north west lines,
    draw=black
  }
}

\makeatletter
\def\gSetStdExtTypes#1{\presetkeys{gTypes}{eA=#1,eB=#1,eC=#1,eD=#1,eE=#1}{}}
\def\gSetStdIntTypes#1{\presetkeys{gTypes}{iA=#1,iB=#1,iC=#1,iD=#1,iE=#1,iF=#1,iG=#1,iH=#1,iI=#1,iJ=#1}{}}
\def\gSetStdTypes#1{\gSetStdExtTypes{#1}\gSetStdIntTypes{#1}}

\define@cmdkeys{general}{scale}
\define@cmdkeys{general}{rotate}
\define@cmdkeys{general}{font}
\define@cmdkeys{general}{yshift}
\define@cmdkeys{general}{yscale}
\presetkeys{general}{scale=1, rotate=0, font=\scriptsize, yshift=0pt, yscale=1}{}

\define@cmdkeys{gTypes}{eA,eB,eC,eD,eE,iA,iB,iC,iD,iE,iF,iG,iH,iI,iJ}
\gSetStdTypes{gluon}
\define@key{pre}{all}{\gSetStdTypes{#1}}
\define@key{pre}{allE}{\gSetStdExtTypes{#1}}
\define@key{pre}{allI}{\gSetStdIntTypes{#1}}

\define@cmdkeys{gLabels}{eLA,eLB,eLC,eLD,eLE,iLA,iLB,iLC,iLD,iLE,iLF,iLG,iLH,iLI,iLJ}
\presetkeys{gLabels}{eLA={},eLB={},eLC={},eLD={},eLE={},iLA={},iLB={},iLC={},iLD={},iLE={},iLF={},iLG={},iLH={},iLI={},iLJ={}}{}

\def\gTreeTri{\@ifnextchar[\@gTreeTri{\@gTreeTri[]}}
\def\@gTreeTri[#1]#2{%
  \setkeys*{pre}{#1}%
  \setrmkeys{general,gTypes,gLabels}%
  \begin{tikzpicture}
    [line width=1pt,
    baseline={([yshift=-0.5ex+\cmdKV@general@yshift]current bounding box.center)},
    scale=\cmdKV@general@scale,
    rotate=\cmdKV@general@rotate,
    font=\cmdKV@general@font,
    yscale=\cmdKV@general@yscale]
    \draw[\cmdKV@gTypes@eA] (0.7,0.3) -- (0.9,0.6) node[label=(\cmdKV@general@rotate):\cmdKV@gLabels@eLA] {};
    \draw[\cmdKV@gTypes@eB] (0.7,0.3) -- (0.9,0) node[label=(\cmdKV@general@rotate):\cmdKV@gLabels@eLB] {};
    \draw[\cmdKV@gTypes@eC] (0.7,0.3) -- (0.4,0.3) node[label=(-180+\cmdKV@general@rotate):\cmdKV@gLabels@eLC] {};
    #2
  \end{tikzpicture}
  \gSetStdTypes{gluon}
}

\def\gTreeS{\@ifnextchar[\@gTreeS{\@gTreeS[]}}
\def\@gTreeS[#1]#2{%
  \setkeys*{pre}{#1}%
  \setrmkeys{general,gTypes,gLabels}%
  \begin{tikzpicture}
    [line width=1pt,
    baseline={([yshift=-0.5ex+\cmdKV@general@yshift]current bounding box.center)},
    scale=\cmdKV@general@scale,
    rotate=\cmdKV@general@rotate,
    font=\cmdKV@general@font,
    yscale=\cmdKV@general@yscale]
    \draw[\cmdKV@gTypes@eA] (0.7,0.3) -- (0.9,0.6) node[label=(\cmdKV@general@rotate):\cmdKV@gLabels@eLA] {};
    \draw[\cmdKV@gTypes@eB] (0.7,0.3) -- (0.9,0) node[label=(\cmdKV@general@rotate):\cmdKV@gLabels@eLB] {};
    \draw[\cmdKV@gTypes@eC] (0.2,0.3) -- (0,0) node[label=(-180+\cmdKV@general@rotate):\cmdKV@gLabels@eLC] {};
    \draw[\cmdKV@gTypes@eD] (0.2,0.3) -- (0,0.6) node[label=(180+\cmdKV@general@rotate):\cmdKV@gLabels@eLD] {};
    \draw[label distance=0,\cmdKV@gTypes@iA] (0.7,0.3) -- node[label=(90+\cmdKV@general@rotate):\cmdKV@gLabels@iLA] {} (0.2,0.3);
    #2
  \end{tikzpicture}
  \gSetStdTypes{gluon}
}

\def\gTreeT{\@ifnextchar[\@gTreeT{\@gTreeT[]}}
\def\@gTreeT[#1]#2{%
  \setkeys*{pre}{#1}%
  \setrmkeys{general,gTypes,gLabels}%
  \begin{tikzpicture}
    [line width=1pt,
    baseline={([yshift=-0.5ex+\cmdKV@general@yshift]current bounding box.center)},
    scale=\cmdKV@general@scale,
    rotate=\cmdKV@general@rotate,
    font=\cmdKV@general@font,
    yscale=\cmdKV@general@yscale]
    \draw[\cmdKV@gTypes@eA] (0.45,0.5) -- (0.9,0.6) node[label=(\cmdKV@general@rotate):\cmdKV@gLabels@eLA] {};
    \draw[\cmdKV@gTypes@eB] (0.45,0.1) -- (0.9,0) node[label=(\cmdKV@general@rotate):\cmdKV@gLabels@eLB] {};
    \draw[\cmdKV@gTypes@eC] (0.45,0.1) -- (0,0) node[label=(-180+\cmdKV@general@rotate):\cmdKV@gLabels@eLC] {};
    \draw[\cmdKV@gTypes@eD] (0.45,0.5) -- (0,0.6) node[label=(180+\cmdKV@general@rotate):\cmdKV@gLabels@eLD] {};
    \draw[label distance=0,\cmdKV@gTypes@iA] (0.45,0.5) -- node[label=(\cmdKV@general@rotate):\cmdKV@gLabels@iLA] {} (0.45,0.1);
    #2
  \end{tikzpicture}
  \gSetStdTypes{gluon}
}

\def\gTreeU{\@ifnextchar[\@gTreeU{\@gTreeU[]}}
\def\@gTreeU[#1]#2{%
  \setkeys*{pre}{#1}%
  \setrmkeys{general,gTypes,gLabels}%
  \begin{tikzpicture}
    [line width=1pt,
    baseline={([yshift=-0.5ex+\cmdKV@general@yshift]current bounding box.center)},
    scale=\cmdKV@general@scale,
    rotate=\cmdKV@general@rotate,
    font=\cmdKV@general@font,
    yscale=\cmdKV@general@yscale]
    \draw[\cmdKV@gTypes@eA] (0.45,0.5) -- (0.9,0.6) node[label=(\cmdKV@general@rotate):\cmdKV@gLabels@eLA] {};
    \draw[\cmdKV@gTypes@eB] (0.45,0.1) -- (0.9,0) node[label=(\cmdKV@general@rotate):\cmdKV@gLabels@eLB] {};
    \draw[\cmdKV@gTypes@eC] (0.45,0.5) -- (0,0) node[label=(-180+\cmdKV@general@rotate):\cmdKV@gLabels@eLC] {};
    \draw[\cmdKV@gTypes@eD] (0.45,0.1) -- (0,0.6) node[label=(180+\cmdKV@general@rotate):\cmdKV@gLabels@eLD] {};
    \draw[label distance=0,\cmdKV@gTypes@iA] (0.45,0.5) -- node[label=(\cmdKV@general@rotate):\cmdKV@gLabels@iLA] {} (0.45,0.1);
    #2
  \end{tikzpicture}
  \gSetStdTypes{gluon}
}

\def\gTadA{\@ifnextchar[\@gTadA{\@gTadA[]}}
\def\@gTadA[#1]#2{%
  \setkeys*{pre}{#1}%
  \setrmkeys{general,gTypes,gLabels}%
  \begin{tikzpicture}
    [line width=1pt,
    baseline={([yshift=-0.5ex+\cmdKV@general@yshift]current bounding box.center)},
    scale=\cmdKV@general@scale,
    rotate=\cmdKV@general@rotate,
    font=\cmdKV@general@font,
    yscale=\cmdKV@general@yscale]
    \draw[\cmdKV@gTypes@eA] (0.7,0.3) -- (0.9,0.6) node[label=(\cmdKV@general@rotate):\cmdKV@gLabels@eLA] {};
    \draw[\cmdKV@gTypes@eB] (0.7,0.3) -- (0.9,0) node[label=(\cmdKV@general@rotate):\cmdKV@gLabels@eLB] {};
    \draw[\cmdKV@gTypes@eC] (0.3,0.3) -- (0.2,0) node[label=(-180+\cmdKV@general@rotate):\cmdKV@gLabels@eLC] {};
    \draw[\cmdKV@gTypes@eD] (-0.15,0.45) -- (-0.5,0.6) node[label=(180+\cmdKV@general@rotate):\cmdKV@gLabels@eLD] {};
    \draw[label distance=0,\cmdKV@gTypes@iA] (0.7,0.3) -- node[label=(90+\cmdKV@general@rotate):\cmdKV@gLabels@iLA] {} (0.3,0.3);
    \draw[label distance=0,\cmdKV@gTypes@iB] (0.3,0.3) -- node[label=(-100+\cmdKV@general@rotate):\cmdKV@gLabels@iLB] {} (-0.15,0.45);
    \draw[label distance=0,\cmdKV@gTypes@iC] (-0.15,0.45) -- node[label=(135+\cmdKV@general@rotate):\cmdKV@gLabels@iLC] {} (0.18,0.69);
    \draw[label distance=-0.5,\cmdKV@gTypes@iD] (0.18,0.69) .. controls ($(0.18,0.69) + (-0.12,0.12)$) and ($(0.35,0.87) + (-0.12,0.12)$) .. (0.35,0.87) node[label=(\cmdKV@general@rotate):\cmdKV@gLabels@iLD] {} .. controls ($(0.35,0.87) + (0.12,-0.12)$) and ($(0.18,0.69) + (0.12,-0.12)$) .. (0.18,0.69);
    #2
  \end{tikzpicture}
  \gSetStdTypes{gluon}
}

\def\gTadB{\@ifnextchar[\@gTadB{\@gTadB[]}}
\def\@gTadB[#1]#2{%
  \setkeys*{pre}{#1}%
  \setrmkeys{general,gTypes,gLabels}%
  \begin{tikzpicture}
    [line width=1pt,
    baseline={([yshift=-0.5ex+\cmdKV@general@yshift]current bounding box.center)},
    scale=\cmdKV@general@scale,
    rotate=\cmdKV@general@rotate,
    font=\cmdKV@general@font,
    yscale=\cmdKV@general@yscale]
    \draw[\cmdKV@gTypes@eA] (0.7,0.3) -- (0.9,0.6) node[label=(\cmdKV@general@rotate):\cmdKV@gLabels@eLA] {};
    \draw[\cmdKV@gTypes@eB] (0.7,0.3) -- (0.9,0) node[label=(\cmdKV@general@rotate):\cmdKV@gLabels@eLB] {};
    \draw[\cmdKV@gTypes@eC] (0,0.3) -- (-0.2,0) node[label=(-180+\cmdKV@general@rotate):\cmdKV@gLabels@eLC] {};
    \draw[\cmdKV@gTypes@eD] (0,0.3) -- (-0.2,0.6) node[label=(180+\cmdKV@general@rotate):\cmdKV@gLabels@eLD] {};
    \draw[label distance=0,\cmdKV@gTypes@iA] (0.7,0.3) -- node[label=(-90+\cmdKV@general@rotate):\cmdKV@gLabels@iLA] {} (0.35,0.3);
    \draw[label distance=0,\cmdKV@gTypes@iB] (0.35,0.3) -- node[label=(\cmdKV@general@rotate):\cmdKV@gLabels@iLB] {} (0.35,0.6);
    \draw[label distance=-0.5,\cmdKV@gTypes@iC] (0.35,0.6) .. controls (0.18,0.6) and (0.18,0.85) .. (0.35,0.85) .. controls (0.52,0.85) and (0.52,0.6) .. node[label=(\cmdKV@general@rotate):\cmdKV@gLabels@iLC] {} (0.35,0.6);
    \draw[label distance=0,\cmdKV@gTypes@iD] (0.35,0.3) -- node[label=(-90+\cmdKV@general@rotate):\cmdKV@gLabels@iLD] {} (0,0.3);
    #2
  \end{tikzpicture}
  \gSetStdTypes{gluon}
}
\def\gBubA{\@ifnextchar[\@gBubA{\@gBubA[]}}
\def\@gBubA[#1]#2{%
  \setkeys*{pre}{#1}%
  \setrmkeys{general,gTypes,gLabels}%
  \begin{tikzpicture}
    [line width=1pt,
    baseline={([yshift=-0.5ex+\cmdKV@general@yshift]current bounding box.center)},
    scale=\cmdKV@general@scale,
    rotate=\cmdKV@general@rotate,
    font=\cmdKV@general@font,
    yscale=\cmdKV@general@yscale]
    \draw[\cmdKV@gTypes@eA] (0.9,0.7) -- (1.2,0.7) node[label=(\cmdKV@general@rotate):\cmdKV@gLabels@eLA] {};
    \draw[\cmdKV@gTypes@eB] (0.5,0.7) -- (0.2,0.7) node[label=(180+\cmdKV@general@rotate):\cmdKV@gLabels@eLB] {};
    \draw[label distance=0,\cmdKV@gTypes@iA] (0.9,0.7) .. controls (0.9,0.45) and (0.5,0.45) .. node[label=(-90+\cmdKV@general@rotate):\cmdKV@gLabels@iLA] {} (0.5,0.7);
    \draw[label distance=0,\cmdKV@gTypes@iB] (0.5,0.7) .. controls (0.5,0.95) and (0.9,0.95) .. node[label=(90+\cmdKV@general@rotate):\cmdKV@gLabels@iLB] {} (0.9,0.7);
    #2
  \end{tikzpicture}
  \gSetStdTypes{gluon}
}
\def\gBubB{\@ifnextchar[\@gBubB{\@gBubB[]}}
\def\@gBubB[#1]#2{%
  \setkeys*{pre}{#1}%
  \setrmkeys{general,gTypes,gLabels}%
  \begin{tikzpicture}
    [line width=1pt,
    baseline={([yshift=-0.5ex+\cmdKV@general@yshift]current bounding box.center)},
    scale=\cmdKV@general@scale,
    rotate=\cmdKV@general@rotate,
    font=\cmdKV@general@font,
    yscale=\cmdKV@general@yscale]
    \draw[\cmdKV@gTypes@eA] (1.2,0.7) -- (1.4,1) node[label=(\cmdKV@general@rotate):\cmdKV@gLabels@eLA] {};
    \draw[\cmdKV@gTypes@eB] (1.2,0.7) -- (1.4,0.4) node[label=(\cmdKV@general@rotate):\cmdKV@gLabels@eLB] {};
    \draw[\cmdKV@gTypes@eC] (0.2,0.7) -- (0,0.4) node[label=(-180+\cmdKV@general@rotate):\cmdKV@gLabels@eLC] {};
    \draw[\cmdKV@gTypes@eD] (0.2,0.7) -- (0,1) node[label=(180+\cmdKV@general@rotate):\cmdKV@gLabels@eLD] {};
    \draw[label distance=0,\cmdKV@gTypes@iA] (1.2,0.7) -- node[label=(90+\cmdKV@general@rotate):\cmdKV@gLabels@iLA] {} (0.9,0.7);
    \draw[label distance=0,\cmdKV@gTypes@iB] (0.9,0.7) .. controls (0.9,0.45) and (0.5,0.45) .. node[label=(-90+\cmdKV@general@rotate):\cmdKV@gLabels@iLB] {} (0.5,0.7);
    \draw[label distance=0,\cmdKV@gTypes@iC] (0.5,0.7) .. controls (0.5,0.95) and (0.9,0.95) .. node[label=(90+\cmdKV@general@rotate):\cmdKV@gLabels@iLC] {} (0.9,0.7);
    \draw[label distance=0,\cmdKV@gTypes@iD] (0.5,0.7) -- node[label=(90+\cmdKV@general@rotate):\cmdKV@gLabels@iLD] {} (0.2,0.7);
    #2
  \end{tikzpicture}
  \gSetStdTypes{gluon}
}

\def\gBubC{\@ifnextchar[\@gBubC{\@gBubC[]}}
\def\@gBubC[#1]#2{%
  \setkeys*{pre}{#1}%
  \setrmkeys{general,gTypes,gLabels}%
  \begin{tikzpicture}
    [line width=1pt,
    baseline={([yshift=-0.5ex+\cmdKV@general@yshift]current bounding box.center)},
    scale=\cmdKV@general@scale,
    rotate=\cmdKV@general@rotate,
    font=\cmdKV@general@font,
    yscale=\cmdKV@general@yscale]
    \draw[\cmdKV@gTypes@eA] (0.7,0.3) -- (0.9,0.6) node[label=(\cmdKV@general@rotate):\cmdKV@gLabels@eLA] {};
    \draw[\cmdKV@gTypes@eB] (0.7,0.3) -- (0.9,0) node[label=(\cmdKV@general@rotate):\cmdKV@gLabels@eLB] {};
    \draw[\cmdKV@gTypes@eC] (0.3,0.3) -- (0.2,0) node[label=(-180+\cmdKV@general@rotate):\cmdKV@gLabels@eLC] {};
    \draw[\cmdKV@gTypes@eD] (-0.3,0.5) -- (-0.5,0.6) node[label=(180+\cmdKV@general@rotate):\cmdKV@gLabels@eLD] {};
    \draw[label distance=0,\cmdKV@gTypes@iA] (0.7,0.3) -- node[label=(90+\cmdKV@general@rotate):\cmdKV@gLabels@iLA] {} (0.3,0.3);
    \draw[label distance=0,\cmdKV@gTypes@iB] (0.3,0.3) -- node[label=(80+\cmdKV@general@rotate):\cmdKV@gLabels@iLB] {} (0,0.4);
    \draw[label distance=0,\cmdKV@gTypes@iC] (0,0.4) .. controls (-0.08,0.21) and (-0.38,0.3) .. node[label=(-90+\cmdKV@general@rotate):\cmdKV@gLabels@iLC] {} (-0.3,0.5);
    \draw[label distance=0,\cmdKV@gTypes@iD] (-0.3,0.5) .. controls (-0.22,0.65) and (0.03,0.6) .. node[label=(90+\cmdKV@general@rotate):\cmdKV@gLabels@iLD] {} (0,0.4);
    #2
  \end{tikzpicture}
  \gSetStdTypes{gluon}
}
\def\gTriA{\@ifnextchar[\@gTriA{\@gTriA[]}}
\def\@gTriA[#1]#2{%
  \setkeys*{pre}{#1}%
  \setrmkeys{general,gTypes,gLabels}%
  \begin{tikzpicture}
    [line width=1pt,
    baseline={([yshift=-0.5ex+\cmdKV@general@yshift]current bounding box.center)},
    scale=\cmdKV@general@scale,
    rotate=\cmdKV@general@rotate,
    font=\cmdKV@general@font,
    yscale=\cmdKV@general@yscale]
    \draw[\cmdKV@gTypes@eA] (0:0.3) -- (0:0.55) node[label=(\cmdKV@general@rotate):\cmdKV@gLabels@eLA] {};
    \draw[\cmdKV@gTypes@eB] (-120:0.3) -- (-120:0.55) node[label=(-180+\cmdKV@general@rotate):\cmdKV@gLabels@eLB] {};
    \draw[\cmdKV@gTypes@eC] (120:0.3) -- (120:0.55) node[label=(180+\cmdKV@general@rotate):\cmdKV@gLabels@eLC] {};
    \draw[label distance=0,\cmdKV@gTypes@iA] (0:0.3) -- node[label=(-60+\cmdKV@general@rotate):\cmdKV@gLabels@iLA] {} (-120:0.3);
    \draw[label distance=0,\cmdKV@gTypes@iB] (-120:0.3) -- node[label=(180+\cmdKV@general@rotate):\cmdKV@gLabels@iLB] {} (120:0.3);
    \draw[label distance=0,\cmdKV@gTypes@iC] (120:0.3) -- node[label=(60+\cmdKV@general@rotate):\cmdKV@gLabels@iLC] {} (0:0.3);
    #2
  \end{tikzpicture}
  \gSetStdTypes{gluon}
}

\def\gTriC{\@ifnextchar[\@gTriC{\@gTriC[]}}
\def\@gTriC[#1]#2{%
  \setkeys*{pre}{#1}%
  \setrmkeys{general,gTypes,gLabels}%
  \begin{tikzpicture}
    [line width=1pt,
    baseline={([yshift=-0.5ex+\cmdKV@general@yshift]current bounding box.center)},
    scale=\cmdKV@general@scale,
    rotate=\cmdKV@general@rotate,
    font=\cmdKV@general@font,
    yscale=\cmdKV@general@yscale]
    \draw[\cmdKV@gTypes@eA] (1,0.7) -- (1.3,1.1) node[label=(\cmdKV@general@rotate):\cmdKV@gLabels@eLA] {};
    \draw[\cmdKV@gTypes@eB] (1,0.7) -- (1.3,0.3) node[label=(\cmdKV@general@rotate):\cmdKV@gLabels@eLB] {};
    \draw[\cmdKV@gTypes@eC] (0.3,0.45) -- (0,0.3) node[label=(-180+\cmdKV@general@rotate):\cmdKV@gLabels@eLC] {};
    \draw[\cmdKV@gTypes@eD] (0.3,0.95) -- (0,1.1) node[label=(180+\cmdKV@general@rotate):\cmdKV@gLabels@eLD] {};
    \draw[label distance=0,\cmdKV@gTypes@iA] (1,0.7) -- node[label=(90+\cmdKV@general@rotate):\cmdKV@gLabels@iLA] {} (0.7,0.7);
    \draw[label distance=0,\cmdKV@gTypes@iB] (0.7,0.7) -- node[label=(-70+\cmdKV@general@rotate):\cmdKV@gLabels@iLB] {} (0.3,0.45);
    \draw[label distance=0,\cmdKV@gTypes@iC] (0.3,0.45) -- node[label=(180+\cmdKV@general@rotate):\cmdKV@gLabels@iLC] {} (0.3,0.95);
    \draw[label distance=0,\cmdKV@gTypes@iD] (0.3,0.95) -- node[label=(70+\cmdKV@general@rotate):\cmdKV@gLabels@iLD] {} (0.7,0.7);
    #2
  \end{tikzpicture}
  \gSetStdTypes{gluon}
}
\def\gBox{\@ifnextchar[\@gBox{\@gBox[]}}
\def\@gBox[#1]#2{%
  \setkeys*{pre}{#1}%
  \setrmkeys{general,gTypes,gLabels}%
  \begin{tikzpicture}
    [line width=1pt,
    baseline={([yshift=-0.5ex+\cmdKV@general@yshift]current bounding box.center)},
    scale=\cmdKV@general@scale,
    rotate=\cmdKV@general@rotate,
    font=\cmdKV@general@font,
    yscale=\cmdKV@general@yscale]
    \draw[\cmdKV@gTypes@eA] (0.75,0.75) -- (1,1) node[label=(\cmdKV@general@rotate):\cmdKV@gLabels@eLA] {};
    \draw[\cmdKV@gTypes@eB] (0.75,0.25) -- (1,0) node[label=(\cmdKV@general@rotate):\cmdKV@gLabels@eLB] {};
    \draw[\cmdKV@gTypes@eC] (0.25,0.25) -- (0,0) node[label=(-180+\cmdKV@general@rotate):\cmdKV@gLabels@eLC] {};
    \draw[\cmdKV@gTypes@eD] (0.25,0.75) -- (0,1) node[label=(180+\cmdKV@general@rotate):\cmdKV@gLabels@eLD] {};
    \draw[label distance=0,\cmdKV@gTypes@iA] (0.75,0.75) -- node[label=(\cmdKV@general@rotate):\cmdKV@gLabels@iLA] {} (0.75,0.25);
    \draw[label distance=0,\cmdKV@gTypes@iB] (0.75,0.25) -- node[label=(-90+\cmdKV@general@rotate):\cmdKV@gLabels@iLB] {} (0.25,0.25);
    \draw[label distance=0,\cmdKV@gTypes@iC] (0.25,0.25) -- node[label=(180+\cmdKV@general@rotate):\cmdKV@gLabels@iLC] {} (0.25,0.75);
    \draw[label distance=0,\cmdKV@gTypes@iD] (0.25,0.75) -- node[label=(90+\cmdKV@general@rotate):\cmdKV@gLabels@iLD] {} (0.75,0.75);
    #2
  \end{tikzpicture}
  \gSetStdTypes{gluon}
}
\def\gPenta{\@ifnextchar[\@gPenta{\@gPenta[]}}
\def\@gPenta[#1]#2{%
  \setkeys*{pre}{#1}%
  \setrmkeys{general,gTypes,gLabels}%
  \begin{tikzpicture}
    [line width=1pt,
    baseline={([yshift=-0.5ex+\cmdKV@general@yshift]current bounding box.center)},
    scale=\cmdKV@general@scale,
    rotate=\cmdKV@general@rotate,
    font=\cmdKV@general@font,
    yscale=\cmdKV@general@yscale]
    \draw[\cmdKV@gTypes@eA] (72:0.35) -- (72:0.75) node[label=(72+\cmdKV@general@rotate):\cmdKV@gLabels@eLA] {};
    \draw[\cmdKV@gTypes@eB] (0:0.35) -- (0:0.75) node[label=(\cmdKV@general@rotate):\cmdKV@gLabels@eLB] {};
    \draw[\cmdKV@gTypes@eC] (-72:0.35) -- (-72:0.75) node[label=(-72+\cmdKV@general@rotate):\cmdKV@gLabels@eLC] {};
    \draw[\cmdKV@gTypes@eD] (-144:0.35) -- (-144:0.75) node[label=(-144+\cmdKV@general@rotate):\cmdKV@gLabels@eLD] {};
    \draw[\cmdKV@gTypes@eE] (144:0.35) -- (144:0.75) node[label=(144+\cmdKV@general@rotate):\cmdKV@gLabels@eLE] {};
    \draw[label distance=0,\cmdKV@gTypes@iA] (72:0.35) -- node[label=(36+\cmdKV@general@rotate):\cmdKV@gLabels@iLA] {} (0:0.35);
    \draw[label distance=0,\cmdKV@gTypes@iB] (0:0.35) -- node[label=(-36+\cmdKV@general@rotate):\cmdKV@gLabels@iLB] {} (-72:0.35);
    \draw[label distance=0,\cmdKV@gTypes@iC] (-72:0.35) -- node[label=(-108+\cmdKV@general@rotate):\cmdKV@gLabels@iLC] {} (-144:0.35);
    \draw[label distance=0,\cmdKV@gTypes@iD] (-144:0.35) -- node[label=(180+\cmdKV@general@rotate):\cmdKV@gLabels@iLD] {} (144:0.35);
    \draw[label distance=0,\cmdKV@gTypes@iE] (144:0.35) -- node[label=(108+\cmdKV@general@rotate):\cmdKV@gLabels@iLE] {} (72:0.35);
    #2
  \end{tikzpicture}
  \gSetStdTypes{gluon}
}

\def\gBoxBox{\@ifnextchar[\@gBoxBox{\@gBoxBox[]}}
\def\@gBoxBox[#1]#2{%
  \setkeys*{pre}{#1}%
  \setrmkeys{general,gTypes,gLabels}%
  \begin{tikzpicture}
    [line width=1pt,
    baseline={([yshift=-0.5ex+\cmdKV@general@yshift]current bounding box.center)},
    scale=\cmdKV@general@scale,
    rotate=\cmdKV@general@rotate,
    font=\cmdKV@general@font,
    yscale=\cmdKV@general@yscale]
    \draw[\cmdKV@gTypes@eA] (1.5,0.9) -- (1.8,1.2) node[label=(\cmdKV@general@rotate):\cmdKV@gLabels@eLA] {};
    \draw[\cmdKV@gTypes@eB] (1.5,0.3) -- (1.8,0) node[label=(\cmdKV@general@rotate):\cmdKV@gLabels@eLB] {};
    \draw[\cmdKV@gTypes@eC] (0.3,0.3) -- (0,0) node[label=(-180+\cmdKV@general@rotate):\cmdKV@gLabels@eLC] {};
    \draw[\cmdKV@gTypes@eD] (0.3,0.9) -- (0,1.2) node[label=(180+\cmdKV@general@rotate):\cmdKV@gLabels@eLD] {};
    \draw[label distance=0,\cmdKV@gTypes@iA] (1.5,0.9) -- node[label=(\cmdKV@general@rotate):\cmdKV@gLabels@iLA] {} (1.5,0.3);
    \draw[label distance=0,\cmdKV@gTypes@iB] (1.5,0.3) -- node[label=(-90+\cmdKV@general@rotate):\cmdKV@gLabels@iLB] {} (0.9,0.3);
    \draw[label distance=0,\cmdKV@gTypes@iC] (0.9,0.3) -- node[label=(-90+\cmdKV@general@rotate):\cmdKV@gLabels@iLC] {} (0.3,0.3);
    \draw[label distance=0,\cmdKV@gTypes@iD] (0.3,0.3) -- node[label=(180+\cmdKV@general@rotate):\cmdKV@gLabels@iLD] {} (0.3,0.9);
    \draw[label distance=0,\cmdKV@gTypes@iE] (0.3,0.9) -- node[label=(90+\cmdKV@general@rotate):\cmdKV@gLabels@iLE] {} (0.9,0.9);
    \draw[label distance=0,\cmdKV@gTypes@iF] (0.9,0.9) -- node[label=(90+\cmdKV@general@rotate):\cmdKV@gLabels@iLF] {} (1.5,0.9);
    \draw[label distance=0,\cmdKV@gTypes@iG] (0.9,0.9) -- node[label=(\cmdKV@general@rotate):\cmdKV@gLabels@iLG] {} (0.9,0.3);
    #2
  \end{tikzpicture}
  \gSetStdTypes{gluon}
}
\def\gBoxBoxNP{\@ifnextchar[\@gBoxBoxNP{\@gBoxBoxNP[]}}
\def\@gBoxBoxNP[#1]#2{%
  \setkeys*{pre}{#1}%
  \setrmkeys{general,gTypes,gLabels}%
  \begin{tikzpicture}
    [line width=1pt,
    baseline={([yshift=-0.5ex+\cmdKV@general@yshift]current bounding box.center)},
    scale=\cmdKV@general@scale,
    rotate=\cmdKV@general@rotate,
    font=\cmdKV@general@font,
    yscale=\cmdKV@general@yscale]
    \draw[\cmdKV@gTypes@eA] (1.9,1.1) -- (2.2,1.4) node[label=(\cmdKV@general@rotate):\cmdKV@gLabels@eLA] {};
    \draw[\cmdKV@gTypes@eB] (1.9,0.3) -- (2.2,0) node[label=(\cmdKV@general@rotate):\cmdKV@gLabels@eLB] {};
    \draw[\cmdKV@gTypes@eC] (0.3,0.7) -- (0,0.7) node[label=(180+\cmdKV@general@rotate):\cmdKV@gLabels@eLC] {};
    \draw[\cmdKV@gTypes@eD] (1.1,0.7) -- (1.4,0.7) node[label=(\cmdKV@general@rotate):\cmdKV@gLabels@eLD] {};
    \draw[label distance=0,\cmdKV@gTypes@iA] (1.9,1.1) -- node[label=(\cmdKV@general@rotate):\cmdKV@gLabels@iLA] {} (1.9,0.3);
    \draw[label distance=0,\cmdKV@gTypes@iB] (1.9,0.3) -- node[label=(-90+\cmdKV@general@rotate):\cmdKV@gLabels@iLB] {} (0.7,0.3);
    \draw[label distance=0,\cmdKV@gTypes@iC] (0.7,0.3) -- node[label=(-135+\cmdKV@general@rotate):\cmdKV@gLabels@iLC] {} (0.3,0.7);
    \draw[label distance=0,\cmdKV@gTypes@iD] (0.3,0.7) -- node[label=(135+\cmdKV@general@rotate):\cmdKV@gLabels@iLD] {} (0.7,1.1);
    \draw[label distance=0,\cmdKV@gTypes@iE] (0.7,1.1) -- node[label=(90+\cmdKV@general@rotate):\cmdKV@gLabels@iLE] {} (1.9,1.1);
    \draw[label distance=0,\cmdKV@gTypes@iF] (0.7,1.1) -- node[label=(45+\cmdKV@general@rotate):\cmdKV@gLabels@iLF] {} (1.1,0.7);
    \draw[label distance=0,\cmdKV@gTypes@iG] (1.1,0.7) -- node[label=(-45+\cmdKV@general@rotate):\cmdKV@gLabels@iLG] {} (0.7,0.3);
    #2
  \end{tikzpicture}
  \gSetStdTypes{gluon}
}

\def\gTriBoxA{\@ifnextchar[\@gTriBoxA{\@gTriBoxA[]}}
\def\@gTriBoxA[#1]#2{%
  \setkeys*{pre}{#1}%
  \setrmkeys{general,gTypes,gLabels}%
  \begin{tikzpicture}
    [line width=1pt,
    baseline={([yshift=-0.5ex+\cmdKV@general@yshift]current bounding box.center)},
    scale=\cmdKV@general@scale,
    rotate=\cmdKV@general@rotate,
    font=\cmdKV@general@font,
    yscale=\cmdKV@general@yscale]
    \draw[\cmdKV@gTypes@eA] (1.5,0.9) -- (1.8,1.2) node[label=(\cmdKV@general@rotate):\cmdKV@gLabels@eLA] {};
    \draw[\cmdKV@gTypes@eB] (1.5,0.3) -- (1.8,0) node[label=(\cmdKV@general@rotate):\cmdKV@gLabels@eLB] {};
    \draw[\cmdKV@gTypes@eC] (0.5,0.6) -- (0,0.6) node[label=(-180+\cmdKV@general@rotate):\cmdKV@gLabels@eLC] {};
    \draw[label distance=0,\cmdKV@gTypes@iA] (1.5,0.9) -- node[label=(\cmdKV@general@rotate):\cmdKV@gLabels@iLA] {} (1.5,0.3);
    \draw[label distance=0,\cmdKV@gTypes@iB] (1.5,0.3) -- node[label=(-90+\cmdKV@general@rotate):\cmdKV@gLabels@iLB] {} (0.9,0.3);
    \draw[label distance=0,\cmdKV@gTypes@iC] (0.9,0.3) -- node[label=(-120+\cmdKV@general@rotate):\cmdKV@gLabels@iLC] {} (0.5,0.6);
    \draw[label distance=0,\cmdKV@gTypes@iD] (0.5,0.6) -- node[label=(120+\cmdKV@general@rotate):\cmdKV@gLabels@iLD] {} (0.9,0.9);
    \draw[label distance=0,\cmdKV@gTypes@iE] (0.9,0.9) -- node[label=(90+\cmdKV@general@rotate):\cmdKV@gLabels@iLE] {} (1.5,0.9);
    \draw[label distance=0,\cmdKV@gTypes@iF] (0.9,0.9) -- node[label=(\cmdKV@general@rotate):\cmdKV@gLabels@iLF] {} (0.9,0.3);
    #2
  \end{tikzpicture}
  \gSetStdTypes{gluon}
}
\def\gTriPenta{\@ifnextchar[\@gTriPenta{\@gTriPenta[]}}
\def\@gTriPenta[#1]#2{%
  \setkeys*{pre}{#1}%
  \setrmkeys{general,gTypes,gLabels}%
  \begin{tikzpicture}
    [line width=1pt,
    baseline={([yshift=-0.5ex+\cmdKV@general@yshift]current bounding box.center)},
    scale=\cmdKV@general@scale,
    rotate=\cmdKV@general@rotate,
    font=\cmdKV@general@font,
    yscale=\cmdKV@general@yscale]
    \draw[\cmdKV@gTypes@eA] (72:0.5) -- (45:1) node[label=(35+\cmdKV@general@rotate):\cmdKV@gLabels@eLA] {};
    \draw[\cmdKV@gTypes@eB] (0:0.5) -- (0:1) node[label=(\cmdKV@general@rotate):\cmdKV@gLabels@eLB] {};
    \draw[\cmdKV@gTypes@eC] (-72:0.5) -- (-45:1) node[label=(-35+\cmdKV@general@rotate):\cmdKV@gLabels@eLC] {};
    \draw[\cmdKV@gTypes@eD] (-180:1.2) -- (-180:1.5) node[label=(180+\cmdKV@general@rotate):\cmdKV@gLabels@eLD] {};
    \draw[label distance=0,\cmdKV@gTypes@iA] (72:0.5) -- node[label=(45+\cmdKV@general@rotate):\cmdKV@gLabels@iLA] {} (0:0.5);
    \draw[label distance=0,\cmdKV@gTypes@iB] (0:0.5) -- node[label=(-45+\cmdKV@general@rotate):\cmdKV@gLabels@iLB] {} (-72:0.5);
    \draw[label distance=0,\cmdKV@gTypes@iC] (-72:0.5) -- node[label=(-108+\cmdKV@general@rotate):\cmdKV@gLabels@iLC] {} (-144:0.5);
    \draw[label distance=0,\cmdKV@gTypes@iD] (-144:0.5) -- node[label=(-108+\cmdKV@general@rotate):\cmdKV@gLabels@iLD] {} (-180:1.2);
    \draw[label distance=2,\cmdKV@gTypes@iE] (-180:1.2) -- node[label=(90+\cmdKV@general@rotate):\cmdKV@gLabels@iLE] {} (144:0.5);
    \draw[label distance=2,\cmdKV@gTypes@iF] (144:0.5) -- node[label=(90+\cmdKV@general@rotate):\cmdKV@gLabels@iLF] {} (72:0.5);
    \draw[label distance=0,\cmdKV@gTypes@iG] (144:0.5) -- node[label=(\cmdKV@general@rotate):\cmdKV@gLabels@iLG] {} (-144:0.5);
    #2
  \end{tikzpicture}
  \gSetStdTypes{gluon}
}
\def\gTriTri{\@ifnextchar[\@gTriTri{\@gTriTri[]}}
\def\@gTriTri[#1]#2{%
  \setkeys*{pre}{#1}%
  \setrmkeys{general,gTypes,gLabels}%
  \begin{tikzpicture}
    [line width=1pt,
    baseline={([yshift=-0.5ex+\cmdKV@general@yshift]current bounding box.center)},
    scale=\cmdKV@general@scale,
    rotate=\cmdKV@general@rotate,
    font=\cmdKV@general@font,
    yscale=\cmdKV@general@yscale]
    \draw[\cmdKV@gTypes@eA] (1.9,1.1) -- (2.2,1.4) node[label=(\cmdKV@general@rotate):\cmdKV@gLabels@eLA] {};
    \draw[\cmdKV@gTypes@eB] (1.9,0.3) -- (2.2,0) node[label=(\cmdKV@general@rotate):\cmdKV@gLabels@eLB] {};
    \draw[\cmdKV@gTypes@eC] (0.3,0.3) -- (0,0) node[label=(-180+\cmdKV@general@rotate):\cmdKV@gLabels@eLC] {};
    \draw[\cmdKV@gTypes@eD] (0.3,1.1) -- (0,1.4) node[label=(180+\cmdKV@general@rotate):\cmdKV@gLabels@eLD] {};
    \draw[label distance=0,\cmdKV@gTypes@iA] (1.9,1.1) -- node[label=(\cmdKV@general@rotate):\cmdKV@gLabels@iLA] {} (1.9,0.3);
    \draw[label distance=0,\cmdKV@gTypes@iB] (1.9,0.3) -- node[label=(-135+\cmdKV@general@rotate):\cmdKV@gLabels@iLB] {} (1.3,0.7);
    \draw[label distance=0,\cmdKV@gTypes@iC] (1.3,0.7) -- node[label=(135+\cmdKV@general@rotate):\cmdKV@gLabels@iLC] {} (1.9,1.1);
    \draw[label distance=0,\cmdKV@gTypes@iD] (1.3,0.7) -- node[label=(90+\cmdKV@general@rotate):\cmdKV@gLabels@iLD] {} (0.9,0.7);
    \draw[label distance=0,\cmdKV@gTypes@iE] (0.9,0.7) -- node[label=(-45+\cmdKV@general@rotate):\cmdKV@gLabels@iLE] {} (0.3,0.3);
    \draw[label distance=0,\cmdKV@gTypes@iF] (0.3,0.3) -- node[label=(180+\cmdKV@general@rotate):\cmdKV@gLabels@iLF] {} (0.3,1.1);
    \draw[label distance=0,\cmdKV@gTypes@iG] (0.3,1.1) -- node[label=(45+\cmdKV@general@rotate):\cmdKV@gLabels@iLG] {} (0.9,0.7);
    #2
  \end{tikzpicture}
  \gSetStdTypes{gluon}
}

\def\gBoxBubA{\@ifnextchar[\@gBoxBubA{\@gBoxBubA[]}}
\def\@gBoxBubA[#1]#2{%
  \setkeys*{pre}{#1}%
  \setrmkeys{general,gTypes,gLabels}%
  \begin{tikzpicture}
    [line width=1pt,
    baseline={([yshift=-0.5ex+\cmdKV@general@yshift]current bounding box.center)},
    scale=\cmdKV@general@scale,
    rotate=\cmdKV@general@rotate,
    font=\cmdKV@general@font,
    yscale=\cmdKV@general@yscale]
    \draw[\cmdKV@gTypes@eA] (0.9,0.9) -- (1.1,1.1) node[label=(\cmdKV@general@rotate):\cmdKV@gLabels@eLA] {};
    \draw[\cmdKV@gTypes@eB] (0.9,0.1) -- (1.1,-0.1) node[label=(\cmdKV@general@rotate):\cmdKV@gLabels@eLB] {};
    \draw[\cmdKV@gTypes@eC] (0.1,0.1) -- (-0.1,-0.1) node[label=(-180+\cmdKV@general@rotate):\cmdKV@gLabels@eLC] {};
    \draw[\cmdKV@gTypes@eD] (0.1,0.9) -- (-0.1,1.1) node[label=(180+\cmdKV@general@rotate):\cmdKV@gLabels@eLD] {};
    \draw[label distance=0,\cmdKV@gTypes@iA] (0.9,0.9) -- node[label=(\cmdKV@general@rotate):\cmdKV@gLabels@iLA] {} (0.9,0.1);
    \draw[label distance=0,\cmdKV@gTypes@iB] (0.9,0.1) -- node[label=(-90+\cmdKV@general@rotate):\cmdKV@gLabels@iLB] {} (0.1,0.1);
    \draw[label distance=0,\cmdKV@gTypes@iC] (0.1,0.1) -- node[label=(180+\cmdKV@general@rotate):\cmdKV@gLabels@iLC] {} (0.1,0.9);
    \draw[label distance=0,\cmdKV@gTypes@iD] (0.1,0.9) -- node[label=(90+\cmdKV@general@rotate):\cmdKV@gLabels@iLD] {} (0.32,0.9);
    \draw[label distance=0,\cmdKV@gTypes@iE] (0.32,0.9) .. controls (0.32,1.15) and (0.68,1.15) .. node[label=(90+\cmdKV@general@rotate):\cmdKV@gLabels@iLE] {} (0.68,0.9);
    \draw[label distance=0,\cmdKV@gTypes@iF] (0.68,0.9) .. controls (0.68,0.65) and (0.32,0.65) .. node[label=(-90+\cmdKV@general@rotate):\cmdKV@gLabels@iLF] {} (0.32,0.9);
    \draw[label distance=0,\cmdKV@gTypes@iG] (0.68,0.9) -- node[label=(90+\cmdKV@general@rotate):\cmdKV@gLabels@iLG] {} (0.9,0.9);
    #2
  \end{tikzpicture}
  \gSetStdTypes{gluon}
}

\def\gBoxBubB{\@ifnextchar[\@gBoxBubB{\@gBoxBubB[]}}
\def\@gBoxBubB[#1]#2{%
  \setkeys*{pre}{#1}%
  \setrmkeys{general,gTypes,gLabels}%
  \begin{tikzpicture}
    [line width=1pt,
    baseline={([yshift=-0.5ex+\cmdKV@general@yshift]current bounding box.center)},
    scale=\cmdKV@general@scale,
    rotate=\cmdKV@general@rotate,
    font=\cmdKV@general@font,
    yscale=\cmdKV@general@yscale]
    \draw[label distance=-1,\cmdKV@gTypes@eA] (0,0.4) -- (0,0.7) node[label=(\cmdKV@general@rotate):\cmdKV@gLabels@eLA] {};
    \draw[\cmdKV@gTypes@eB] (0.4,0) -- (0.7,0) node[label=(\cmdKV@general@rotate):\cmdKV@gLabels@eLB] {};
    \draw[label distance=-1,\cmdKV@gTypes@eC] (0,-0.4) -- (0,-0.7) node[label=(\cmdKV@general@rotate):\cmdKV@gLabels@eLC] {};
    \draw[\cmdKV@gTypes@eD] (-1.1,0) -- (-1.4,0) node[label=(180+\cmdKV@general@rotate):\cmdKV@gLabels@eLD] {};
    \draw[label distance=0,\cmdKV@gTypes@iA] (0,0.4) -- node[label=(45+\cmdKV@general@rotate):\cmdKV@gLabels@iLA] {} (0.4,0);
    \draw[label distance=0,\cmdKV@gTypes@iB] (0.4,0) -- node[label=(-45+\cmdKV@general@rotate):\cmdKV@gLabels@iLB] {} (0,-0.4);
    \draw[label distance=0,\cmdKV@gTypes@iC] (0,-0.4) -- node[label=(-135+\cmdKV@general@rotate):\cmdKV@gLabels@iLC] {} (-0.4,0);
    \draw[label distance=0,\cmdKV@gTypes@iD] (-0.4,0) -- node[label=(135+\cmdKV@general@rotate):\cmdKV@gLabels@iLD] {} (0,0.4);
    \draw[label distance=0,\cmdKV@gTypes@iE] (-0.4,0) --  node[label=(90+\cmdKV@general@rotate):\cmdKV@gLabels@iLE] {} (-0.7,0);
    \draw[label distance=0,\cmdKV@gTypes@iF] (-0.7,0) .. controls (-0.7,-0.26) and (-1.1,-0.26) .. node[label=(-90+\cmdKV@general@rotate):\cmdKV@gLabels@iLF] {} (-1.1,0);
    \draw[label distance=0,\cmdKV@gTypes@iG] (-1.1,0) .. controls (-1.1,0.26) and (-0.7,0.26) ..  node[label=(90+\cmdKV@general@rotate):\cmdKV@gLabels@iLG] {} (-0.7,0);
    #2
  \end{tikzpicture}
  \gSetStdTypes{gluon}
}

\def\gBoxBubC{\@ifnextchar[\@gBoxBubC{\@gBoxBubC[]}}
\def\@gBoxBubC[#1]#2{%
  \setkeys*{pre}{#1}%
  \setrmkeys{general,gTypes,gLabels}%
  \begin{tikzpicture}
    [line width=1pt,
    baseline={([yshift=-0.5ex+\cmdKV@general@yshift]current bounding box.center)},
    scale=\cmdKV@general@scale,
    rotate=\cmdKV@general@rotate,
    font=\cmdKV@general@font,
    yscale=\cmdKV@general@yscale]
    \draw[\cmdKV@gTypes@eA] (0.9,0.9) -- (1.1,1.1) node[label=(\cmdKV@general@rotate):\cmdKV@gLabels@eLA] {};
    \draw[\cmdKV@gTypes@eB] (0.9,0.1) -- (1.1,-0.1) node[label=(\cmdKV@general@rotate):\cmdKV@gLabels@eLB] {};
    \draw[\cmdKV@gTypes@eC] (0.1,0.1) -- (-0.1,-0.1) node[label=(-180+\cmdKV@general@rotate):\cmdKV@gLabels@eLC] {};
    \draw[\cmdKV@gTypes@eD] (0.1,0.9) -- (-0.1,1.1) node[label=(180+\cmdKV@general@rotate):\cmdKV@gLabels@eLD] {};
    \draw[label distance=0,\cmdKV@gTypes@iA] (0.9,0.9) -- node[label=(\cmdKV@general@rotate):\cmdKV@gLabels@iLA] {} (0.9,0.1);
    \draw[label distance=0,\cmdKV@gTypes@iB] (0.9,0.1) -- node[label=(-90+\cmdKV@general@rotate):\cmdKV@gLabels@iLB] {} (0.1,0.1);
    \draw[label distance=0,\cmdKV@gTypes@iC] (0.1,0.1) -- node[label=(180+\cmdKV@general@rotate):\cmdKV@gLabels@iLC] {} (0.1,0.9);
    \draw[label distance=0,\cmdKV@gTypes@iD] (0.1,0.9) -- node[label=(90+\cmdKV@general@rotate):\cmdKV@gLabels@iLD] {} (0.5,0.9);
    \draw[label distance=0,\cmdKV@gTypes@iE] (0.5,0.9) .. controls (0.5,1.15) and (0.9,1.15) .. node[label=(90+\cmdKV@general@rotate):\cmdKV@gLabels@iLE] {} (0.9,0.9);
    \draw[label distance=0,\cmdKV@gTypes@iF] (0.9,0.9) .. controls (0.9,0.65) and (0.5,0.65) .. node[label=(-90+\cmdKV@general@rotate):\cmdKV@gLabels@iLF] {} (0.5,0.9);
    #2
  \end{tikzpicture}
  \gSetStdTypes{gluon}
}

\def\gPentaTad{\@ifnextchar[\@gPentaTad{\@gPentaTad[]}}
\def\@gPentaTad[#1]#2{%
  \setkeys*{pre}{#1}%
  \setrmkeys{general,gTypes,gLabels}%
  \begin{tikzpicture}
    [line width=1pt,
    baseline={([yshift=-0.5ex+\cmdKV@general@yshift]current bounding box.center)},
    scale=\cmdKV@general@scale,
    rotate=\cmdKV@general@rotate,
    font=\cmdKV@general@font,
    yscale=\cmdKV@general@yscale]
    \draw[label distance=-0.5,\cmdKV@gTypes@eA] (108:0.35) -- (108:0.75) node[label=(180+\cmdKV@general@rotate):\cmdKV@gLabels@eLA] {};
    \draw[\cmdKV@gTypes@eB] (36:0.35) -- (36:0.75) node[label=(36+\cmdKV@general@rotate):\cmdKV@gLabels@eLB] {};
    \draw[\cmdKV@gTypes@eC] (-36:0.35) -- (-36:0.75) node[label=(-36+\cmdKV@general@rotate):\cmdKV@gLabels@eLC] {};
    \draw[label distance=-0.5,\cmdKV@gTypes@eD] (-108:0.35) -- (-108:0.75) node[label=(-180+\cmdKV@general@rotate):\cmdKV@gLabels@eLD] {};
    \draw[label distance=0,\cmdKV@gTypes@iA] (108:0.35) -- node[label=(72+\cmdKV@general@rotate):\cmdKV@gLabels@iLA] {} (36:0.35);
    \draw[label distance=0,\cmdKV@gTypes@iB] (36:0.35) -- node[label=(\cmdKV@general@rotate):\cmdKV@gLabels@iLB] {} (-36:0.35);
    \draw[label distance=0,\cmdKV@gTypes@iC] (-36:0.35) -- node[label=(-72+\cmdKV@general@rotate):\cmdKV@gLabels@iLC] {} (-108:0.35);
    \draw[label distance=0,\cmdKV@gTypes@iD] (-108:0.35) -- node[label=(-144+\cmdKV@general@rotate):\cmdKV@gLabels@iLD] {} (-180:0.35);
    \draw[label distance=0,\cmdKV@gTypes@iE] (180:0.35) -- node[label=(144+\cmdKV@general@rotate):\cmdKV@gLabels@iLE] {} (108:0.35);
    \draw[label distance=0,\cmdKV@gTypes@iF] (180:0.35) -- node[label=(90+\cmdKV@general@rotate):\cmdKV@gLabels@iLF] {} (180:0.75);
    \draw[label distance=-0.5,\cmdKV@gTypes@iG] (180:0.75) .. controls ($(180:0.75)+(0,-0.18)$) and ($(180:1)+(0,-0.18)$) .. (180:1) node[label=(180+\cmdKV@general@rotate):\cmdKV@gLabels@iLG] {} .. controls ($(180:1)+(0,0.18)$) and ($(180:0.75)+(0,0.18)$) .. (180:0.75);
    #2
  \end{tikzpicture}
  \gSetStdTypes{gluon}
}

\def\gBoxTadA{\@ifnextchar[\@gBoxTadA{\@gBoxTadA[]}}
\def\@gBoxTadA[#1]#2{%
  \setkeys*{pre}{#1}%
  \setrmkeys{general,gTypes,gLabels}%
  \begin{tikzpicture}
    [line width=1pt,
    baseline={([yshift=-0.5ex+\cmdKV@general@yshift]current bounding box.center)},
    scale=\cmdKV@general@scale,
    rotate=\cmdKV@general@rotate,
    font=\cmdKV@general@font,
    yscale=\cmdKV@general@yscale]
    \draw[label distance=-0.5,\cmdKV@gTypes@eA] (0,0.4) -- (0,0.7) node[label=(\cmdKV@general@rotate):\cmdKV@gLabels@eLA] {};
    \draw[\cmdKV@gTypes@eB] (0.4,0) -- (0.7,0) node[label=(\cmdKV@general@rotate):\cmdKV@gLabels@eLB] {};
    \draw[label distance=-0.5,\cmdKV@gTypes@eC] (0,-0.4) -- (0,-0.7) node[label=(\cmdKV@general@rotate):\cmdKV@gLabels@eLC] {};
    \draw[\cmdKV@gTypes@eD] (-0.7,0) -- (-0.7,-0.3) node[label=(-90+\cmdKV@general@rotate):\cmdKV@gLabels@eLD] {};
    \draw[label distance=0,\cmdKV@gTypes@iA] (0,0.4) -- node[label=(45+\cmdKV@general@rotate):\cmdKV@gLabels@iLA] {} (0.4,0);
    \draw[label distance=0,\cmdKV@gTypes@iB] (0.4,0) -- node[label=(-45+\cmdKV@general@rotate):\cmdKV@gLabels@iLB] {} (0,-0.4);
    \draw[label distance=0,\cmdKV@gTypes@iC] (0,-0.4) -- node[label=(-135+\cmdKV@general@rotate):\cmdKV@gLabels@iLC] {} (-0.4,0);
    \draw[label distance=0,\cmdKV@gTypes@iD] (-0.4,0) -- node[label=(135+\cmdKV@general@rotate):\cmdKV@gLabels@iLD] {} (0,0.4);
    \draw[label distance=0,\cmdKV@gTypes@iE] (-0.4,0) --  node[label=(90+\cmdKV@general@rotate):\cmdKV@gLabels@iLE] {} (-0.7,0);
    \draw[label distance=0,\cmdKV@gTypes@iF] (-0.7,0) -- node[label=(90+\cmdKV@general@rotate):\cmdKV@gLabels@iLF] {} (-1.1,0);
    \draw[label distance=-0.5,\cmdKV@gTypes@iG] (-1.1,0) .. controls (-1.1,-0.18) and (-1.35,-0.18) .. (-1.35,0) node[label=(180+\cmdKV@general@rotate):\cmdKV@gLabels@iLG] {} .. controls (-1.35,0.18) and (-1.1,0.18) .. (-1.1,0);
    #2
  \end{tikzpicture}
  \gSetStdTypes{gluon}
}

\def\gTriBubA{\@ifnextchar[\@gTriBubA{\@gTriBubA[]}}
\def\@gTriBubA[#1]#2{%
  \setkeys*{pre}{#1}%
  \setrmkeys{general,gTypes,gLabels}%
  \begin{tikzpicture}
    [line width=1pt,
    baseline={([yshift=-0.5ex+\cmdKV@general@yshift]current bounding box.center)},
    scale=\cmdKV@general@scale,
    rotate=\cmdKV@general@rotate,
    font=\cmdKV@general@font,
    yscale=\cmdKV@general@yscale]
    \draw[\cmdKV@gTypes@eA] (0:0.3) -- (0:0.55) node[label=(\cmdKV@general@rotate):\cmdKV@gLabels@eLA] {};
    \draw[\cmdKV@gTypes@eB] (-120:0.3) -- (-120:0.55) node[label=(-180+\cmdKV@general@rotate):\cmdKV@gLabels@eLB] {};
    \draw[\cmdKV@gTypes@eC] (120:0.3) -- (120:0.55) node[label=(180+\cmdKV@general@rotate):\cmdKV@gLabels@eLC] {};
    \draw[label distance=0,\cmdKV@gTypes@iA] (0:0.3) -- node[label=(-60+\cmdKV@general@rotate):\cmdKV@gLabels@iLA] {} (-120:0.3);
    \draw[label distance=0,\cmdKV@gTypes@iD] (180:0.15) to[bend right=70] (-120:0.3);
    \draw[label distance=0,\cmdKV@gTypes@iE] (180:0.15) to[bend left=70] (-120:0.3);
    \draw[label distance=0,\cmdKV@gTypes@iB] (180:0.15) -- node[label=(180+\cmdKV@general@rotate):\cmdKV@gLabels@iLB] {} (120:0.3);
    \draw[label distance=0,\cmdKV@gTypes@iC] (120:0.3) -- node[label=(60+\cmdKV@general@rotate):\cmdKV@gLabels@iLC] {} (0:0.3);
    #2
  \end{tikzpicture}
  \gSetStdTypes{gluon}
}

\def\gTriBubB{\@ifnextchar[\@gTriBubB{\@gTriBubB[]}}
\def\@gTriBubB[#1]#2{%
  \setkeys*{pre}{#1}%
  \setrmkeys{general,gTypes,gLabels}%
  \begin{tikzpicture}
    [line width=1pt,
    baseline={([yshift=-0.5ex+\cmdKV@general@yshift]current bounding box.center)},
    scale=\cmdKV@general@scale,
    rotate=\cmdKV@general@rotate,
    font=\cmdKV@general@font,
    yscale=\cmdKV@general@yscale]
    \draw[\cmdKV@gTypes@eA] (0:0.3) -- (0:0.55) node[label=(\cmdKV@general@rotate):\cmdKV@gLabels@eLA] {};
    \draw[\cmdKV@gTypes@eB] (-120:0.3) -- (-120:0.55) node[label=(-180+\cmdKV@general@rotate):\cmdKV@gLabels@eLB] {};
    \draw[\cmdKV@gTypes@eC] (120:0.3) -- (120:0.55) node[label=(180+\cmdKV@general@rotate):\cmdKV@gLabels@eLC] {};
    \draw[label distance=0,\cmdKV@gTypes@iA] (0:0.3) -- node[label=(-60+\cmdKV@general@rotate):\cmdKV@gLabels@iLA] {} (-60:0.15);
    \draw[label distance=0,\cmdKV@gTypes@iD] (-120:0.3) to[bend right=70] (-60:0.15);
    \draw[label distance=0,\cmdKV@gTypes@iE] (-120:0.3) to[bend left=70] (-60:0.15);
    \draw[label distance=0,\cmdKV@gTypes@iB] (-120:0.3) -- node[label=(180+\cmdKV@general@rotate):\cmdKV@gLabels@iLB] {} (120:0.3);
    \draw[label distance=0,\cmdKV@gTypes@iC] (120:0.3) -- node[label=(60+\cmdKV@general@rotate):\cmdKV@gLabels@iLC] {} (0:0.3);
    
    #2
  \end{tikzpicture}
  \gSetStdTypes{gluon}
}

\def\gBoxBoxBoxA{\@ifnextchar[\@gBoxBoxBoxA{\@gBoxBoxBoxA[]}}
\def\@gBoxBoxBoxA[#1]#2{%
  \setkeys*{pre}{#1}%
  \setrmkeys{general,gTypes,gLabels}%
  \begin{tikzpicture}
    [line width=1pt,
    baseline={([yshift=-0.5ex+\cmdKV@general@yshift]current bounding box.center)},
    scale=\cmdKV@general@scale,
    rotate=\cmdKV@general@rotate,
    font=\cmdKV@general@font,
    yscale=\cmdKV@general@yscale]
    \draw[\cmdKV@gTypes@eA] (1.5,0.9) -- (1.8,1.2) node[label=(\cmdKV@general@rotate):\cmdKV@gLabels@eLA] {};
    \draw[\cmdKV@gTypes@eB] (1.5,0.3) -- (1.8,0) node[label=(\cmdKV@general@rotate):\cmdKV@gLabels@eLB] {};
    \draw[\cmdKV@gTypes@eC] (-0.3,0.3) -- (-0.6,0) node[label=(-180+\cmdKV@general@rotate):\cmdKV@gLabels@eLC] {};
    \draw[\cmdKV@gTypes@eD] (-0.3,0.9) -- (-0.6,1.2) node[label=(180+\cmdKV@general@rotate):\cmdKV@gLabels@eLD] {};
    \draw[label distance=0,\cmdKV@gTypes@iA] (1.5,0.9) -- node[label=(\cmdKV@general@rotate):\cmdKV@gLabels@iLA] {} (1.5,0.3);
    \draw[label distance=0,\cmdKV@gTypes@iB] (1.5,0.3) -- node[label=(-90+\cmdKV@general@rotate):\cmdKV@gLabels@iLB] {} (0.9,0.3);
    \draw[label distance=0,\cmdKV@gTypes@iC] (0.9,0.3) -- node[label=(-90+\cmdKV@general@rotate):\cmdKV@gLabels@iLC] {} (0.3,0.3);
    \draw[label distance=0,\cmdKV@gTypes@iD] (0.3,0.3) -- node[label=(-90+\cmdKV@general@rotate):\cmdKV@gLabels@iLD] {} (-0.3,0.3);
    \draw[label distance=0,\cmdKV@gTypes@iE] (-0.3,0.3) -- node[label=(180+\cmdKV@general@rotate):\cmdKV@gLabels@iLE] {} (-0.3,0.9);
    \draw[label distance=0,\cmdKV@gTypes@iF] (-0.3,0.9) -- node[label=(90+\cmdKV@general@rotate):\cmdKV@gLabels@iLF] {} (0.3,0.9);
    \draw[label distance=0,\cmdKV@gTypes@iG] (0.3,0.9) -- node[label=(90+\cmdKV@general@rotate):\cmdKV@gLabels@iLG] {} (0.9,0.9);
    \draw[label distance=0,\cmdKV@gTypes@iH] (0.9,0.9) -- node[label=(90+\cmdKV@general@rotate):\cmdKV@gLabels@iLH] {} (1.5,0.9);
    \draw[label distance=0,\cmdKV@gTypes@iI] (0.9,0.9) -- node[label=(\cmdKV@general@rotate):\cmdKV@gLabels@iLI] {} (0.9,0.3);
    \draw[label distance=0,\cmdKV@gTypes@iJ] (0.3,0.9) -- node[label=(\cmdKV@general@rotate):\cmdKV@gLabels@iLJ] {} (0.3,0.3);
    #2
  \end{tikzpicture}
  \gSetStdTypes{gluon}
}

\def\gBoxBoxBoxB{\@ifnextchar[\@gBoxBoxBoxB{\@gBoxBoxBoxB[]}}
\def\@gBoxBoxBoxB[#1]#2{%
  \setkeys*{pre}{#1}%
  \setrmkeys{general,gTypes,gLabels}%
  \begin{tikzpicture}
    [line width=1pt,
    baseline={([yshift=-0.5ex+\cmdKV@general@yshift]current bounding box.center)},
    scale=\cmdKV@general@scale,
    rotate=\cmdKV@general@rotate,
    font=\cmdKV@general@font,
    yscale=\cmdKV@general@yscale]
    \draw[\cmdKV@gTypes@eA] (1.5,1.5) -- (1.8,1.8) node[label=(\cmdKV@general@rotate):\cmdKV@gLabels@eLA] {};
    \draw[\cmdKV@gTypes@eB] (1.5,0.3) -- (1.8,0) node[label=(\cmdKV@general@rotate):\cmdKV@gLabels@eLB] {};
    \draw[\cmdKV@gTypes@eC] (0.3,0.3) -- (0,0) node[label=(-180+\cmdKV@general@rotate):\cmdKV@gLabels@eLC] {};
    \draw[\cmdKV@gTypes@eD] (0.3,1.5) -- (0,1.8) node[label=(180+\cmdKV@general@rotate):\cmdKV@gLabels@eLD] {};
    \draw[label distance=0,\cmdKV@gTypes@iA] (1.5,1.5) -- node[label=(\cmdKV@general@rotate):\cmdKV@gLabels@iLA] {} (1.5,0.3);
    \draw[label distance=0,\cmdKV@gTypes@iB] (1.5,0.3) -- node[label=(-90+\cmdKV@general@rotate):\cmdKV@gLabels@iLB] {} (0.9,0.3);
    \draw[label distance=0,\cmdKV@gTypes@iC] (0.9,0.3) -- node[label=(-90+\cmdKV@general@rotate):\cmdKV@gLabels@iLC] {} (0.3,0.3);
    \draw[label distance=0,\cmdKV@gTypes@iD] (0.3,0.3) -- node[label=(180+\cmdKV@general@rotate):\cmdKV@gLabels@iLD] {} (0.3,0.9);
    \draw[label distance=0,\cmdKV@gTypes@iE] (0.3,0.9) -- node[label=(180+\cmdKV@general@rotate):\cmdKV@gLabels@iLE] {} (0.3,1.5);
    \draw[label distance=0,\cmdKV@gTypes@iF] (0.3,1.5) -- node[label=(90+\cmdKV@general@rotate):\cmdKV@gLabels@iLF] {} (0.9,1.5);
    \draw[label distance=0,\cmdKV@gTypes@iG] (0.9,1.5) -- node[label=(90+\cmdKV@general@rotate):\cmdKV@gLabels@iLG] {} (1.5,1.5);
    \draw[label distance=0,\cmdKV@gTypes@iH] (0.9,1.5) -- node[label=(\cmdKV@general@rotate):\cmdKV@gLabels@iLH] {} (0.9,0.9);
    \draw[label distance=0,\cmdKV@gTypes@iI] (0.9,0.9) -- node[label=(\cmdKV@general@rotate):\cmdKV@gLabels@iLI] {} (0.9,0.3);
    \draw[label distance=0,\cmdKV@gTypes@iJ] (0.9,0.9) -- node[label=(90+\cmdKV@general@rotate):\cmdKV@gLabels@iLJ] {} (0.3,0.9);
    #2
  \end{tikzpicture}
  \gSetStdTypes{gluon}
}

\def\cBoxA{\@ifnextchar[\@cBoxA{\@cBoxA[]}}
\def\@cBoxA[#1]#2{%
  \setkeys*{pre}{#1}%
  \setrmkeys{general,gTypes,gLabels}%
  \begin{tikzpicture}
    [line width=1pt,
    baseline={([yshift=-0.5ex+\cmdKV@general@yshift]current bounding box.center)},
    scale=\cmdKV@general@scale,
    rotate=\cmdKV@general@rotate,
    font=\cmdKV@general@font,
    yscale=\cmdKV@general@yscale]
    \fill[blob] (0.4,0) ellipse (0.2 and 0.4);
    \fill[blob] (-0.4,0) ellipse (0.2 and 0.4);
    \draw[\cmdKV@gTypes@eA] ($ (0.4,0) + (45:0.26) $) -- ($ (0.4,0) + (45:0.7) $) node[label=(\cmdKV@general@rotate):\cmdKV@gLabels@eLA] {};
    \draw[\cmdKV@gTypes@eB] ($ (0.4,0) + (-45:0.26) $) -- ($ (0.4,0) + (-45:0.7) $) node[label=(\cmdKV@general@rotate):\cmdKV@gLabels@eLB] {};
    \draw[\cmdKV@gTypes@eC] ($ (-0.4,0) + (-135:0.26) $) -- ($ (-0.4,0) + (-135:0.7) $) node[label=(-180+\cmdKV@general@rotate):\cmdKV@gLabels@eLC] {};
    \draw[\cmdKV@gTypes@eD] ($ (-0.4,0) + (135:0.26) $) -- ($ (-0.4,0) + (135:0.7) $) node[label=(180+\cmdKV@general@rotate):\cmdKV@gLabels@eLD] {};
    \draw[label distance=1,\cmdKV@gTypes@iA] ($ (0.4,0) + (-135:0.26) $) -- node[label=(-90+\cmdKV@general@rotate):\cmdKV@gLabels@iLA] {}($ (-0.4,0) + (-45:0.26) $);
    \draw[label distance=1,\cmdKV@gTypes@iB] ($ (-0.4,0) + (45:0.26) $) -- node[label=(90+\cmdKV@general@rotate):\cmdKV@gLabels@iLB] {}($ (0.4,0) + (135:0.26) $);
    \node[label=(\cmdKV@general@rotate):\cmdKV@gLabels@iLC] at (0.6,0) {};
    \node[label=(180+\cmdKV@general@rotate):\cmdKV@gLabels@iLD] at (-0.6,0) {};
    #2
  \end{tikzpicture}
  \gSetStdTypes{gluon}
}

\def\cBoxB{\@ifnextchar[\@cBoxB{\@cBoxB[]}}
\def\@cBoxB[#1]#2{%
  \setkeys*{pre}{#1}%
  \setrmkeys{general,gTypes,gLabels}%
  \begin{tikzpicture}
    [line width=1pt,
    baseline={([yshift=-0.5ex+\cmdKV@general@yshift]current bounding box.center)},
    scale=\cmdKV@general@scale,
    rotate=\cmdKV@general@rotate,
    font=\cmdKV@general@font,
    yscale=\cmdKV@general@yscale]
    \fill[blob] (0,0.3) ellipse (0.4 and 0.2);
    \fill[blob] (0,-0.3) ellipse (0.4 and 0.2);
    \draw[\cmdKV@gTypes@eA] ($ (0,0.3) + (0:0.4) $) -- ($ (0,0.3) + (10:0.7) $) node[label=(\cmdKV@general@rotate):\cmdKV@gLabels@eLA] {};
    \draw[\cmdKV@gTypes@eB] ($ (0,-0.3) + (0:0.4) $) -- ($ (0,-0.3) + (-10:0.7) $) node[label=(\cmdKV@general@rotate):\cmdKV@gLabels@eLB] {};
    \draw[\cmdKV@gTypes@eC] ($ (0,-0.3) + (180:0.4) $) -- ($ (0,-0.3) + (-170:0.7) $) node[label=(180+\cmdKV@general@rotate):\cmdKV@gLabels@eLC] {};
    \draw[\cmdKV@gTypes@eD] ($ (0,0.3) + (180:0.4) $) -- ($ (0,0.3) + (170:0.7) $) node[label=(180+\cmdKV@general@rotate):\cmdKV@gLabels@eLD] {};
    \draw[label distance=1,\cmdKV@gTypes@iA] ($ (0,0.3) + (-45:0.26) $) -- node[label=(\cmdKV@general@rotate):\cmdKV@gLabels@iLA] {}($ (0,-0.3) + (45:0.26) $);
    \draw[label distance=1,\cmdKV@gTypes@iA] ($ (0,-0.3) + (135:0.26) $) -- node[label=(-180+\cmdKV@general@rotate):\cmdKV@gLabels@iLB] {}($ (0,0.3) + (-135:0.26) $);
    \node[label=(-90+\cmdKV@general@rotate):\cmdKV@gLabels@iLC] at (0,-0.5) {};
    \node[label=(90+\cmdKV@general@rotate):\cmdKV@gLabels@iLD] at (0,0.5) {};
    #2
  \end{tikzpicture}
  \gSetStdTypes{gluon}
}

\def\cBoxBoxA{\@ifnextchar[\@cBoxBoxA{\@cBoxBoxA[]}}
\def\@cBoxBoxA[#1]#2{%
  \setkeys*{pre}{#1}%
  \setrmkeys{general,gTypes,gLabels}%
  \begin{tikzpicture}
    [line width=1pt,
    baseline={([yshift=-0.5ex+\cmdKV@general@yshift]current bounding box.center)},
    scale=\cmdKV@general@scale,
    rotate=\cmdKV@general@rotate,
    font=\cmdKV@general@font,
    yscale=\cmdKV@general@yscale]
    \fill[blob] (0.4,0) ellipse (0.2 and 0.4);
    \fill[blob] (-0.4,0) ellipse (0.2 and 0.4);
    \fill[blob] (1.2,0) ellipse (0.2 and 0.4);
    \draw[\cmdKV@gTypes@eA] ($ (1.2,0) + (45:0.26) $) -- ($ (1.2,0) + (45:0.7) $) node[label=(\cmdKV@general@rotate):\cmdKV@gLabels@eLA] {};
    \draw[\cmdKV@gTypes@eB] ($ (1.2,0) + (-45:0.26) $) -- ($ (1.2,0) + (-45:0.7) $) node[label=(\cmdKV@general@rotate):\cmdKV@gLabels@eLB] {};
    \draw[\cmdKV@gTypes@eC] ($ (-0.4,0) + (-135:0.26) $) -- ($ (-0.4,0) + (-135:0.7) $) node[label=(-180+\cmdKV@general@rotate):\cmdKV@gLabels@eLC] {};
    \draw[\cmdKV@gTypes@eD] ($ (-0.4,0) + (135:0.26) $) -- ($ (-0.4,0) + (135:0.7) $) node[label=(180+\cmdKV@general@rotate):\cmdKV@gLabels@eLD] {};
    \draw[label distance=1,\cmdKV@gTypes@iA] ($ (1.2,0) + (-135:0.26) $) -- node[label=(-90+\cmdKV@general@rotate):\cmdKV@gLabels@iLA] {}($ (0.4,0) + (-45:0.26) $);
    \draw[label distance=1,\cmdKV@gTypes@iB] ($ (0.4,0) + (-135:0.26) $) -- node[label=(-90+\cmdKV@general@rotate):\cmdKV@gLabels@iLB] {}($ (-0.4,0) + (-45:0.26) $);
    \draw[label distance=1,\cmdKV@gTypes@iC] ($ (-0.4,0) + (45:0.26) $) -- node[label=(90+\cmdKV@general@rotate):\cmdKV@gLabels@iLC] {}($ (0.4,0) + (135:0.26) $);
    \draw[label distance=1,\cmdKV@gTypes@iD] ($ (0.4,0) + (45:0.26) $) -- node[label=(90+\cmdKV@general@rotate):\cmdKV@gLabels@iLD] {}($ (1.2,0) + (135:0.26) $);
    \node[label=(\cmdKV@general@rotate):\cmdKV@gLabels@iLE] at (1.4,0) {};
    \node[label=(\cmdKV@general@rotate):\cmdKV@gLabels@iLF] at (0.6,0) {};
    \node[label=(180+\cmdKV@general@rotate):\cmdKV@gLabels@iLG] at (-0.6,0) {};
    #2
  \end{tikzpicture}
  \gSetStdTypes{gluon}
}

\def\cBoxBoxB{\@ifnextchar[\@cBoxBoxB{\@cBoxBoxB[]}}
\def\@cBoxBoxB[#1]#2{%
  \setkeys*{pre}{#1}%
  \setrmkeys{general,gTypes,gLabels}%
  \begin{tikzpicture}
    [line width=1pt,
    baseline={([yshift=-0.5ex+\cmdKV@general@yshift]current bounding box.center)},
    scale=\cmdKV@general@scale,
    rotate=\cmdKV@general@rotate,
    font=\cmdKV@general@font,
    yscale=\cmdKV@general@yscale]
    \fill[blob] (0,0.3) ellipse (0.4 and 0.2);
    \fill[blob] (0,-0.3) ellipse (0.4 and 0.2);
    \fill[blob] (1,0) ellipse (0.2 and 0.4);
    \draw[\cmdKV@gTypes@eA] ($ (1,0) + (45:0.26) $) -- ($ (1,0) + (45:0.7) $) node[label=(\cmdKV@general@rotate):\cmdKV@gLabels@eLA] {};
    \draw[\cmdKV@gTypes@eB] ($ (1,0) + (-45:0.26) $) -- ($ (1,0) + (-45:0.7) $) node[label=(\cmdKV@general@rotate):\cmdKV@gLabels@eLB] {};
    \draw[\cmdKV@gTypes@eC] ($ (0,-0.3) + (180:0.4) $) -- ($ (0,-0.3) + (-170:0.7) $) node[label=(180+\cmdKV@general@rotate):\cmdKV@gLabels@eLC] {};
    \draw[\cmdKV@gTypes@eD] ($ (0,0.3) + (180:0.4) $) -- ($ (0,0.3) + (170:0.7) $) node[label=(180+\cmdKV@general@rotate):\cmdKV@gLabels@eLD] {};
    \draw[label distance=1,\cmdKV@gTypes@iA] ($ (0,-0.3) + (0:0.4) $) -- node[label=(-90+\cmdKV@general@rotate):\cmdKV@gLabels@iLA] {} ($ (1,0) + (-135:0.26) $);
    \draw[label distance=1,\cmdKV@gTypes@iB] ($ (0,-0.3) + (135:0.26) $) -- node[label=(-180+\cmdKV@general@rotate):\cmdKV@gLabels@iLB] {}($ (0,0.3) + (-135:0.26) $);
    \draw[label distance=1,\cmdKV@gTypes@iC] ($ (0,0.3) + (0:0.4) $) -- node[label=(90+\cmdKV@general@rotate):\cmdKV@gLabels@iLC] {} ($ (1,0) + (135:0.26) $);
    \draw[label distance=1,\cmdKV@gTypes@iD] ($ (0,0.3) + (-45:0.26) $) -- node[label=(\cmdKV@general@rotate):\cmdKV@gLabels@iLD] {}($ (0,-0.3) + (45:0.26) $);
    \node[label=(\cmdKV@general@rotate):\cmdKV@gLabels@iLE] at (1.2,0) {};
    \node[label=(-90+\cmdKV@general@rotate):\cmdKV@gLabels@iLF] at (0,-0.5) {};
    \node[label=(90+\cmdKV@general@rotate):\cmdKV@gLabels@iLG] at (0,0.5) {};
    #2
  \end{tikzpicture}
  \gSetStdTypes{gluon}
}

\makeatother

\usepackage{tikzexternal}
\tikzsetexternalprefix{figures/}
\tikzsetfigurename{main-figure}

\def\Res{\mathop{\textrm{Res}}}

\def\braket#1{\langle #1 \rangle}

\def\la{\lambda}

\def\bT{\mathbf{T}}

\def\Res_#1{\operatorname*{Res}_{#1}}
\def\Tr{\operatorname*{tr}}
\def\trp{\operatorname*{tr}\,\!\!_{+}}
\def\trm{\operatorname*{tr}\,\!\!_{-}}
\def\trpm{\operatorname*{tr}\,\!\!_{\pm}}
\def\trmp{\operatorname*{tr}\,\!\!_{\mp}}
\def\tr5{\operatorname*{tr}\,\!\!_{5}}

\def\d{\mathrm{d}}

\def\ie{i.e. }
\def\eg{e.g. }
\def\etc{etc. }

\def\eqn#1{eq.~\eqref{#1}}
\def\eqns#1#2{eqs.~\eqref{#1} and~\eqref{#2}}

\def\Tab#1{Table~{\ref{#1}}}

\def\sec#1{section~{\ref{#1}}}
\def\secs#1#2{sections~{\ref{#1}} and~{\ref{#2}}}

\def\app#1{appendix~{\ref{#1}}}

\def\rcite#1{ref.~\cite{#1}}
\def\rcites#1{refs.~\cite{#1}}

\def\be{\begin{equation}}
\def\ee{\end{equation}}
\def\bea{\begin{eqnarray}}
\def\eea{\end{eqnarray}}
\def\beal{\begin{equation}\begin{aligned}}
\def\eeal{\end{aligned}\end{equation}}
\def\nn{\nonumber}

\def\bI{\mathrm{I}}

\def\cH{\mathcal{H}}

\def\cM{\mathcal{M}}
\def\cN{\mathcal{N}}
\def\cO{\mathcal{O}}
\def\cW{\mathcal{W}}
\def\cR{\mathcal{R}}
\def\cS{\mathcal{S}}
\def\cT{\mathcal{T}}
\def\cU{\mathcal{U}}
\def\tf{\tilde{f}}

\def\eps{\epsilon}

\usepackage{stackengine}
\newcommand{\oset}[3][0.3ex]{\stackon[#1]{$#3$}{$\scriptstyle#2$}}
\newcommand{\uset}[3][0.3ex]{\stackunder[#1]{$#3$}{$\scriptstyle#2$}}

\begin{document}
\preprint{UUITP-47/19 \\ \phantom{~} \hfill SLAC-PUB-17488}

\title{Infrared and transcendental structure of two-loop supersymmetric QCD amplitudes}

\author[a,b]{Gregor K\"alin,}
\author[a]{Gustav Mogull,}
\author[c]{Alexander Ochirov,}
\author[a]{and Bram Verbeek}
\affiliation[a]{Department of Physics and Astronomy, Uppsala University, \\
   Box 516, 75108 Uppsala, Sweden}
\affiliation[b]{SLAC National Accelerator Laboratory, Stanford University,
   Stanford, CA 94309, USA}
\affiliation[c]{ETH Z\"urich, Institut f\"ur Theoretische Physik, \\
   Wolfgang-Pauli-Str. 27, 8093 Z\"urich, Switzerland}
\emailAdd{gregor.kaelin@physics.uu.se, gustav.mogull@physics.uu.se, aochirov@phys.ethz.ch, bram.verbeek@physics.uu.se}

\abstract{
Using a careful choice of infrared (IR) subtraction scheme, we demonstrate cancellation of all terms with transcendental weights 0, 1, 2 from the finite part of the full-color two-loop four-gluon $\mathcal{N}=2$ supersymmetric QCD amplitude, with $N_f$ massless supersymmetric quarks. This generalizes the previously observed cancellation of weight-2 terms in the superconformal theory, where $N_f=2N_c$ for gauge group SU$(N_c)$. The subtraction scheme follows naturally both from general IR factorization principles and from an integrand-level analysis of divergences in this amplitude. The divergences are written in terms of scalar triangle integrals whose expressions are known to all orders in the dimensional regulator $\epsilon=(4-D)/2$. We also present integrated expressions for the full-color two-loop four-point amplitudes with both matter and vectors on external legs in which lower-weight terms also cancel using an appropriate IR scheme. This provides us with values for the two-loop cusp, gluonic, and quark anomalous dimensions in $\mathcal{N}=2$ supersymmetric QCD, which are cross-checked between the three different amplitudes.
}

\maketitle

\addtocontents{toc}{\protect\setcounter{tocdepth}{2}}

\section{Introduction}
\label{sec:intro}

Infrared (IR) divergences are a vital aspect of
the physics of scattering amplitudes in gauge theory,
and in recent years our understanding of both
has improved considerably
\cite{Gardi:2018arz,Elvang:2015rqa,Dixon:2011xs,Dixon:1996wi}.
In an amplitude with loop momenta $\{\ell_i\}$ there are two kinds of IR divergence:
soft, where $\ell_i\to0$, and collinear,
where $\ell_i\to\tau p_j$ for an external momentum $p_j$.
Most crucial has been the observation that for massless parton scattering,
in the fixed-angle limit
where all kinematic invariants $s_{ij}=2p_i\cdot p_j$ are large,
soft and collinear divergences factorize away from an IR-finite hard function
\cite{Akhoury:1978vq,Sen:1982bt}.
The divergences take an exponential form involving an anomalous dimension
\cite{Sterman:2002qn,Aybat:2006wq,Aybat:2006mz,Gardi:2009qi,Gardi:2009zv,Becher:2009cu,Becher:2009qa,Moult:2019uhz},
which at two loops justifies the form of the divergences
predicted decades ago by Catani~\cite{Catani:1998bh}.

Despite this progress in understanding the IR
behavior of multi-loop scattering amplitudes,
modern computational methods often obscure IR structure, leading to it reappearing
only after ultraviolet (UV) renormalization of final integrated results.
To bridge this gap, efforts have been made to make the IR behavior apparent
already at the integrand level. A notable example of this is the planar all-plus sector,
where the IR structure --- reduced in this case to the one-loop complexity ---
has been exploited to obtain compact
two-loop integrands~\cite{Badger:2016ozq,Badger:2016egz}
and amplitudes~\cite{Dunbar:2016aux,Dunbar:2016cxp,Dunbar:2016gjb,Dunbar:2017nfy,Dunbar:2019fcq}
at five and higher points.
In particular, the unrenormalized all-plus integrands were built
from loop variables tailored to control IR divergences in specific regions.

Another important aspect of scattering amplitudes
is their transcendental structure.
Massless loop amplitudes can often be expressed
in terms of multiple polylogarithms~\cite{Goncharov:1998kja,
GoncharovMixedTate}, the algebra of which
is conjecturally graded by a property called transcendental weight.
It corresponds to the number of integrations over
rational kernels involved in a functions' definition:
a logarithm (and $i\pi=\log(-1)$) has unit transcendental weight,
a dilogarithm (and $\zeta_2$) has weight two, \etc
$L$-loop amplitudes are observed to have an upper bound of weight~$2L$.
In the case of amplitudes in the ``simplest'' gauge theory
\cite{ArkaniHamed:2008gz,Dixon:2011xs}
--- $\cN=4$ supersymmetric Yang-Mills (SYM) ---
this bound has so far been observed to be saturated
(see \eg \rcites{Bern:1997nh,Anastasiou:2003kj,
Bern:2005iz,Naculich:2008ys,DelDuca:2010zg,
Goncharov:2010jf,Golden:2014xqa,
Henn:2016jdu,Drummond:2018caf,Abreu:2018aqd,Chicherin:2018yne,Caron-Huot:2019vjl}).
This property is commonly referred to as maximal transcendentality
\cite{Kotikov:2001sc,Kotikov:2002ab},
and the maximally supersymmetric theory has thus far
remained unique in this regard among Yang-Mills theories.
The origin of this uniform-transcendentality property is not fully understood,
and finding how and why it is violated in theories with $\cN<4$ supersymmetries remains an open question.

In this paper,
we further explore the connection between transcendental and IR structure in
$\cN=2$ supersymmetric theories initially observed in~\rcite{Duhr:2019ywc}.
In particular, we identify an IR subtraction
that makes this connection maximally apparent.
We approach the problem from two opposite sides.
On the one hand, the two-loop IR divergence formulae,
which follow from general factorization principles,
are rewritten in \sec{sec:IRfactorization}
so as to facilitate the analysis of our unrenormalized amplitudes.
On the other hand, in \secs{sec:oneloop}{sec:twoloop}
we examine the divergent parts
of specific one- and two-loop amplitude integrands.
Both analyses suggest a certain IR subtraction scheme
in which divergences are written in terms of scalar triangle integrals,
thereby extending the approach of \rcites{Badger:2016ozq,Badger:2016egz}
to a genuinely two-loop setting.
In \sec{sec:integrated}
we verify that this scheme choice results in a cancellation of lower-weight terms
from the finite, hard part of the two-loop four-gluon and two-gluon-two-matter
amplitudes in $\cN=2$ supersymmetric QCD (SQCD).

$\cN=2$ SQCD consists of $\cN=2$ supersymmetric Yang-Mills (SYM) theory
coupled to $N_f$ massless supersymmetric quarks (hypermultiplets).
As a model for quantum chromodynamics (QCD), the theory
has significantly richer physics than the maximally constrained $\cN=4$ SYM.
In particular, it has an arbitrary number of matter flavors~$N_f$.
For $N_f=C_A/T_F$ it develops a weakly-coupled superconformal phase,
in which case its gluonic amplitudes become very similar
to those in $\cN=4$ SYM~\cite{Dixon2008talk,Duhr:2019ywc}.
The theory of $\cN=2$ superconformal QCD (SCQCD),
which is the aforementioned special case of $\cN=2$ SQCD,
has previously been seen to have interesting transcendental structure.
Its planar four-point amplitudes were computed up through two loops in
\rcites{Dixon2008talk,Andree:2010na,Leoni:2014fja,Leoni:2015zxa}.
Moreover, in \rcite{Duhr:2019ywc} integration of
the full-color two-loop four-gluon amplitude in general $\cN=2$ SQCD revealed that,
for a particular choice of IR subtraction scheme in the superconformal theory,
all terms with weight less than three cancel from the finite part of the amplitude.
In this paper we extend this property to the full QCD-like theory
using a judicious choice of IR subtraction scheme.

Achieving these results has been facilitated by remarkable properties
of the $\cN=2$ SQCD integrands obtained previously
in \rcites{Johansson:2017bfl,Kalin:2018thp}.
In particular, diagrams containing internal matter lines
are observed to diverge in fewer IR regions,
as soft and collinear divergences arise only from virtual gluon exchanges.
Purely gluonic diagrams (or those related to them by supersymmetry),
which have the strongest IR divergences,
are naturally eliminated from the remainder function
$\cW_n^{(L)}=\cM_n^{(L)}-\cM_n^{(L)[\cN=4]}$
that measures scattering amplitudes as a ``correction'' to those in $\cN=4$ SYM.

Finally, we also take this opportunity to complete the work
begun in \rcite{Kalin:2018thp}
by presenting fully integrated two-loop four-point
amplitudes with matter on external legs.
The results provide us with expressions for the cusp, gluonic, and quark anomalous
dimensions in $\cN=2$ SQCD up to two-loop order,
which are conveniently cross-checked by comparison between the different amplitudes.

\subsection{Notation and conventions}
\label{sec:notation}

\paragraph{Integration and normalizations.}
We use dimensional regularization in $D=4-2\eps$ dimensions.
Recurring conventional prefactors are
\be
   S_\eps = (4\pi)^{\eps} e^{-\eps\gamma_{\rm E}} \,, \qquad
   r_\Gamma = e^{\eps\gamma_{\rm E}}
      \frac{\Gamma(1+\eps)\Gamma^2(1-\eps)}{\Gamma(1-2\eps)}
    = 1 - \frac{1}{2} \zeta_2\eps^2 - \frac{7}{3} \zeta_3\eps^3 + \cO(\eps^4) \,,
\ee
where $\gamma_{\rm E}=-\Gamma'(1)$ is Euler's constant.
Our $L$-loop integration operator $\bI$ is accordingly normalized as
\be
   \bI[f(\ell)] = e^{\eps\gamma_{\rm E}L}\!\int\!\frac{\d^{DL}\ell}{(i\pi^{D/2})^L}
   \frac{f(\ell)}{D(f)}\,.
\ee
where $D(f)$ are the quadratic propagators associated with $f$.
Unrenormalized (bare) $n$-point amplitudes $\cM_n$ are expanded
in powers of the bare coupling $\alpha_{\rm s}^0$:
\be
   \cM_n = \big(4\pi\alpha_{\rm s}^0\big)^{\frac{n-2}{2}}
      \sum_{L=0}^{\infty}
      \bigg(\!\frac{\alpha_{\rm s}^0 S_\eps}{4\pi}\!\bigg)^{\!\!L} \cM_n^{(L)} \,.
\label{LoopExpansion0}
\ee

\paragraph{Color.}
The gauge group $G$ is arbitrary unless stated otherwise.
We rely on a color operator~$c$ which extracts the color factor of
a given Feynman-like diagram expressed in terms of the structure
constants~$\tilde{f}^{abc}$ and generators~$T^a$
for gluonic and quark-gluon vertices, respectively,
\be
   c\Big(\!\gTreeTri[scale=1.2,eLA=$b$,eLB=$c$,eLC=$a$]{}\Big)
 = \tilde{f}^{abc} \,, \qquad \quad
   c\Big(\!\gTreeTri[scale=1.2,eLB=$i$,eLA=$\bar{\jmath}$,eLC=$a$,
                     eA=aquark,eB=quark,eC=gluon]{}\Big)
 = T_{i\bar{\jmath}}^a \,,
\ee
obeying the commutation relation
$[T^a,T^b]_{i\bar{\jmath}}=\tf^{abc}T^c_{i\bar{\jmath}}$.
The Casimirs are defined as
\be
   c\big(\!\gBubA[eLA=$b$,eLB=$a$]{}\!\big)
    = C_A \delta^{ab} \,, \qquad \quad
   c\big(\!\gBubA[eLA=$b$,eLB=$a$,iA=aquark,iB=aquark]{}\!\big)
    = T_F \delta^{ab} \,, \qquad \quad
   c\big(\!\gBubA[eLA=$\bar\jmath$,eLB=$i$,eA=aquark,eB=quark,iA=quark]{}\!\big)
    = C_F \delta_{i\bar\jmath} \,.
\ee
For ${\rm SU}(N_c)$ we normalize them as
$C_A=2N_c$, $T_F=1$, and $C_F=(N_c^2-1)/N_c$.

We also use the symbolic color generator $\bT_i^a$
which belongs to the gauge-group representation of the $i$-th parton.
For instance, $(\bT^a)_{i\bar{\jmath}}=T^a_{i\bar{\jmath}}$,
$(\bT^a)_{\bar{\imath}j}=-T^a_{j\bar{\imath}}$,
and $(\bT^a)_{bc}=\tf^{bac}$ for external-state quarks,
antiquarks, and gluons respectively.
The dipole $\bT_i\cdot\bT_j$ is a conventional shorthand
for the contraction $\sum_a \bT_i^a \otimes \bT_j^a$ over the adjoint indices ---
with the color indices of partons $i$ and $j$ still implicit and free.

\paragraph{Kinematics.}
All of our external momenta $p_i$ are taken outgoing.
Kinematic invariants are denoted by $s_{ij} = 2p_i \cdot p_j$.
At four points we use the Mandelstam variables
$s=s_{12}$, $t=s_{23}$, and $u=s_{13}$.
We split the $D$-dimensional loop momenta
$\ell_i = \bar{\ell}_i + \ell^{[-2\eps]}$
into four- and extra-dimensional parts, so that we can define the invariants
$\mu_{ij} = -\ell_i^{[-2\eps]}\!\cdot\ell_j^{[-2\eps]}$.
When $\ell_i$ and $p_j$ become collinear we write $\ell_i || p_j$.

We often use Dirac traces to represent
the kinematic dependence of amplitude numerators.
They are defined via the spinor products (see \eg \rcite{Dixon:1996wi})
\begin{equation}
\begin{aligned} \!\!\!
   [i_1i_2]\braket{i_2i_3}\cdots[i_{k-1}i_k]\braket{i_ki_1} &
    = p_{i_1}^{\mu_1} p_{i_2}^{\mu_2} \cdots p_{i_k}^{\mu_k}
      \Tr(\bar{\sigma}_{\mu_1}\sigma_{\mu_2}\cdots\sigma_{\mu_k})
    = \trp(i_1i_2\cdots i_k)\,,\! \\ \!\!\!
   \braket{i_1i_2}[i_2i_3]\cdots\braket{i_{k-1}i_k}[i_ki_1] &
    = p_{i_1}^{\mu_1} p_{i_2}^{\mu_2} \cdots p_{i_k}^{\mu_k}
      \Tr(\sigma_{\mu_1}\bar{\sigma}_{\mu_2}\cdots\bar\sigma_{\mu_k})
    = \trm(i_1i_2\cdots i_k)\,,
\end{aligned}
\end{equation}
where $\sigma$ are the usual four-dimensional Pauli spin matrices.
If an argument
is a loop momentum~$\ell_i$ it needs to be projected
to its four-dimensional part $\bar{\ell}_i$.

\paragraph{Integrands.}
We present an $L$-loop full-color amplitude
as a sum over a set $\Gamma_n^{(L)}$ of purely trivalent diagrams
\be
  i\cM_n^{(L)} = (-1)^Le^{\eps\gamma_{\rm E}L}\!
  \sum_{i\in \Gamma_n^{(L)}}\int\!\frac{\d^{LD}\ell}{(i\pi^{D/2})^L}
  \frac{(N_f)^{|i|}}{S_i} \frac{n_i c_i}{D_i}\,.
\ee
For each diagram we have
\begin{itemize}
\item a symmetry factor~$S_i$;
\item an overall factor~$(N_f)^{|i|}$, where $|i|$ is the number of matter loops;
\item a denominator~$D_i$, absorbing the quadratic propagator denominators
      for the exposed internal edges of the diagram;
\item a color factor~$c_i$, corresponding to the color operator~$c$ applied
      to the diagram;
\item a numerator~$n_i$, capturing all remaining kinematic dependence.
\end{itemize}
Different representations of the same integrand differ insofar as
they assign different numerator factors to the diagrams.
This way of organizing a full-color amplitude is tailored to exploit
color-kinematics duality
\cite{Bern:2008qj,Bern:2010ue,Johansson:2014zca,Johansson:2015oia,Bern:2019prr}
and is alternative to the method of
\rcites{Badger:2015lda,Ochirov:2016ewn,Ochirov:2019mtf}.

\paragraph{State configurations.}
To organize the particle content of four-point MHV amplitudes in a QCD-like theory,
we use $\kappa_{(ij)(kl)}$ introduced in~\rcites{Johansson:2014zca, Kalin:2018thp}
which carries the helicity weight of different external state configurations.
It is defined to absorb
the appropriate color-ordered tree amplitude as $ist M_4^{(0)}(1,2,3,4)$
and is given by\footnote{To obtain supersymmetric partners
related by the Ward identities in $\cN=1,2$ SQCD theories,
one can promote $\braket{ij}^\cN$ to $\delta^{2\cN}(Q)$,
a supermomentum-conserving delta function, see \eg\rcite{Elvang:2015rqa}.}
\be
  \kappa_{(ij)(kl)}=\frac{[12][34]}{\braket{12}\braket{34}}\braket{ij}^3\braket{kl}\,,
\label{eq:kappaDef}
\ee
where $i<j$ and $k<l$.
A particle label appearing in both parentheses of the subscript
corresponds to a ne\-ga\-tive-he\-li\-city gluon~$g^-$;
a label not appearing at all corresponds to a positive-helicity gluon~$g^+$;
a label appearing only in the first parenthesis is a quark~$q$;
a label appearing in the second is an antiquark~$\bar{q}$.
When all external states are gluons, we also abbreviate
\be
   \kappa_{ij}\equiv\kappa_{(ij)(ij)} .
\ee
For example, $\kappa_{12}=\kappa_{(12)(12)}$ corresponds to the
state configuration $(g_1^-\!,g_2^-\!,g_3^+\!,g_4^+)$,
and $\kappa_{(13)(14)}$ encodes $(g_1^-\!,g_2^+\!,q_3,\bar{q}_4)$.
The use of symbolic $\kappa$-prefactors allows us to add amplitudes
with different state configurations, which meshes well
with their on-shell superspace interpretation
in \rcites{Johansson:2014zca,Johansson:2017bfl,Kalin:2018thp}.

\section{IR factorization}
\label{sec:IRfactorization}

In this section we review the IR factorization of UV-renormalized amplitudes
in gauge theory
and derive factorization formulae for their unrenormalized counterparts.
The latter will be more useful for our subsequent analysis of ${\cal N}=2$ SQCD amplitudes.

\subsection{Soft-collinear exponentiation}
\label{sec:exponentiation}

Through two loops
the IR singularities of renormalized gauge theory amplitudes are entirely encoded
by an anomalous dimension~\cite{Sterman:2002qn,Aybat:2006wq,Aybat:2006mz,
Gardi:2009qi,Gardi:2009zv,Becher:2009cu,Becher:2009qa}\footnote{Starting
at three loops the complete anomalous dimension
$\mathbf{\Gamma}\big(\frac{p_i}{\mu},\alpha_{\rm s}\big)=
\mathbf{\Gamma}_\text{dip}\big(\frac{p_i}{\mu},\alpha_{\rm s}\big)+
\mathbf{\Delta}(\rho_{ijkl},\alpha_{\rm s})$
is corrected by a function $\mathbf{\Delta}$ of conformal-invariant cross ratios
$\rho_{ijkl}=\frac{s_{ij}s_{kl}}{s_{ik}s_{jl}}$
\cite{Almelid:2015jia,Gardi:2016ttq,Almelid:2017qju},
which has been confirmed by explicit calculation of the three-loop
four-point $\cN=4$ SYM amplitude~\cite{Henn:2016jdu}.}
\be
   \mathbf{\Gamma}_\text{dip}\!\left(\frac{p_i}{\mu},\alpha_{\rm s}\:\!\!\right)
    = - \frac{\gamma_K(\alpha_{\rm s})}{4}
        \sum_{i<j}^{n} \log\!\bigg(\!\frac{-s_{ij}}{\mu^2}\!\bigg) \bT_i\cdot\bT_j
      + \sum_{i=1}^{n} \gamma_i(\alpha_{\rm s}) \,.
\label{DipoleOperator}
\ee
Apart from depending on the $n$-parton kinematic space,
it involves a dipole operator $\bT_i\cdot\bT_j$
in the corresponding color space.
That $\bT_i\cdot\bT_j$ only involves adjoint color in the intermediate
state is important for later discussion of the IR finiteness of matter loop sub-diagrams.
Moreover, it depends on the strong coupling constant exclusively
through the (light-like) cusp anomalous dimension $\gamma_K$,
as well as the field anomalous dimensions $\gamma_i$ of the partons
which control hard collinear singularities.
A renormalized amplitude $\widetilde{\cM}_n$ factorizes as
\be
   \widetilde{\cM}_n(p_i,\mu,\alpha_{\rm s}(\mu))
    = \mathbf{Z}(p_i,\mu,\alpha_{\rm s}(\mu)) \cH_n(p_i,\mu,\alpha_{\rm s}(\mu)) \,,
\label{IRfactorization}
\ee
where the IR-divergent color-space operator\footnote{We
omit the usual path-ordering sign as $\mathbf{\Gamma}_\text{dip}$
has only one non-trivial color operator and a constant that commutes with it.
}
\be
   \mathbf{Z}(p_i,\mu,\alpha_{\rm s}(\mu))
    = \exp\!\left\{-\!\int^\mu_0\!\frac{\d\la}{\la}\:\!\mathbf{\Gamma}_\text{dip}\!
                    \left(\frac{p_i}{\la},\alpha_{\rm s}(\la)\:\!\!\right)
            \right\}
\label{IRfactor}
\ee
acts on the hard-scattering amplitude $\cH_n$,
which is UV- and IR-finite by definition.

The IR poles in the dimensional-regularization parameter $\eps$
arise in $\mathbf{Z}$ from the integrated scale dependence
of the strong coupling constant~$\alpha_{\rm s}(\la)$.
Let us see how this works at two loops,
where only the first two orders in~$\alpha_{\rm s}(\mu)$ are needed
out of $\log \mathbf{Z}$.
We expand all ingredients of the exponent
in powers of the coupling constant:
\be
   \gamma_K(\alpha_{\rm s}) = \sum_{L=1}^{\infty} \gamma_K^{(L)}
      \bigg(\!\frac{\alpha_{\rm s}}{2\pi}\!\bigg)^{\!L} \,, \qquad \quad
   \gamma_i(\alpha_{\rm s}) = \sum_{L=1}^{\infty} \gamma_i^{(L)}
      \bigg(\!\frac{\alpha_{\rm s}}{2\pi}\!\bigg)^{\!L} \,.
\ee
To integrate the anomalous dimension,
we need the scale dependence of the coupling constant $\alpha_{\rm s}(\lambda)$.
It is convenient to expand it as power series in $\alpha_{\rm s}(\mu)$:
\be
   \alpha_{\rm s}(\la) = \alpha_{\rm s}(\mu) \Big(\frac{\mu}{\la}\Big)^{\!2\eps}
      \sum_{l=0}^\infty a_l(\lambda,\mu)
      \bigg(\!\frac{\alpha_{\rm s}(\mu)}{2\pi}\!\bigg)^{\!l} \,,
\label{couplingconstant}
\ee
where the coefficient functions $a_l(\la,\mu)$ obey
the initial conditions $a_l(\mu,\mu) = \delta_{0l}$.
The prefactor $(\mu/\la)^{2\eps}$ is due to the leading term
$-2\eps\:\!\alpha_{\rm s}$ in the beta function,
\be
   \beta(\alpha_{\rm s}) = -\alpha_{\rm s}
      \bigg\{ 2\eps + \sum_{l=0}^{\infty} \beta_l
                      \bigg(\!\frac{\alpha_{\rm s}}{2\pi}\!\bigg)^{\!\!l+1} \bigg\}
    =-2\eps\:\!\alpha_{\rm s} - \frac{\beta_0}{2\pi} \alpha_{\rm s}^2
    + \cO(\alpha_{\rm s}^3) \,.
\label{betafunction}
\ee
The renormalization group equation
\beal
      \beta(\alpha_{\rm s}(\la)) = \la \frac{\d\alpha_{\rm s}(\la)}{\d\la} &
    = -2\eps\:\!\alpha_{\rm s}(\la)
    + \alpha_{\rm s}(\mu) \Big(\frac{\mu}{\la}\Big)^{\!2\eps}
      \sum_{l=0}^\infty \bigg(\!\frac{\alpha_{\rm s}(\mu)}{2\pi}\!\bigg)^{\!l}
      \la \frac{\partial a_l(\lambda,\mu)}{\partial \la}
\eeal
then implies a system of linear differential equations
that can be solved for $a_l(\la,\mu)$:
\be
   \left\{
   \begin{aligned}
   \la \frac{\partial a_0(\lambda,\mu)}{\partial \la} & = 0 \,, \\
   \la \frac{\partial a_1(\lambda,\mu)}{\partial \la} &
    = -\beta_0 \Big(\frac{\mu}{\la}\Big)^{\!2\eps} a_0^2(\lambda,\mu) \,, \\
   &\!\dots
   \end{aligned}
   \right.
   \qquad \Rightarrow \qquad
   \left\{
   \begin{aligned}
   a_0(\la,\mu) & = 1 \,, \\
   a_1(\la,\mu) & = \frac{\beta_0}{2\eps}
      \Big[\!\left(\frac{\mu}{\la}\right)^{\!2\eps}\!\!- 1\Big] \,, \\
   &\!\dots
   \end{aligned}
   \right.
\ee
Therefore, the terms in \eqn{couplingconstant}
that are relevant for the two-loop IR structure are
\be
   \alpha_{\rm s}(\la) = \alpha_{\rm s}(\mu) \Big(\frac{\mu}{\la}\Big)^{\!2\eps}
      \bigg\{ 1 + \frac{\beta_0\alpha_{\rm s}(\mu)}{4\pi\eps}
                  \Big[\!\left(\frac{\mu}{\la}\right)^{\!2\eps}\!\!- 1\Big] \bigg\}
    + \cO(\alpha_{\rm s}(\mu)^3) \,.
\label{couplingconstant2}
\ee
It is now straightforward to integrate the needed
$\la$-dependent terms in the anomalous dimension.\footnote{Two elementary integrals
that are used in the derivation of \eqn{IRfactorLog2} are
\begin{equation*}
   \int_0^\mu\!\frac{\d\la}{\la} \Big(\frac{\mu}{\la}\Big)^{\!2k\eps}\!
    = -\frac{1}{2k\eps} \,, \qquad\!\quad
   \int_0^\mu\!\frac{\d\la}{\la} \Big(\frac{\mu}{\la}\Big)^{\!2k\eps}\!
      \log\!\bigg(\frac{-s}{\la^2}\bigg)
    = \frac{1}{2k^2\eps^2}
      \bigg[ 1 - k\eps\log\!\bigg(\frac{-s}{\mu^2}\bigg)\!\bigg] \,.
\end{equation*}}
In this way, we obtain the IR poles
encoded by $\mathbf{Z}$ in an explicit form:
\begin{align}
   \log \mathbf{Z} = \frac{\alpha_{\rm s}(\mu)}{4\pi}
      \Bigg\{ \frac{\gamma_K^{(1)}}{4} \sum_{i<j}^{n} &\,\bT_i\cdot\bT_j
              \bigg[ \frac{1}{\eps^2}
                   - \frac{1}{\eps}
                     \log\!\bigg(\!\frac{-s_{ij}}{\mu^2}\!\bigg)\bigg]
            + \frac{1}{\eps} \sum_{i=1}^{n} \gamma_i^{(1)}
      \Bigg\} \nn \\*
    - \bigg(\!\frac{\alpha_{\rm s}(\mu)}{4\pi}\!\bigg)^{\!\!2}
      \Bigg\{ \frac{1}{4} \sum_{i<j}^{n} &\,\bT_i\cdot\bT_j
              \Bigg[ \frac{3\beta_0\gamma_K^{(1)}}{4\eps^3}
                   - \frac{\gamma_K^{(2)}}{2\eps^2}
                   - \bigg( \frac{\beta_0\gamma_K^{(1)}}{2\eps^2}
                          - \frac{\gamma_K^{(2)}}{\eps} \bigg)
                     \log\!\bigg(\!\frac{-s_{ij}}{\mu^2}\!\bigg)\bigg]\nn\\*
            + \sum_{i=1}^{n} &
              \bigg( \frac{\beta_0\gamma_i^{(1)}}{2\eps^2}
                   - \frac{\gamma_i^{(2)}}{\eps} \bigg)
      \Bigg\} + \cO(\alpha_{\rm s}^3) \,.
\label{IRfactorLog2}
\end{align}

To see how the IR divergences are organized perturbatively,
we expand the complete and hard-scattering amplitudes
in powers of the coupling:
\be
   \widetilde{\cM}_n = \big(4\pi\alpha_{\rm s}\big)^{\frac{n-2}{2}}
      \sum_{L=0}^{\infty} \bigg(\!\frac{\alpha_{\rm s}}{4\pi}\!\bigg)^{\!\!L}
      \widetilde{\cM}_n^{(L)} \,, \qquad \quad
   \cH_n = \big(4\pi\alpha_{\rm s}\big)^{\frac{n-2}{2}}
      \sum_{L=0}^{\infty} \bigg(\!\frac{\alpha_{\rm s}}{4\pi}\!\bigg)^{\!\!L}
      \cH_n^{(L)} \,.
\label{LoopExpansion}
\ee
We also introduce a convenient notation for the loop coefficients of $\log \mathbf{Z}$:
\be
   \log \mathbf{Z} = \frac{\alpha_{\rm s}(\mu)}{4\pi} \mathbf{Y}^{(1)}(\eps)
    + \bigg(\!\frac{\alpha_{\rm s}(\mu)}{4\pi}\!\bigg)^{\!\!2} \mathbf{Y}^{(2)}(\eps)
    + \cO(\alpha_{\rm s}^3) \,.
\ee
Substituting these expansions into
the factorization formula~\eqref{IRfactorization}, we find
\begin{subequations}
\begin{align}
\label{eq:IRfactorization2loop1}
 \widetilde{\cM}_n^{(1)}
   & = \mathbf{Y}^{(1)}(\eps) \cM_n^{(0)} + \cH_n^{(1)} \,, \\
\label{eq:IRfactorization2loop2}
 \widetilde{\cM}_n^{(2)}
   & = \bigg[ \mathbf{Y}^{(2)}(\eps)\!
           - \frac{1}{2} \big[ \mathbf{Y}^{(1)}(\eps) \big]^2
      \bigg] \cM_n^{(0)}
    + \mathbf{Y}^{(1)}(\eps) \widetilde{\cM}_n^{(1)}
    + \cH_n^{(2)} \,,
\end{align} \label{eq:IRfactorization2loop}%
\end{subequations}
where $\cH_n^{(0)}=\cM_n^{(0)}$
(we drop the tilde as the tree amplitude is not renormalized).
The IR operators $\mathbf{Y}^{(L)}(\eps)$ are then explicitly given as
\begin{subequations}
\begin{align}
\label{CatanilikeOperator1}
   \mathbf{Y}^{(1)}(\eps) &
    = \sum_{i<j}^{n} \bT_i\cdot\bT_j
              \bigg[ \frac{1}{\eps^2}
                   - \frac{1}{\eps}
                     \log\!\bigg(\!\frac{-s_{ij}}{\mu^2}\!\bigg)\bigg]
            + \frac{1}{\eps} \sum_{i=1}^{n} \gamma_i^{(1)} \,, \! \\ \!\!\!
\label{CatanilikeOperator2}
   \mathbf{Y}^{(2)}(\eps) &
    =-\frac{\beta_0}{\eps} \mathbf{Y}^{(1)}(\eps)
    + \bigg( \frac{\beta_0}{\eps} + \frac{\gamma_K^{(2)}}{2} \bigg)
      \mathbf{Y}^{(1)}(2\eps)
    + \frac{1}{4\eps} \sum_{i=1}^n
      \big(4\gamma_i^{(2)}\!- \gamma_K^{(2)} \gamma_i^{(1)}\big) \,.
\end{align}
\label{eq:CatanilikeOperators}%
\end{subequations}
We have plugged in the value $\gamma_K^{(1)}=4$ of the cusp anomalous dimension,
which at the lowest order is regularization-scheme independent~\cite{Gnendiger:2014nxa}.

It is illuminating to compare these factorization formulae
with the widely used expressions due to Catani and Seymour
\cite{Catani:1996jh,Catani:1998bh}:
\be
   \widetilde{\cM}_n^{(1)}
    = \mathbf{I}^{(1)}(\eps) \cM_n^{(0)} + \cO(\eps^0) \,, \qquad \quad
   \widetilde{\cM}_n^{(2)}
    = \mathbf{I}^{(2)}(\eps) \cM_n^{(0)}
    + \mathbf{I}^{(1)}(\eps) \widetilde{\cM}_n^{(1)} + \cO(\eps^0) \,,
\label{eq:IRfactorizationCatani}
\ee
in which the IR operators read~\cite{Sterman:2002qn,Bern:2004kq}
\begin{subequations}
\begin{align}
\label{eq:CataniOperator1}
   \mathbf{I}^{(1)}(\eps) &
    = \frac{e^{\eps\gamma_{\rm E}}}{2\Gamma(1-\eps)} \sum_{i=1}^n
      \left(\frac1{\eps^2}-\frac{2\gamma_i^{(1)}}{\eps \bT_i^2}\right)
      \sum_{j\neq i}^n \bigg(\!\frac{-s_{ij}}{\mu^2}\!\bigg)^{\!-\eps}
      \bT_i\cdot\bT_j
    = \mathbf{Y}^{(1)}(\eps) + \cO(\eps^0) \,, \\
\label{eq:CataniOperator2}
   \mathbf{I}^{(2)}(\eps) &
    =-\frac{1}{2} \mathbf{I}^{(1)}(\eps)
      \bigg(\mathbf{I}^{(1)}(\eps)+\frac{2\beta_0}{\eps}\bigg)
    + \frac{e^{-\eps\gamma_{\rm E}}\Gamma(1-2\eps)}{\Gamma(1-\eps)}
      \bigg(\frac{\beta_0}{\eps}+K_\text{R.S.}\bigg)
      \mathbf{I}^{(1)}(2\eps) \\* &~\:\:\quad
    + \frac{e^{\eps\gamma_{\rm E}}}{4\eps \Gamma(1-\eps)}
      \bigg[{-\sum_{i=1}^n} \frac{H_{i,\text{R.S.}}}{\bT_i^2} \sum_{j\neq i}^n
             \bigg(\!\frac{-s_{ij}}{\mu^2}\!\bigg)^{\!-2\eps}
             \bT_i\cdot\bT_j
           + \mathbf{H}_\text{R.S.}
      \bigg] \,. \nn
\end{align} \label{eq:CataniOperators}%
\end{subequations}
The specific form of the operator
$\mathbf{H}_\text{R.S.} = \cO(\eps^0)$ will be irrelevant to us,
but we note that it is known to involve
color structures of the form $\tf^{abc} \bT_i^a \bT_j^b \bT_k^c$
(see \eg \rcites{Bern:2002tk,Bern:2003ck,Becher:2009cu})
that are absent from \eqn{eq:CatanilikeOperators}.
Moreover, notice that the $\epsilon$-dependence of $\mathbf{I}^{(L)}(\eps)$
is somewhat more involved compared to $\mathbf{Y}^{(L)}(\eps)$.
The one-loop operators $\mathbf{I}^{(1)}(\eps)$ and $\mathbf{Y}^{(1)}(\eps)$
begin to differ after two orders in~$\eps$.
Although it is possible to identify
\be
   K_\text{R.S.} = \frac{\gamma_K^{(2)}}{2} \,, \qquad \quad
   H_{i,\text{R.S.}} = 4 \gamma_i^{(2)} - \gamma_K^{(2)} \gamma_i^{(1)}
    + \frac{3}{4} \beta_0 \zeta_2 \bT_i^2 \,,
\label{eq:CataniTermsConversion}
\ee
this discrepancy makes the regularization-scheme dependence
and color structure of $\mathbf{H}_\text{R.S.}$ significantly more complicated
than the explicit soft structure~\eqref{IRfactorLog2}
warrants it~\cite{Aybat:2006wq,Aybat:2006mz,Becher:2009cu,Becher:2009qa}.
Therefore, in the following we favor the latter and stick to the IR factorization
formulae~\eqref{eq:IRfactorization2loop} and~\eqref{eq:CatanilikeOperators}.

\subsection{Factorization for unrenormalized amplitudes}
\label{sec:unrenormalization}

We continue by studying the factorization properties
fulfilled by unrenormalized amplitudes $\cM_n^{(L)}$.
Our definition of the beta function \eqref{betafunction} is,
in the $\overline{\text{MS}}$ scheme,
equivalent to the following relationship between bare and renormalized couplings:
\be
\label{MSbarScheme}
   \alpha_{\rm s}^0 S_\eps\!
    = \alpha_{\rm s} \mu^{2\eps}\!
      \bigg\{ 1 - \frac{\alpha_{\rm s}\beta_0}{4\pi\eps}
              + \bigg(\!\frac{\alpha_{\rm s}}{4\pi}\!\bigg)^{\!\!2}
                \bigg[ \frac{\beta_0^2}{\eps^2} - \frac{\beta_1}{\eps} \bigg]
              + \cO(\alpha_{\rm s}^3) \bigg\} \,.
\ee
As we will not be needing the scale dependence anymore,
for simplicity we have set $\mu=1$.
This implies the usual renormalization relations
\begin{subequations}
\begin{align}
\label{Renormalization2loop1}
   \widetilde{\cM}_n^{(1)} &
    = \cM_n^{(1)} - \frac{(n-2)\beta_0}{2\eps} \cM_n^{(0)} \,, \\
\label{Renormalization2loop2}
   \widetilde{\cM}_n^{(2)} &
    = \cM_n^{(2)} - \frac{n\beta_0}{2\eps} \cM_n^{(1)}
    + \frac{(n-2)}{2}
      \bigg[ \frac{n\beta_0^2}{4\eps^2} - \frac{\beta_1}{\eps} \bigg]
      \cM_n^{(0)} \,.
\end{align} \label{Renormalization2loop}%
\end{subequations}
Notice that the two-loop beta-function coefficient $\beta_1$,
which is present in the UV renormalization above,
does not appear in the IR divergence structure of renormalized amplitudes
until three-loop order~\cite{Aybat:2006wq,Aybat:2006mz,Almelid:2015jia,Caron-Huot:2017fxr}.

Combining \eqns{eq:IRfactorization2loop1}{Renormalization2loop1}
at one loop is simple:
\be
   \cM_n^{(1)}
    = \widetilde{\cM}_n^{(1)} + \frac{(n-2)\beta_0}{2\eps} \cM_n^{(0)}
    = \bigg[ \mathbf{Y}^{(1)}(\eps)
           + \frac{(n-2)\beta_0}{2\eps} \bigg] \cM_n^{(0)} + \cH_n^{(1)} .
\ee
A similar exercise at two loops is more cumbersome but straightforward,
and we present the answer below.
In order to make the pole structure of the amplitude more transparent,
we separate the one-loop IR operator~$\mathbf{Y}^{(1)}(\eps)$
into its $1/\eps$ monopole (collinear) and $1/\eps^2$ dipole (soft) parts:
\be\label{eq:SoftOperator}
   \mathbf{Y}^{(1)}(\eps)
    = \mathbf{S}(\eps) + \frac{1}{\eps} \sum_{i=1}^{n} \gamma_i^{(1)}\,, \qquad \quad
   \mathbf{S}(\eps)
    = \frac{1}{\eps^2} \sum_{i<j}^{n} \bT_i\cdot\bT_j
      \left[ 1 - \eps \log\!\left(-s_{ij}\right) \right] \,.
\ee
In terms of this dipole operator~$\mathbf{S}(\eps)$,
we find that the unrenormalized amplitudes factorize as
\begin{subequations}
\begin{align}
\label{Factorization2loop1}
   \cM_n^{(1)}
    = \mathbf{S}(\eps) \cM_n^{(0)} &
    + \frac{1}{\eps}
      \bigg[ \frac{n-2}{2} \beta_0 + \sum_{i=1}^n \gamma_i^{(1)}
      \bigg] \cM_n^{(0)}
    + \cH_n^{(1)} , \\
   \cM_n^{(2)}
    = \mathbf{S}(\eps)\cM_n^{(1)} &
    - \frac{1}{2} \mathbf{S}(\eps) \mathbf{S}(\eps)
      \cM_n^{(0)}
    + \bigg[ \frac{\beta_0}{\eps} + \frac{\gamma_K^{(2)}}{2} \bigg]
      \mathbf{S}(2\eps) \cM_n^{(0)} \nn \\* &
\label{Factorization2loop2}
    + \frac{1}{2\eps^2}
      \bigg[ \frac{n-2}{2} \beta_0 + \sum_{i=1}^n \gamma_i^{(1)} \bigg]
      \bigg[ \frac{n}{2} \beta_0 + \sum_{i=1}^n \gamma_i^{(1)}
      \bigg] \cM_n^{(0)} \\* &
    + \frac{1}{\eps}
      \bigg[ \frac{n-2}{2} \beta_1 + \sum_{i=1}^n \gamma_i^{(2)}
      \bigg] \cM_n^{(0)}
    + \frac{1}{\eps}
      \bigg[ \frac{n}{2} \beta_0 + \sum_{i=1}^n \gamma_i^{(1)}
      \bigg] \cH_n^{(1)}
    + \cH_n^{(2)} . \nn
\end{align} \label{Factorization2loop}%
\end{subequations}
These formulae capture all $\eps$-divergences in dimensional regularization
through two loops.
They hold in a general gauge theory with massless matter,
since at this point we have not specialized to $\cN=2$ SQCD in any way.

The above factorization formulae contain dependence on
the scheme of dimensional regularization~\cite{tHooft:1972tcz,Collins:1984xc,
Siegel:1979wq,Bern:1991aq,Bern:2002zk}
(see \rcite{Gnendiger:2017pys} for a recent overview),
which is hardly surprising.
In fact, the beta-function coefficients~$\beta_{L-1}$
and the anomalous dimensions $\gamma_K^{(L)}$ and $\gamma_i^{(L)}$
may also depend on the specific subtraction scheme
used for removing the UV divergences~\cite{Kilgore:2011ta,Gnendiger:2014nxa}.
It is, however, clear that the latter dependence should be
spurious in the factorization properties~\eqref{Factorization2loop}
of unrenormalized amplitudes.\footnote{For a non-supersymmetric theory
considered in the dimensional-reduction (DRED~\cite{Siegel:1979wq})
or the four-dimensional-helicity (FDH~\cite{Bern:1991aq,Bern:2002zk}) scheme,
\eqn{Factorization2loop} would need to take into account
a set of additional evanescent scalars
with two separate running couplings $\alpha_{4\eps}$ and $\alpha_{\rm e}$
for their self- and gauge interactions~\cite{vanDamme:1984ig,Jack:1993ws,Jack:1994bn}.
However, the principal purpose of these schemes is to preserve supersymmetry,
which pegs the evanescent couplings to the gauge coupling.
So in $\cN=2$ SQCD we can use the precise formulae~\eqref{Factorization2loop}
in a close analogue of the FDH scheme,
which we will discuss in \sec{sec:N2review}.}

\subsection{Subtracting IR structure of \texorpdfstring{$\cN=4$}{N=4} SYM}
\label{sec:N4diff}

In this section we specialize to the case where all external partons are gluons
(or more generally vector multiplets that include gluons)
and define the discrepancy between the $n$-gluon amplitude in question
and its $\cN=4$ counterpart.
The remainder function has a simpler divergence structure
than the original amplitude.

Due to a simple relation between the gluonic collinear anomalous dimension
and the $\beta$-function coefficient (see \eg \rcite{Gnendiger:2014nxa}),
\be
   \gamma_g^{(1)} = -\frac{\beta_0}{2} \,,
\label{gColl1}
\ee
the factorization formulae~\eqref{Factorization2loop}
for purely gluonic amplitudes can be rewritten in a streamlined form:
\small
\begin{align}
\label{Factorization2loopGluonic}
   \cM_n^{(1)} &
    = \mathbf{S}(\eps) \cM_n^{(0)}
    - \frac{\beta_0}{\eps} \cM_n^{(0)} + \cH_n^{(1)} \,, \\
   \cM_n^{(2)} &
    = \mathbf{S}(\eps)\cM_n^{(1)}
    + \bigg\{\!\!
   -\!\frac{1}{2} \mathbf{S}(\eps) \mathbf{S}(\eps)
    + \bigg[ \frac{\beta_0}{\eps} + \frac{\gamma_K^{(2)}}{2}
      \bigg] \mathbf{S}(2\eps)
    + \frac{1}{\eps}
      \bigg[ \frac{n\!-\!2}{2} \beta_1 + n \gamma_g^{(2)} \bigg]
      \bigg\} \cM_n^{(0)}
    + \cH_n^{(2)} \,. \nn
\end{align}
\normalsize
Let us turn for a moment to the special case of $\cN=4$ SYM.
Its gluonic tree-level amplitudes $\cM_n^{(0)}$ trivially
coincide with those in massless QCD
(or any four-dimensional gauge theory for that matter);
the specifics of the matter content start to enter at one loop.
Moreover, the theory is UV finite (hence $\beta_i=0$) and only has IR divergences.
Despite being considered in many ways
the simplest gauge theory~\cite{ArkaniHamed:2008gz},
its scattering amplitudes contain the most IR-divergent kinematic regions.
The above formulae simplify to
\small
\begin{align}
\label{Factorization2loopN4}%
   \cM_n^{(1)[\cN=4]}\:\!\!&
    = \mathbf{S}(\eps) \cM_n^{(0)}\!+ \cH_n^{(1)[\cN=4]} , \\
   \cM_n^{(2)[\cN=4]}\:\!\!&
    = \mathbf{S}(\eps) \cM_n^{(1)[\cN=4]}\!
    +\!\!\bigg\{
    {\!-\frac{1}{2}}\mathbf{S}(\eps) \mathbf{S}(\eps)
    + \frac{1}{2} \gamma_K^{(2)[\cN=4]} \mathbf{S}(2\eps)
    + \frac{n}{\eps}\gamma_g^{(2)[\cN=4]}\!
      \bigg\} \cM_n^{(0)}\!
    + \cH_n^{(2)[\cN=4]} , \nn
\end{align}
\normalsize
where the anomalous dimensions are
\cite{Bern:2004kq,Moch:2005tm,Gehrmann:2010ue,Caron-Huot:2017fxr}
\be
   \gamma_K^{(2)[\cN=4]} = -2\zeta_2 C_A \,, \qquad \quad
   \gamma_g^{(2)[\cN=4]} = \frac{1}{8} \zeta_3 C_A^2 \,.
\label{eq:N4AnomDimensions2}
\ee
These formulae are consistent with
the iterative construction of \rcites{Anastasiou:2003kj,Bern:2005iz},
as verified in \app{sec:BDS}.

We wish to study the remainder function $\cW_n^{(L)}$,
which we define perturbatively as the difference between the (gluonic) amplitude
in an arbitrary gauge theory and that of $\cN=4$ SYM:
\be
   \cM_n^{(L)} = \cM_n^{(L)[\cN=4]} + \cW_n^{(L)} \,.
\label{N4diff}
\ee
Assuming the factorization~\eqref{Factorization2loopN4}
for the amplitudes in $\cN=4$ SYM,
we find that these subtracted amplitudes factorize in a much simpler way:
\begin{subequations}
\begin{align}
\label{eq:Factorization2loopN4diff1}
   \cW_n^{(1)} &
    = - \frac{\beta_0}{\eps} \cM_n^{(0)}
    + \cH_n^{(1)} - \cH_n^{(1)[\cN=4]} \,, \\
\label{eq:Factorization2loopN4diff2}
   \cW_n^{(2)} &
    = \mathbf{S}(\eps) \cW_n^{(1)}
    + \bigg( \frac{\beta_0}{\eps}
           + \frac{1}{2} \big[ \gamma_K^{(2)}\!- \gamma_K^{(2)[\cN=4]} \big]
      \bigg) \mathbf{S}(2\eps)
      \cM_n^{(0)} \\* & \qquad \qquad\:\quad
    + \frac{1}{\eps}
      \bigg( \frac{n-2}{2} \beta_1
           + n \big[ \gamma_g^{(2)}\!- \gamma_g^{(2)[\cN=4]} \big]
      \bigg) \cM_n^{(0)}
    + \cH_n^{(2)} - \cH_n^{(2)[\cN=4]} \,, \nn
\end{align}
\label{eq:Factorization2loopN4diff}%
\end{subequations}
where the hierarchy of $1/\eps^k$ poles is maximally transparent.
Namely,
\begin{itemize}
\item
The most singular $1/\eps^{2L}$ amplitude pieces are absent
from \eqn{eq:Factorization2loopN4diff}
as they have been absorbed by the $\cN=4$ amplitudes.
\item
The presence of $\beta_0$ in front of the $1/\eps$ terms
in \eqn{eq:Factorization2loopN4diff1}
suggests that the remaining one-loop divergences are exclusively UV,
\ie the complete IR structure of purely gluonic amplitudes
is at one loop captured by $\cN=4$ SYM.
\item
The leading $1/\eps^3$ amplitude pieces in \eqn{eq:Factorization2loopN4diff2}
are also accompanied by $\beta_0$. Therefore, they occur due to the multiplication
of the UV divergence $1/\eps$ in one of the two loops
by the IR divergences $1/\eps^2$ in the other.
\item
All of the $1/\eps^2$ terms in $\cW_n^{(2)}$
are caused by the overlap of soft and collinear
divergences in only one of the two loops, while the other loop stays finite.
\item
The remaining $1/\eps$ terms arise due to a single type of
UV, soft or collinear divergence occurring in one of the loops.
\end{itemize}
The emergence of this divergence hierarchy from the one-loop integrands
in $\cN=2$ SQCD is discussed in more detail in \sec{sec:oneloop},
and the two-loop integrands in \sec{sec:twoloop}.

\subsection{Alternative IR subtraction}
\label{sec:triangles}

In anticipation of specializing to $\cN=2$ SQCD (but not doing so yet)
we find it convenient to rewrite \eqn{eq:Factorization2loopN4diff2}
in terms of scalar triangle integrals.\footnote{Rewriting
  IR operators in terms of scalar triangles has already
  proved fruitful in \rcites{Badger:2015lda,Badger:2016ozq}.}
For instance, the one-loop dipole operator
\eqref{eq:SoftOperator} may be represented by
\begin{equation}
   \mathbf{S}(\eps) = \sum_{i<j}^n s_{ij}
   \gTriA[rotate=180,eLB=$i$,eLC=$j$,
          eA=double,eB=line,eC=line,iA=line,iB=line,iC=line]{} \bT_i\cdot\bT_j
    + \cO(\eps^0)\,,
\end{equation}
where the missing higher-order terms come from one-loop triangle integral:
\be
   \gTriA[rotate=180,eLB=$i$,eLC=$j$,
          eA=double,eB=line,eC=line,iA=line,iB=line,iC=line]{}
 = -\frac{r_\Gamma}{\eps^2}
   \left(-s_{ij}\right)^{-1-\eps}
 = \frac{1}{\eps^2s_{ij}}
   \bigg[1-\eps\log(-s_{ij})+\frac{\eps^2}{2}\big(\log^2(-s_{ij})-\zeta_2\big)
   \bigg] + \cO(\eps) \,.
\label{eq:1Ltriangle}
\ee
Similarly, using the two-loop triangle integrals
(see \eg \rcite{Gehrmann:1999as})
\begin{subequations} \begin{align}
\label{eq:2LtriangleA}
\gTriBubA[rotate=180,all=line,eA=doubleline,eLB=$i$,eLC=$j$]{}&=
-\frac{r_{\Gamma} e^{\eps\gamma_{\rm E}} \Gamma^2(1-2\eps) \Gamma(1+2\eps)}
{4 \eps^3 (1-2\eps) \Gamma(1-3\eps)}
(-s_{ij})^{-1-2\eps} \,,\\
\label{eq:2LtriangleB}
\gTriBubB[rotate=180,all=line,eA=doubleline,eLB=$i$,eLC=$j$]{}&=
-\frac{e^{2\eps\gamma_{\rm E}}\Gamma(1+2\eps)\Gamma^3(1-\eps)}
{2\eps^3(1-2\eps)\Gamma(1-3\eps)}
(-s_{ij})^{-1-2\eps} \,,
\end{align} \label{eq:2Ltriangles}%
\end{subequations}
the two-loop factorization formula~\eqref{eq:Factorization2loopN4diff2}
may be non-trivially rearranged into
\begin{align}\label{eq:Factorization2loopN4int}
   \cW_n^{(2)} &
    = \left( \sum_{i<j}^n s_{ij}
             \gTriA[scale=0.8,rotate=180,eLB=$i$,eLC=$j$,
                    eA=doubleline,eB=line,eC=line,iA=line,iB=line,iC=line]{}
             \bT_i\cdot\bT_j\!
      \right) \cW_n^{(1)}
    + \beta_0\!
      \left( \sum_{i<j}^ns_{ij}
             \left[\!\gTriBubB[scale=0.8,rotate=180,all=line,
                              eA=doubleline,eLB=$i$,eLC=$j$]{}\!\!
                  - \gTriBubA[scale=0.8,rotate=180,all=line,
                              eA=doubleline,eLB=$i$,eLC=$j$]{}\right]
            \bT_i\cdot\bT_j\!
      \right) \cM_n^{(0)} \nn \\
  &\qquad
  -2\left(\beta_0-\frac14\big[\gamma_K^{(2)}-\gamma_K^{(2)[\cN=4]}\big]\right)
  \mathbf{S}(2\eps)\cM_n^{(0)}\\
  &\qquad
  +\frac1{\eps}\left(\frac{n}8\big(4-\zeta_2\big) \beta_0 C_A
    +\frac{n-2}2\beta_1
    +n\big[\gamma_g^{(2)}-\gamma_g^{(2)[\cN=4]}\big]\right)\cM_n^{(0)}
  + \cW_n^{(2)\text{fin}} \,. \nn
\end{align}
To verify this, one requires color conservation $\sum_i \bT_i=0$,
which for external gluons implies $\sum_{i,j\neq i}^n\bT_i\cdot\bT_j=-n C_A$.
Moreover, finite $\cO(\eps^0)$ terms
have been absorbed into the hard function,
so $\cW_n^{(2){\rm fin}}\neq\cH_n^{(2)}-\cH_n^{(2)[\cN=4]}$.
We have therefore mandated a new IR subtraction scheme for $\cW_n^{(2)}$
which differs from the minimal scheme given in~\eqn{eq:Factorization2loopN4diff2}.
As we shall see in the next subsection,
when the two-loop anomalous dimensions $\gamma_K^{(2)}$ and $\gamma_g^{(2)}$
for $\cN=2$ SQCD are inserted,
the second two lines in \eqn{eq:Factorization2loopN4int} cancel away.

\subsection{Divergence structure of \texorpdfstring{$\cN=2$}{N=2} SQCD}
\label{sec:N2review}

\begin{table}[t]
\centering
\vspace{-7pt}
\includegraphics[scale=1.25]{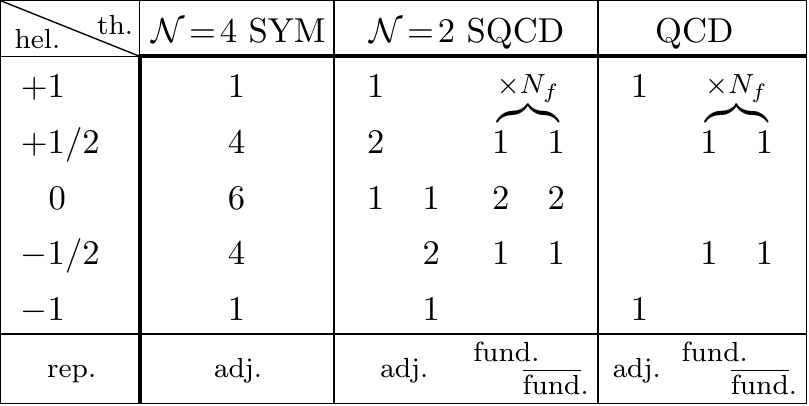}
\caption{\small On-shell states of $\cN=2$ SQCD in comparison with
$\cN=4$ SYM and conventional massless QCD.
The helicities are listed in the left column,
and ${\rm SU}(N_c)$-group representations of the particles
are shown in the lower row.}
\label{tab:SQCD}
\end{table}

We discuss in more detail the $\cN=2$ SQCD theory,
whose amplitudes we study in this paper.
The theory is built from $\cN=2$ SYM coupled to $N_f$ supersymmetric quarks
transforming under the fundamental representation of the gauge group~$G$.
One can view it as a two-fold supersymmetry enhancement of QCD,
or as an orbifold truncation of $\cN=4$ SYM~\cite{Kachru:1998ys,Lawrence:1998ja,Bershadsky:1998mb,Bershadsky:1998cb,Chiodaroli:2013upa},
where half of the particle content is promoted to
the fundamental or antifundamental representation of the gauge group~\cite{Johansson:2017bfl}.
In \Tab{tab:SQCD} we summarize its particle-helicity content
while displaying its intermediate position
between the physical theory of massless QCD and the
maximally supersymmetric $\cN=4$ SYM theory
(in which no matter content is permitted).

The intermediate complexity of $\cN=2$ SQCD is reflected in its UV structure.
Recall that the one-loop beta-function coefficient for a gauge theory
minimally coupled to $n_f$~Weyl fermions and $n_s$~real scalars is
\be
   \beta_0 = \frac{11}{6} C_A - \frac{1}{3} T_f n_f - \frac{1}{12} T_s n_s \,.
\label{QCDbeta}
\ee
$\cN=2$ SYM theories with matter are known to have a
one-loop exact beta-function~\cite{Novikov:1983uc,Seiberg:1988ur},
so for $\cN=2$ SQCD we get
\beal
   \beta(\alpha_{\rm s}) = -\alpha_{\rm s} \bigg( 2\eps + \beta_0 \frac{\alpha_{\rm s}}{2\pi} \bigg) \,, \qquad \quad
   \beta_0 = C_A-T_F N_f\,\overset{{\rm SU}(N_c)}{=}\,2N_c - N_f \,.
\label{N2beta}
\eeal
This $\beta_0$ value can be computed from the general formula~\eqref{QCDbeta}
by substituting $n_f=n_s=2$, $T_f=T_s=C_A$ for the adjoint fermions and scalars
and $n_f=2N_f$, $n_s=4N_f$, $T_f=T_s=T_F$ for
(the parity-even combination of)
the fundamental and antifundamental matter multiplets.

As is often done in QCD,
we leave the number of matter flavors $N_f$ arbitrary.
The notable special cases are the pure $\cN=2$ SYM theory for $N_f=0$
and the $\cN=2$ superconformal (SCQCD) theory for $N_f=C_A/T_F$~\cite{Seiberg:1988ur}.
Moreover, it is evident from \Tab{tab:SQCD} that if we switch
the representation of one of the matter multiplets to adjoint,
in combination with the $\cN=2$ vector multiplet
it will enhance it to the $\cN=4$ one.
This allows us to impose certain $\cN=4$ matching conditions
\cite{Johansson:2017bfl,Kalin:2018thp} on the kinematic numerators
(illustrated by \eqns{eq:N4matching1loop}{eq:N4matching2loop} below)
which will significantly facilitate our integrand-level analysis
of the two-loop IR structure in \sec{sec:twoloop}.

Like $\cN=4$ SYM, the $\cN=2$ SCQCD theory is UV finite.
However, it also contains matter amplitudes
with a non-trivial IR divergence structure.
Therefore, if we expand $\cN=2$ SQCD theory
around the conformal fixed point $N_f=C_A/T_F$,
we observe a clean separation between different kinds of divergences.
By explicit calculation of three different two-loop amplitudes
(to be expanded upon in \sec{sec:integrated})
we find that the cusp and collinear anomalous dimensions can be written as
\begin{subequations} \begin{align}
\label{eq:N2AnomDimensionsCusp}
   \gamma_K^{(1)} & = 4 \,, & \qquad &
   \gamma_K^{(2)} = -2\zeta_2 C_A +4\beta_0 \,, \\*
\label{eq:N2AnomDimensionsGluon}
   \gamma_g^{(1)} & = -\frac{\beta_0}{2} \,, & \qquad &
   \gamma_g^{(2)}
    = \frac{1}{8} C_A \big( \zeta_3 C_A + \beta_0(\zeta_2-4) \big) \,, \\*
\label{eq:N2AnomDimensionsQuark}
   \gamma_q^{(1)} & = \gamma_{\bar{q}}^{(1)} = 0 \,, & \qquad &
   \gamma_q^{(2)} = \gamma_{\bar{q}}^{(2)}
    = \frac{1}{8} C_F
      \big(13\zeta_3 C_A - 12\zeta_3 C_F - \beta_0(3\zeta_2+4)\big) \,.
\end{align} \label{eq:N2AnomDimensions}%
\end{subequations}
When $\beta_0=0$ both $\gamma_K^{(2)}$ and $\gamma_g^{(2)}$ coincide
with their $\cN=4$ values given in \eqn{eq:N4AnomDimensions2}.
This is consistent with the known observation that cusp anomalous dimensions
in $\cN=2$ SCQCD and $\cN=4$ SYM only start to differ
at three loops~\cite{Pomoni:2013poa,Mitev:2014yba,Mitev:2015oty}.
If we furthermore promote the SCQCD matter to the adjoint representation
(by substituting $T_F, C_F \to C_A$)
it becomes a part of the larger $\cN=4$ multiplet,
such that $\gamma_q^{(2)}$ coincides with $\gamma_g^{(2)[\cN=4]}$.
The same identification happens for the gauge group $G={\rm SO}(3)$,
in which case $C_A=N_c-2=1$ and $C_F=\frac12(N_c-1)=1$ ---
setting $\beta_0=0$ gives $N_f=C_A/T_F=1$,
and we recover the $\cN=4$ SYM theory.

We invite the reader to confirm the cancellation of
the second two lines in \eqn{eq:Factorization2loopN4int},
which occurs when the specific values of $\gamma_K^{(2)}$ and $\gamma_g^{(2)}$
quoted above are inserted.
Soft, collinear, and UV divergences are now all exposed:
soft occurring between massless legs of the triangle integrals,
collinear at their vertices,
and UV in the embedded bubble integrals.
Notice the placement of $\beta_0$:
since $\cW_n^{(1)}=-\frac{\beta_0}{\eps}\cM_n^{(0)}+\cO(\eps^0)$,
setting $\beta_0=0$ manifestly eliminates all UV divergences,
and $\cW_n^{(2)}$ diverges only as $\cO(\eps^{-2})$.
As we shall see in \sec{sec:integrated},
this new scheme is privileged as it succeeds in cancelling all
terms from $\cW_4^{(2){\rm fin}}$ with transcendental weight less than three.
To better motivate \eqn{eq:Factorization2loopN4int} we will examine the IR structure
of the two-loop four-gluon integrand in \sec{sec:twoloop},
which exhibits the divergence factorization structure
in a surprisingly transparent way.

Our one- and two-loop integrands are obtained
by matching to generalized unitarity cuts
\cite{Bern:1994zx,Bern:1994cg,Britto:2004nc,Giele:2008ve}.
Extra-dimensional terms $\mu_{ij}$, required by dimensional regularization,
are obtained by evaluating six-dimensional unitarity cuts
\cite{Cheung:2009dc,Bern:2010qa} for chiral $\cN=(1,0)$ SYM
coupled to $N_f$ copies of the $\cN=(0,1)$ hypermultiplet.\footnote{In
   this paper we omit the antisymmetric echo
   of the six-dimensional Levi-Civita tensor $\eps(\mu_i,\mu_j)$
   defined in \rcites{Johansson:2017bfl,Kalin:2018thp},
   as it always integrates to zero.
}
As explained in \rcite{Johansson:2017bfl}
(see also \rcites{deWit:2002vz,Boels:2009bv,Dennen:2009vk,Elvang:2011fx,Huang:2011um}),
this theory is a natural six-dimensional uplift of $\cN=2$ SQCD.
It defines our dimensional regularization scheme:
external states are strictly four-dimensional
(like in the 't Hooft-Veltman scheme~\cite{tHooft:1972tcz}),
and the internal state counting precisely matches
that of the four-dimensional $\cN=2$ SQCD theory.
This is therefore a close analogue of the four-dimensional helicity
(FDH~\cite{Bern:1991aq,Bern:2002zk}) scheme.
The anomalous dimensions~\eqref{eq:N2AnomDimensions} are therefore
consistent with their FDH values~\cite{Bern:2002tk,Bern:2003ck,DeFreitas:2004tk,
Gnendiger:2014nxa}.

\section{IR and transcendental structure of one-loop \texorpdfstring{$\cN=2$}{N=2} SQCD}
\label{sec:oneloop}

We begin with a couple of warm-ups involving the four-point
one-loop $\cN=2$ SQCD amplitudes in the representation derived in \rcite{Kalin:2018thp}.
As the full integration of these amplitudes is straightforward,
we do not dwell on it here.
Instead, we concentrate on interpreting their divergent and transcendental behavior ---
without performing any loop integrals explicitly,
to see how this behavior stems directly from the integrands.
Of course, at one loop such observations are somewhat redundant given
how easily the relevant integrals may be performed.
However, examination of the one-loop integrands serves to illustrate
the concepts discussed in previous sections,
and will prepare us for the more involved two-loop amplitudes in \sec{sec:twoloop}.

\subsection{External vectors + matter}
\label{sec:1lmixed}

As a first illustration of how IR behavior operates prior to integration,
we consider the one-loop $\cN=2$ SQCD amplitude with both vectors and matter on external legs.
In \rcite{Kalin:2018thp} this amplitude was presented in a form
that makes manifest the color-kinematics duality
\cite{Bern:2008qj,Bern:2010ue,Johansson:2014zca,Johansson:2015oia,Bern:2019prr}.
Its color-dual numerators are nonzero for boxes and triangles,
but only the former give rise to non-vanishing integrals:\footnote{Numerators with reversed arrows are given by matter-reversal symmetries, see \rcite{Kalin:2018thp} for details.
}
\begin{subequations}
\begin{align}
\label{eq:1LMixedNum1}
n\Bigg(\!\gBox[scale=.8,yshift=-1,eLA=$1^-$,eLB=$2$,eLC=$3$,eLD=$4^+$,iA=quark,iD=quark,iC=quark,
eB=quark,eC=aquark,iLD=$\uset{\rightarrow}{\ell}$]{}\!\Bigg)&=
\frac{\kappa_{(12)(13)}}{su}\trp(4\ell12)\,,\\
\label{eq:1LMixedNum2}
n\Bigg(\gBox[scale=.8,yshift=-1,eLA=$1^-$,eLB=$2$,eLC=$3^+$,eLD=$4$,iA=quark,iD=quark,
eB=quark,eD=aquark,iLD=$\uset{\rightarrow}{\ell}$]{}\!\Bigg)&=
\frac{\kappa_{(12)(14)}}{st}\trp(3\ell12)\,,\\
\label{eq:1LMixedNum3}
n\Bigg(\gBox[scale=.8,yshift=-1,eLA=$1$,eLB=$2^-$,eLC=$3^+$,eLD=$4$,iD=quark,
eA=quark,eD=aquark,iLD=$\uset{\rightarrow}{\ell}$]{}\!\Bigg)&=
\frac{\kappa_{(12)(24)}}{su}\trp(3\ell21)\,.
\end{align} \label{eq:1LMixedNums}%
\end{subequations}
Diagrammatically, we denote gluons (vector multiplets) by curly lines
and the helicities for external lines.
The matter flow is shown using arrowed lines,
such that the arrow direction corresponds to the chirality
of a quark (matter hypermultiplet).

A remarkable property of the above numerators is that they vanish
whenever the loop momentum associated with any matter edge goes to zero.
Similarly, the numerators vanish
if a loop momentum becomes collinear to one of the external vectors,
but not the external matter.
For instance, the numerator~\eqref{eq:1LMixedNum2} vanishes
when $\ell \to 0$, but not when $\ell+p_4 \to 0$;
it also vanishes when $\ell || p_1$, but not when $\ell || p_4$.
These properties will remain true for all kinematic numerators
explicitly shown in this paper.

The full color-dressed amplitude is expressed in terms of these three diagrams as
\begin{align}\label{eq:1LMixedAmp}
\begin{aligned}
\cM_4^{(1)}
=i\:\!\bI\bigg[
c\!\left(\!\gBox[scale=.7,yshift=-1,eLA=$1$,eLB=$2$,eLC=$3$,eLD=$4$,
iB=quark,iD=quark,iC=quark,eA=quark,eB=aquark]{}\!\right)\!
n\!\left(\!\gBox[scale=.7,yshift=-1,eLA=$1$,eLB=$2$,eLC=$3^-$,eLD=$4^+$,
iB=quark,iD=quark,iC=quark,eA=quark,eB=aquark]{}\!\right)&+
c\!\left(\!\gBox[scale=.7,yshift=-1,eLA=$1$,eLB=$2$,eLC=$4$,eLD=$3$,
iB=quark,iD=quark,iC=quark,eA=quark,eB=aquark]{}\!\right)\!
n\!\left(\!\gBox[scale=.7,yshift=-1,eLA=$1$,eLB=$2$,eLD=$3^-$,eLC=$4^+$,
iB=quark,iD=quark,iC=quark,eA=quark,eB=aquark]{}\!\right)\\+\,
c\!\left(\!\gBox[scale=.7,yshift=-1,eLA=$1$,eLB=$3$,eLC=$2$,eLD=$4$,
iD=quark,iC=quark,eA=quark,eC=aquark]{}\!\right)\!
n\!\left(\!\gBox[scale=.7,yshift=-1,eLA=$1$,eLB=$3^-$,eLC=$2$,eLD=$4^+$,
iD=quark,iC=quark,eA=quark,eC=aquark]{}\!\!\right)&+
c\!\left(\!\gBox[scale=.7,yshift=-1,eLA=$1$,eLB=$4$,eLC=$2$,eLD=$3$,
iD=quark,iC=quark,eA=quark,eC=aquark]{}\!\right)\!
n\!\left(\!\gBox[scale=.7,yshift=-1,eLA=$1$,eLB=$4^+$,eLC=$2$,eLD=$3^-$,
iD=quark,iC=quark,eA=quark,eC=aquark]{}\!\!\right)\\+\,
c\!\left(\!\gBox[scale=.7,yshift=-1,eLA=$1$,eLB=$2$,eLC=$3$,eLD=$4$,
iA=aquark,eA=quark,eB=aquark]{}\!\right)\!
n\!\left(\!\gBox[scale=.7,yshift=-1,eLA=$1$,eLB=$2$,eLC=$3^-$,eLD=$4^+$,
iA=aquark,eA=quark,eB=aquark]{}\!\right)&+
c\!\left(\!\gBox[scale=.7,yshift=-1,eLA=$1$,eLB=$2$,eLC=$4$,eLD=$3$,
iA=aquark,eA=quark,eB=aquark]{}\!\right)\!
n\!\left(\!\gBox[scale=.7,yshift=-1,eLA=$1$,eLB=$2$,eLC=$4^+$,eLD=$3^-$,
iA=aquark,eA=quark,eB=aquark]{}\!\right)\bigg]\,.
\end{aligned}
\end{align}
In the following we will inspect the IR divergence structure of the amplitude~\eqref{eq:1LMixedAmp}
and derive the IR factorization formula~\eqref{Factorization2loop1}.
To do that, we analyze the soft and collinear regions of individual diagrams
in a way that is similar to what underlies the strategy of expansion by regions
\cite{Beneke:1997zp,Smirnov:2002pj,Jantzen:2011nz,Semenova:2018cwy},
as well as the recent IR subtraction method of \rcite{Anastasiou:2018rib}.

Consider for concreteness the first diagram
\be
\bI\Bigg[n\Bigg(\!\gBox[scale=.8,yshift=1,eLA=$1^-$,eLB=$2$,eLC=$3$,eLD=$4^+$,
iA=quark,iD=quark,iC=quark,eB=quark,eC=aquark,iLB=$\overset{\leftarrow}{\ell}$]{}\!\Bigg)\Bigg]
=\frac{\kappa_{(12)(13)}}{su}e^{\eps\gamma_{\rm E}}\!\!\int\!\frac{\d^D\ell}{i\pi^{D/2}}
\frac{\trp(4(\ell+p_{12})12)}{(\ell+p_{12})^2(\ell+p_2)^2\ell^2(\ell-p_3)^2}\,,
\ee
where we have relabeled the loop momentum $\ell$ for convenience.
This integral diverges in three regions:
one soft, $\ell\to0$, and two collinear, $\ell||p_2$ and $\ell||p_3$.
In all of those regions,
we find that the integrand is approximated by a single function:
\begin{subequations} \begin{align}
\frac{\trp(4(\ell+p_{12})12)}{(\ell+p_{12})^2}
\xrightarrow{~\ell\to0~\,}&\,
\frac{\trp(4p_{12}12)}s=u\,,\\
\xrightarrow{\ell\to\tau p_2}&\,
\frac{\trp(4(p_1+(1+\tau)p_2)12)}{(p_1+(1+\tau)p_2)^2}=u\,,\\
\xrightarrow{\ell\to\tau p_3}&\,
\frac{\trp(4(p_{12}+\tau p_3)12)}{(p_{12}+\tau p_3)^2}=u\,.
\end{align} \label{eq:1LMixedNum1limits}%
\end{subequations}
In all other potentially soft or collinear regions
the integral vanishes due to the presence of the trace factor, \eg
\be
  \trp(4(\ell+p_{12})12)\xrightarrow{\ell+p_2\rightarrow0}0
\ee
Focusing on the divergent behavior,
we may replace $\trp(4(\ell+p_{12})12)/(\ell+p_{12})^2\to u$
at the integrand level.
Ignoring finite $\cO(\eps^0)$ terms, we retrieve a scalar triangle diagram:
\begin{subequations}
\be
\bI\bigg[n\bigg(\gBox[scale=.8,yshift=-1,eLA=$1^-$,eLB=$2$,eLC=$3$,eLD=$4^+$,
iA=quark,iD=quark,iC=quark,eB=quark,eC=aquark]{}\bigg)\bigg] =
\frac{\kappa_{(12)(13)}}{s}\times
\gTriA[rotate=180,eLB=$2$,eLC=$3$,eA=double,eB=line,eC=line,iA=line,iB=line,iC=line]{}
+\cO(\eps^0)\,,
\ee
where we have used the integral notation for the triangle integral
previously introduced in \eqn{eq:1Ltriangle}.
The other two diagrams are similarly expressed as
\begin{align}
\label{eq:1LMixedIRb}
\bI\left[n\!\left(\gBox[scale=.8,yshift=-1,eLA=$1^-$,eLB=$2$,eLC=$3^+$,eLD=$4$,iA=quark,iD=quark,
eB=quark,eD=aquark]{}\right)\right]&=
-\frac{\kappa_{(12)(14)}}{st}\left(
t\gTriA[rotate=180,eLB=$2$,eLC=$3$,eA=double,eB=line,eC=line,iA=line,iB=line,iC=line]{}+
s\gTriA[rotate=180,eLB=$3$,eLC=$4$,eA=double,eB=line,eC=line,iA=line,iB=line,iC=line]{}
\right)+\cO(\eps^0)\,,\\
\bI\left[n\!\left(\gBox[scale=.8,yshift=-1,eLA=$1$,eLB=$2^-$,eLC=$3^+$,eLD=$4$,iD=quark,
eA=quark,eD=aquark]{}\right)\right]&=
\frac{\kappa_{(12)(24)}}{st}\left(
s\gTriA[rotate=180,eLB=$1$,eLC=$2$,eA=double,eB=line,eC=line,iA=line,iB=line,iC=line]{}+
t\gTriA[rotate=180,eLB=$2$,eLC=$3$,eA=double,eB=line,eC=line,iA=line,iB=line,iC=line]{}+
s\gTriA[rotate=180,eLB=$3$,eLC=$4$,eA=double,eB=line,eC=line,iA=line,iB=line,iC=line]{}
\right)+\cO(\eps^0)\,.
\end{align} \label{eq:1LMixedIR}%
\end{subequations}
Each triangle is associated with a soft region on the left-hand side,
corresponding to when gluons are exchanged on the outside edge.\footnote{We
have implicitly made a choice in \eqn{eq:1LMixedNum1limits}
when we decided not to apply the kinematic limits to the ``eikonal'' propagators
which in the soft limit would become linear,
$(\ell+p_2)^2 \to 2\ell\cdot p_2$,
$(\ell-p_3)^2 \to -2\ell\cdot p_3$.
This allowed us to avoid spurious UV divergences and
at the same time to implement the collinear divergences
into the same approximating function as the soft ones.
This choice is consistent with the prescription of \rcite{Nagy:2003qn},
also recently used in \rcite{Anastasiou:2018rib}.
}

In terms of color factors the IR divergent regions are
naturally represented by the dipole operator $\bT_i\cdot\bT_j$,
which attaches a gluon bridge to the tree-level factors.
We find it illuminating to use color diagrams to illustrate this point.
For instance,  the three diagrams in \eqn{eq:1LMixedIR}
can be obtained from tree-level color factors as follows:
\begin{align}
-c\bigg(\gBox[scale=.8,yshift=-1,eLA=$1$,eLB=$2$,eLC=$3$,eLD=$4$,
iA=quark,iD=quark,iC=quark,eB=quark,eC=aquark]{}\bigg)&=
(\bT_2\cdot\bT_3)\:\!
c\Big(\gTreeS[eLA=$1$,eLB=$2$,eLC=$3$,eLD=$4$,eB=quark,eC=aquark,iA=aquark]{}\Big)\,,\nn\\
-c\bigg(\gBox[scale=.8,yshift=-1,eLA=$1$,eLB=$2$,eLC=$3$,eLD=$4$,
iA=quark,iD=quark,eB=quark,eD=aquark]{}\bigg)&=
(\bT_2\cdot\bT_3)\:\!
c\Big(\gTreeS[eLA=$1$,eLB=$2$,eLC=$3$,eLD=$4$,eB=quark,eD=aquark,iA=aquark]{}\Big)=
(\bT_3\cdot\bT_4)\:\!
c\!\left(\gTreeT[eLA=$1$,eLB=$2$,eLC=$3$,eLD=$4$,eB=quark,eD=aquark,iA=quark]{}\right)\,,\\
-c\bigg(\gBox[scale=.8,yshift=-1,eLA=$1$,eLB=$2$,eLC=$3$,
eLD=$4$,iD=quark,eA=quark,eD=aquark]{}\bigg)&=
(\bT_1\cdot\bT_2)\:\!
c\Big(\gTreeT[eLA=$1$,eLB=$2$,eLC=$3$,eLD=$4$,eA=quark,eD=aquark]{}\Big)=
(\bT_2\cdot\bT_3)\:\!
c\Big(\gTreeS[eLA=$1$,eLB=$2$,eLC=$3$,eLD=$4$,eA=quark,eD=aquark,iA=aquark]{}\Big)=
(\bT_3\cdot\bT_4)\:\!
c\Big(\gTreeT[eLA=$1$,eLB=$2$,eLC=$3$,eLD=$4$,eA=quark,eD=aquark]{}\Big).\nn
\end{align}
The idea here is that one should insert
the appropriate relation for a particular soft region.
Putting the pieces together, it is then a simple exercise to show that\footnote{One also requires the commutation relation
$$c\!\left(\gTreeS[eLA=$1$,eLB=$2$,eLC=$3$,eLD=$4$,eA=quark,eB=aquark]{}\right)=
c\!\left(\gTreeT[eLA=$1$,eLB=$2$,eLC=$3$,eLD=$4$,eA=quark,eB=aquark,iA=aquark]{}\right)-
c\!\left(\gTreeU[eLA=$1$,eLB=$2$,eLC=$3$,eLD=$4$,eA=quark,eB=aquark,iA=aquark]{}\right).$$}
\be
\cM_4^{(1)}
=\left(\sum_{i<j}^4s_{ij}
\gTriA[rotate=180,eLB=$i$,eLC=$j$,eA=double,eB=line,eC=line,iA=line,iB=line,iC=line]{}
\bT_i\cdot\bT_j\right) \cM_4^{(0)} + \cM_4^{(1){\rm fin}}\,,
\ee
where $\cM_4^{(1){\rm fin}}=\cO(\eps^0)$,
and the tree-level amplitude may be written
in the color basis of Del Duca, Dixon, and Maltoni~\cite{DelDuca:1999rs} as
\be
   \cM_4^{(0)}
    =-i\kappa_{(13)(23)}
       \bigg[ \frac{1}{st}
              c\Big(\gTreeT[eLA=$1$,eLB=$2$,eLC=$3$,eLD=$4$,
                            eA=quark,eB=aquark,iA=aquark]{}\Big)
            + \frac{1}{su}
              c\Big(\gTreeT[eLA=$1$,eLB=$2$,eLC=$4$,eLD=$3$,
                            eA=quark,eB=aquark,iA=aquark]{}\Big)
       \bigg]  \,.
\ee
In this way,
we have derived the one-loop IR factorization formula~\eqref{Factorization2loop1}
for the anomalous dimensions
$\gamma_g^{(1)}=-\beta_0/2$ and $\gamma_q^{(1)}=\gamma_{\bar{q}}^{(1)}=0$,
as given by \eqn{eq:N2AnomDimensions}.
By promoting ${\mathbf S}(\eps)$ to include a full triangle integral
we have incorporated $\cO(\eps^0)$ terms into the finite piece,
so $\cM_4^{(1){\rm fin}}\neq\cH_4^{(1)}$
when comparing with \eqn{Factorization2loop1}.

Let us now comment on the transcendentality structure of this amplitude.
Performing integrand reduction on the three numerators in \eqn{eq:1LMixedNums}
we see that only scalar box and triangle integrals appear.
As both of these have uniform transcendental weight,
$\cM_4^{(1)}$, $\cM_4^{(1){\rm fin}}$, and $\cH_4^{(1)}$
all contain only weight-2 terms at $\cO(\eps^0)$.
So in this example the choice of IR scheme has no bearing on transcendentality.
The uniform weight property is linked to the absence of scalar bubble integrals,
which in this case happens without the need to specialize to the
conformal theory where $N_f=C_A/T_F$.
The dependence on number of flavors $N_f$,
encoded by $\beta_0 = C_A - N_f T_F$,
cancels between $\beta_0$ and $2 \gamma_g^{(1)}$,
which is reflected by the absence of closed matter loops
in the integrand~\eqref{eq:1LMixedAmp}.

In this exercise we have seen how an analysis of soft regions
can be used to expose the IR behavior of one-loop amplitudes.
A similar exercise works for the one-loop four-quark amplitude;
however, in that case the color-dual numerators presented in \rcite{Kalin:2018thp}
have soft and
collinear divergences in undesirable regions associated with matter lines.
Some re-arrangement of the integrand is necessary,
which has the unfortunate side-effect of spoiling color-kinematics duality.
In the next example, involving only external gluons,
we will see how relaxing the duality yields an integrand well suited
for our analysis of the divergence structure.

\subsection{External vectors}
\label{sec:1lvectors}

In \rcites{Carrasco:2012ca,Bern:2013yya,Nohle:2013bfa,Chiodaroli:2013upa,
Ochirov:2013xba,Johansson:2014zca}
the one-loop four-gluon integrand was presented in a form satisfying color-kinematics duality.
By allowing ourselves to violate the duality, we write
down the integrand in a form that more readily exhibits its expected IR behavior:
\begin{subequations}
\begin{align}
\label{eq:1lbox}
n\Bigg(\gBox[scale=.8,yshift=-1,eLA=$1^-$,eLB=$2^+$,eLC=$3^-$,eLD=$4^+$,
iA=aquark,iB=aquark,iD=aquark,iC=aquark,iLD=$\uset{\rightarrow}{\ell}$]{}\!\Bigg) &
= \frac{\kappa_{13}}{u^2} \trm(1\ell(\ell+p_4)3) \,,\qquad \qquad \quad
n\Bigg(\gBox[scale=.8,yshift=-1,eLA=$1^-$,eLB=$2^-$,eLC=$3^+$,eLD=$4^+$,
iA=aquark,iB=aquark,iD=aquark,iC=aquark,iLD=$\uset{\rightarrow}{\ell}$]{}\!\Bigg)
= \frac{\kappa_{12}}{s}\mu^2 \,, \\
n\bigg(\gTriC[scale=.8,eLA=$2^-$,eLB=$3^+$,eLC=$4^+$,eLD=$1^-$,
iB=aquark,iD=aquark,iC=aquark,iLC=$\ell\uparrow$]{}\!\bigg) &
= -\frac{1}{2}
n\bigg(\gTriC[scale=.8,eLD=$1^-$,eLA=$2^-$,eLB=$3^+$,eLC=$4^+$,iLC=$\ell\uparrow$]{}\!\bigg)
= \frac{\kappa_{12}}{s^2} \trp(1\ell43) \,, \\
n\Big(\gBubB[scale=.8,eLA=$1^-$,eLB=$2^+$,eLC=$3^+$,eLD=$4^-$,iB=aquark,iC=aquark]{}\Big) &
= -\frac{1}{2} n\Big(\gBubB[scale=.8,eLA=$1^-$,eLB=$2^+$,eLC=$3^+$,eLD=$4^-$]{}\Big)
= -\frac{s}{t} \kappa_{14} \,.
\end{align} \label{eq:1Lnumerators}%
\end{subequations}
Recall that our external-state notation involving $\kappa_{ij}$ lets us
add multiple helicity configurations within the same object.
For example, the matter bubble numerator can be promoted to
\be
n\Big(\gBubB[scale=.8,eLA=$1$,eLB=$2$,eLC=$3$,eLD=$4$,iB=aquark,iC=aquark]{}\Big)
= -s \bigg( \frac{\kappa_{14}+\kappa_{23}}{t}
          - \frac{\kappa_{13}+\kappa_{24}}{u} \bigg) .
\label{eq:1Lbubble}
\ee
The only non-vanishing numerator absent from \eqn{eq:1Lnumerators}
is the purely vector box; it is given by the $\cN=4$ matching identity
\be
n\bigg(\gBox[scale=.8,eLA=$1$,eLB=$2$,eLC=$3$,eLD=$4$]{}\bigg)
   = n^{[\cN=4]}\bigg(\gBox[scale=.8,eLA=$1$,eLB=$2$,eLC=$3$,eLD=$4$]{}\bigg)
   - n\bigg(\gBox[scale=.8,eLA=$1$,eLB=$2$,eLC=$3$,eLD=$4$,iA=aquark,iB=aquark,
                     iD=aquark,iC=aquark]{}\bigg)
   - n\bigg(\gBox[scale=.8,eLA=$1$,eLB=$2$,eLC=$3$,eLD=$4$,iA=quark,iB=quark,
                     iD=quark,iC=quark]{}\bigg)\,,
\label{eq:N4matching1loop}
\ee
where the two matter numerators are equal.
The $\cN=4$ box numerator is explicitly
\be\label{eq:N4box}
n^{[\cN=4]}\bigg(\gBox[scale=.8,eLA=$1$,eLB=$2$,eLC=$3$,eLD=$4$]{}\bigg)=
\sum_{i<j}\kappa_{ij}\,.
\ee
As discussed in \sec{sec:N2review},
identities of the type~\eqref{eq:N4matching1loop} follow from the fact that
after stripping away the color information,
the states of the $\cN=2$ vector and matter multiplets
add up to a single $\cN=4$ multiplet.

An advantage of the above diagrammatic representation is that
the triangle integrals vanish to all orders in $\eps$
due to basic symmetry arguments.
Using these expressions, let us consider the divergent behavior of the amplitude:
\beal
\cM^{(1)}_4=i\sum_{S_4}\bI\bigg[
\frac18c\bigg(\gBox[scale=.8,eLA=$1$,eLB=$2$,eLC=$3$,eLD=$4$]{}\bigg)
n\bigg(\gBox[scale=.8,eLA=$1$,eLB=$2$,eLC=$3$,eLD=$4$]{}\bigg)&+
\frac{N_f}4c\bigg(\gBox[scale=.8,eLA=$1$,eLB=$2$,eLC=$3$,eLD=$4$,
                  iA=aquark,iB=aquark,iD=aquark,iC=aquark]{}\bigg)
          n\bigg(\gBox[scale=.8,eLA=$1$,eLB=$2$,eLC=$3$,eLD=$4$,
                  iA=aquark,iB=aquark,iD=aquark,iC=aquark]{}\bigg)\\\qquad+
\frac1{16}c\Big(\gBubB[scale=.8,eLA=$1$,eLB=$2$,eLC=$3$,eLD=$4$]{}\Big)
n\Big(\gBubB[scale=.8,eLA=$1$,eLB=$2$,eLC=$3$,eLD=$4$]{}\Big)&+
\frac{N_f}{8}c\Big(\gBubB[scale=.8,eLA=$1$,eLB=$2$,eLC=$3$,eLD=$4$,iB=aquark,iC=aquark]{}\Big)
 n\Big(\gBubB[scale=.8,eLA=$1$,eLB=$2$,eLC=$3$,eLD=$4$,iB=aquark,iC=aquark]{}\Big)\bigg]\,,
\eeal
where the sum is over permutations of external legs.
For comparison, the corresponding $\cN=4$ SYM amplitude consists only of the box \eqref{eq:N4box}:
\be
\cM^{(1)[\cN=4]}_4=\frac{i}{8}\sum_{S_4}
c\bigg(\gBox[scale=.8,eLA=$1$,eLB=$2$,eLC=$3$,eLD=$4$]{}\bigg)\bI\bigg[
n^{[\cN=4]}\bigg(\gBox[scale=.8,eLA=$1$,eLB=$2$,eLC=$3$,eLD=$4$]{}\bigg)\bigg]\,.
\ee
The remainder function~$\cW_4^{(1)}=\cM_4^{(1)}-\cM^{(1)[\cN=4]}_4$
(see \sec{sec:N4diff}) is then given as:
\beal
\cW_4^{(1)}=-\frac{i}4\sum_{S_4}
\bI\bigg[
\left(c\bigg(\gBox[scale=.8,eLA=$1$,eLB=$2$,eLC=$3$,eLD=$4$]{}\bigg)-
N_fc\bigg(\gBox[scale=.8,eLA=$1$,eLB=$2$,eLC=$3$,eLD=$4$,
iA=aquark,iB=aquark,iD=aquark,iC=aquark]{}\bigg)\right) &
n\bigg(\gBox[scale=.8,eLA=$1$,eLB=$2$,eLC=$3$,eLD=$4$,
iA=aquark,iB=aquark,iD=aquark,iC=aquark]{}\bigg)\\
+\,\frac{\beta_0}2
c\Big(\gTreeS[eLA=$1$,eLB=$2$,eLC=$3$,eLD=$4$]{}\Big) &
n\Big(\gBubB[scale=.8,eLA=$1$,eLB=$2$,eLC=$3$,eLD=$4$,iB=aquark,iC=aquark]{}\Big)
\bigg]\,,
\label{eq:1loopRemainder}
\eeal
where we have reinstated $\beta_0=C_A-N_f T_F$
by factoring out the Casimir values from the bubble color factors.

In the box numerators~\eqref{eq:1lbox},
the $\mu^2$ components integrate to $\cO(\eps)$ terms.
In the other helicity configurations
it is the Dirac traces that block all potentially IR-singular regions
without introducing any additional UV divergences,
so all box contributions are finite.
Therefore, $\cW_4^{(1)}$ only diverges in the UV
due to the bubble integrals:
\be
\gBubA[eA=double,eB=double,iA=line,iB=line]{}=
\frac{r_\Gamma}{\eps(1-2\eps)}(-p^2)^{-\eps}=
\frac{1}{\eps}+2-\log(-p^2)+\cO(\eps)\,.\label{eq:bubble}
\ee
One can now easily show that
the permutation sum in \eqn{eq:1loopRemainder} leads to
the correct factorization:
\be
   \cW_4^{(1)} = -\frac{\beta_0}{\eps}\cM_4^{(0)}
   +\cH_4^{(1)} - \cH_4^{(1)[\cN=4]}\,,
\label{eq:1loopUV}
\ee
where the tree-level amplitude in this case is
\be
   \cM_4^{(0)}
    =-i\sum_{i<j}^4\kappa_{ij}
      \bigg[ \frac{1}{st} c\Big(\gTreeT[eLA=$1$,eLB=$2$,eLC=$3$,eLD=$4$]{}\Big)
           + \frac{1}{su} c\Big(\gTreeT[eLA=$1$,eLB=$2$,eLC=$4$,eLD=$3$]{}\Big)
      \bigg] \,.
\ee
As predicted in \sec{sec:N4diff},
the IR behavior of the one-loop four-vector amplitude
is entirely captured by $\cN=4$ SYM.
As for the UV, the bubble integral \eqref{eq:bubble} contributes
weight-0,1 terms to the $\cO(\eps^0)$ part of $\cH_4^{(1)}$
which violate uniform transcendentality.

Might there exist a better UV-subtraction scheme that
ameliorates the transcendentality properties?
One could be tempted to promote the $1/\eps$ divergence
in \eqn{eq:1loopUV} to a full bubble integral
and thus absorb the unwanted extra terms in $\cH_4^{(1)}$.
However, this would introduce additional
kinematic dependence of the form $\log(-p^2)$ into the subtraction,
and whatever color-space operator acts on the tree-level amplitude
must be symmetric on $s$, $t$, and $u$ --- therein lies the problem.
For instance, the only $s$-channel bubble in $\cW_4^{(1)}$ has
a kinematic numerator given in \eqn{eq:1Lbubble};
as it does not contain all six $\kappa_{ij}$ components,
it cannot be made proportional to the tree amplitude
in all external helicity configurations at once.
So no sensible color-space operator,
kinematically depending on bubbles and acting on $\cM_4^{(0)}$,
manages to reproduce the desired behavior.

We must therefore conclude that,
unless we specialize to the conformal theory where $\beta_0=0$,
the lower-weight terms introduced by bubble integrals to the
one-loop amplitudes cannot be subtracted.
This will also be true for less supersymmetry and more external legs.
At two loops, our ability to incorporate bubbles into two-loop triangle
integrals will enable us to subtract them consistently,
and thus radically improve the transcendentality properties of the amplitudes.

\section{IR structure of two-loop \texorpdfstring{$\cN=2$}{N=2} SQCD}
\label{sec:twoloop}

In this section we examine the IR behavior of
the two-loop four-point amplitudes prior to full integration.
This precedes our analysis of their transcendentality properties after integration,
which we will do in \sec{sec:integrated}.
In the four-gluon case,
we use the compact form of the integrand presented in \rcite{Kalin:2018thp},
which is well-suited to the analysis of its pole structure;
here we expose it at the level of individual diagrams before integration.
This provides us with a strong motivation of the IR decomposition
presented in~\eqn{eq:Factorization2loopN4int}.
For the mixed amplitude,
we show how --- by relaxing the constraints imposed by color-kinematics duality ---
the integrand can be reorganized into a form making use of similar Dirac traces,
which again highlights the structure.

\subsection{External vectors}
\label{sec:2Lextvectors}

The full two-loop four-vector amplitude may be expressed in terms of
ten diagrams of four topologies, none of which vanish upon integration:
\beal\label{eq:2loopVectorAmplitude}
\cM^{(2)}_4&=-i\sum_{S_4}\bI\bigg[
\frac14c\bigg(\gBoxBox[scale=.6,eLA=$1$,eLB=$2$,eLC=$3$,eLD=$4$]{}\bigg)
n\bigg(\gBoxBox[scale=.6,eLA=$1$,eLB=$2$,eLC=$3$,eLD=$4$]{}\bigg)+
N_fc\bigg(\gBoxBox[scale=.6,eLA=$1$,eLB=$2$,eLC=$3$,eLD=$4$,
iA=aquark,iB=aquark,iF=aquark,iG=quark]{}\bigg)
n\bigg(\gBoxBox[scale=.6,eLA=$1$,eLB=$2$,eLC=$3$,eLD=$4$,
iA=aquark,iB=aquark,iF=aquark,iG=quark]{}\bigg)
\\&\qquad+
\frac{N_f}2c\bigg(\gBoxBox[scale=.6,eLA=$1$,eLB=$2$,eLC=$3$,eLD=$4$,
iA=aquark,iB=aquark,iF=aquark,iE=aquark,iD=aquark,iC=aquark]{}\bigg)
n\bigg(\gBoxBox[scale=.6,eLA=$1$,eLB=$2$,eLC=$3$,eLD=$4$,
iA=aquark,iB=aquark,iF=aquark,iE=aquark,iD=aquark,iC=aquark]{}\bigg)+
\frac14c\bigg(\!\!\gBoxBoxNP[scale=.6,eLA=$1$,eLB=$2$,eLC=$4$,eLD=$3$]{}\bigg)
n\bigg(\!\!\gBoxBoxNP[scale=.6,eLA=$1$,eLB=$2$,eLC=$4$,eLD=$3$]{}\bigg)
\\&\qquad+
N_fc\bigg(\!\!\gBoxBoxNP[scale=.6,eLA=$1$,eLB=$2$,eLC=$4$,eLD=$3$,
iA=aquark,iB=aquark,iC=aquark,iD=aquark,iE=aquark]{}\bigg)
n\bigg(\!\!\gBoxBoxNP[scale=.6,eLA=$1$,eLB=$2$,eLC=$4$,eLD=$3$,
iA=aquark,iB=aquark,iC=aquark,iD=aquark,iE=aquark]{}\bigg)+
\frac{N_f}2c\bigg(\!\!\gBoxBoxNP[scale=.6,eLA=$1$,eLB=$2$,eLC=$4$,eLD=$3$,
iC=aquark,iD=aquark,iF=aquark,iG=aquark]{}\bigg)
n\bigg(\!\!\gBoxBoxNP[scale=.6,eLA=$1$,eLB=$2$,eLC=$4$,eLD=$3$,
iC=aquark,iD=aquark,iF=aquark,iG=aquark]{}\bigg)
\\&\qquad+
\frac12c\bigg(\!\!\gTriPenta[scale=.6,eLA=$1$,eLB=$2$,eLC=$3$,eLD=$4$]{}\!\bigg)
n\bigg(\!\!\gTriPenta[scale=.6,eLA=$1$,eLB=$2$,eLC=$3$,eLD=$4$]{}\!\bigg)+
N_fc\bigg(\!\!\gTriPenta[scale=.6,eLA=$1$,eLB=$2$,eLC=$3$,eLD=$4$,
iD=aquark,iG=aquark,iE=aquark]{}\!\bigg)
n\bigg(\!\!\gTriPenta[scale=.6,eLA=$1$,eLB=$2$,eLC=$3$,eLD=$4$,
iD=aquark,iG=aquark,iE=aquark]{}\!\bigg)\\&\qquad+
\frac14c\bigg(\gBoxBubA[scale=.6,eLA=$1$,eLB=$2$,eLC=$3$,eLD=$4$]{}\bigg)
n\bigg(\gBoxBubA[scale=.6,eLA=$1$,eLB=$2$,eLC=$3$,eLD=$4$]{}\bigg)+
\frac{N_f}2c\bigg(\gBoxBubA[scale=.6,eLA=$1$,eLB=$2$,eLC=$3$,eLD=$4$,
iE=aquark,iF=aquark]{}\bigg)
n\bigg(\gBoxBubA[scale=.6,eLA=$1$,eLB=$2$,eLC=$3$,eLD=$4$,
iE=aquark,iF=aquark]{}\bigg)
\bigg]\,.
\eeal
We could try to analyze this amplitude using the
minimal factorization formulae~\eqref{Factorization2loop}
or \eqref{Factorization2loopGluonic},
involving poles ranging from $1/\eps$ to $1/\eps^4$.
It is, however, much more appealing to use
the more transparent formula~\eqref{eq:Factorization2loopN4diff},
which holds after we subtract the maximally supersymmetric amplitude
\begin{align}
&\cM^{(2)[\cN=4]}_4\\*
&=-\frac{i}4\sum_{S_4}\bI\bigg[
c\bigg(\gBoxBox[scale=.6,eLA=$1$,eLB=$2$,eLC=$3$,eLD=$4$]{}\bigg)
n^{[\cN=4]}\bigg(\gBoxBox[scale=.6,eLA=$1$,eLB=$2$,eLC=$3$,eLD=$4$]{}\bigg)
+c\bigg(\!\!\gBoxBoxNP[scale=.6,eLA=$1$,eLB=$2$,eLC=$4$,eLD=$3$]{}\bigg)
n^{[\cN=4]}\bigg(\!\!\gBoxBoxNP[scale=.6,eLA=$1$,eLB=$2$,eLC=$4$,eLD=$3$]{}\bigg)
\bigg]\,.\nn
\end{align}
We are aided by the $\cN=4$ matching conditions,
which the kinematic numerators satisfy
by construction~\cite{Johansson:2017bfl,Kalin:2018thp}.
For instance, the following combination of two-loop non-planar numerators
add up to a single $\cN=4$ numerator:
\beal\label{eq:N4matching2loop}\!
n^{[\cN=4]}\bigg(\!\!\gBoxBoxNP[scale=.6,eLA=$1$,eLB=$2$,eLC=$4$,eLD=$3$]{}\bigg) =
n\bigg(\!\!\gBoxBoxNP[scale=.6,eLA=$1$,eLB=$2$,eLC=$4$,eLD=$3$]{}\bigg)+
n\bigg(\!\!\gBoxBoxNP[scale=.6,eLA=$1$,eLB=$2$,eLC=$4$,eLD=$3$,
                      iC=aquark,iD=aquark,iF=aquark,iG=aquark]{}\bigg)+
n\bigg(\!\!\gBoxBoxNP[scale=.6,eLA=$1$,eLB=$2$,eLC=$4$,eLD=$3$,
                      iA=aquark,iB=aquark,iC=aquark,iD=aquark,iE=aquark]{}\bigg)+
n\bigg(\!\!\gBoxBoxNP[scale=.6,eLA=$1$,eLB=$2$,eLC=$3$,eLD=$4$,
                      iA=aquark,iB=aquark,iC=aquark,iD=aquark,iE=aquark]{}\bigg)&\\+\,
n\bigg(\!\!\gBoxBoxNP[scale=.6,eLA=$1$,eLB=$2$,eLC=$4$,eLD=$3$,
                      iC=quark,iD=quark,iF=quark,iG=quark]{}\bigg)+
n\bigg(\!\!\gBoxBoxNP[scale=.6,eLA=$1$,eLB=$2$,eLC=$4$,eLD=$3$,
                      iA=quark,iB=quark,iC=quark,iD=quark,iE=quark]{}\bigg)+
n\bigg(\!\!\gBoxBoxNP[scale=.6,eLA=$1$,eLB=$2$,eLC=$3$,eLD=$4$,
                      iA=quark,iB=quark,iC=quark,iD=quark,iE=quark]{}\bigg)&\,,
\eeal
and similarly for the other three topologies
(two of which are zero in $\cN=4$ SYM).
These identities allow us to express the two-loop remainder
in terms of only six kinematic numerators:
\begin{align}
\label{eq:2loopRemainder}
\!\!\!\cW_4^{(2)}=i\sum_{S_4}\bI\bigg[
\frac12\left(c\bigg(\gBoxBox[scale=.6,eLA=$1$,eLB=$2$,eLC=$3$,eLD=$4$]{}\bigg)
-N_fc\bigg(\gBoxBox[scale=.6,eLA=$1$,eLB=$2$,eLC=$3$,eLD=$4$,
iA=aquark,iB=aquark,iF=aquark,iE=aquark,iD=aquark,iC=aquark]{}\bigg)\right)
&n\bigg(\gBoxBox[scale=.6,eLA=$1$,eLB=$2$,eLC=$3$,eLD=$4$,
iA=aquark,iB=aquark,iF=aquark,iE=aquark,iD=aquark,iC=aquark]{}\bigg)
\qquad\sim\cO(\eps^0)\\\qquad\qquad+
\left(c\bigg(\gBoxBox[scale=.6,eLA=$1$,eLB=$2$,eLC=$3$,eLD=$4$]{}\bigg)
-N_fc\bigg(\gBoxBox[scale=.6,eLA=$1$,eLB=$2$,eLC=$3$,eLD=$4$,
iA=aquark,iB=aquark,iF=aquark,iG=quark]{}\bigg)\right)
&n\bigg(\gBoxBox[scale=.6,eLA=$1$,eLB=$2$,eLC=$3$,eLD=$4$,
iA=aquark,iB=aquark,iF=aquark,iG=quark]{}\bigg)
\qquad\sim\cO(\eps^{-2})\nn\\\qquad\qquad+
\left(c\bigg(\!\!\gBoxBoxNP[scale=.5,eLA=$1$,eLB=$2$,eLC=$4$,eLD=$3$]{}\bigg)-
N_fc\bigg(\!\!\gBoxBoxNP[scale=.6,eLA=$1$,eLB=$2$,eLC=$4$,eLD=$3$,
iA=aquark,iB=aquark,iC=aquark,iD=aquark,iE=aquark]{}\bigg)\right)
&n\bigg(\!\!\gBoxBoxNP[scale=.6,eLA=$1$,eLB=$2$,eLC=$4$,eLD=$3$,
iA=aquark,iB=aquark,iC=aquark,iD=aquark,iE=aquark]{}\bigg)
\quad~\:\,\sim\cO(\eps^{-1})\nn\\\qquad\qquad+
\frac12\left(c\bigg(\!\!\gBoxBoxNP[scale=.5,eLA=$1$,eLB=$2$,eLC=$4$,eLD=$3$]{}\bigg)-
N_fc\bigg(\!\!\gBoxBoxNP[scale=.6,eLA=$1$,eLB=$2$,eLC=$4$,eLD=$3$,
iC=aquark,iD=aquark,iF=aquark,iG=aquark]{}\bigg)\right)
&n\bigg(\!\!\gBoxBoxNP[scale=.6,eLA=$1$,eLB=$2$,eLC=$4$,eLD=$3$,
iC=aquark,iD=aquark,iF=aquark,iG=aquark]{}\bigg)
\quad~\:\,\sim\cO(\eps^{-2})\nn\\\qquad\qquad
+\beta_0c\bigg(\gBox[scale=.8,eLA=$1$,eLB=$2$,eLC=$3$,eLD=$4$]{}\bigg)
\bigg(n\bigg(\!\!\gTriPenta[scale=.6,eLA=$1$,eLB=$2$,eLC=$3$,eLD=$4$,
iD=aquark,iG=aquark,iE=aquark]{}\!\bigg)+\,
&n\bigg(\gBoxBubA[scale=.6,eLA=$1$,eLB=$2$,eLC=$3$,eLD=$4$,
iE=aquark,iF=aquark]{}\bigg)\bigg)
\bigg]\,,\quad\:\sim\cO(\eps^{-3})\nn
\end{align}
where to the right of each line we have displayed its divergent behavior.
Ahead of the detailed analysis below,
let us point out the prominent features of the remainder~\eqref{eq:2loopRemainder}.

Due to the IR-blocking numerator properties
there is no overlap of different soft, collinear, or UV regions
in any of the remaining diagrams, as all of them contain a closed matter loop.
Indeed, by using the $\cN=4$ matching identities
we have eliminated exactly the four diagrams that did not have any such loops
and therefore incorporated the leading $1/\eps^4$ divergences ---
shared with the subtracted $\cN=4$ amplitude.

Note that the color factors for the two topologies
in the last line of \eqn{eq:2loopRemainder} have been rearranged
with respect to \eqn{eq:2loopVectorAmplitude}.
They coincide up to certain Casimir factors,
which leads to a natural appearance of the beta-function coefficient.
This is consistent with the fact that these topologies,
which we dub ``pentagon-triangle'' and ``box-bubble'',
are the only ones found to be UV-divergent.
Their UV divergences come from the closed matter loops
and enhance the $1/\eps^2$ IR divergence behavior of the gluonic part of the diagrams
to an overall $1/\eps^3$.

The leading IR divergence rates of the other four diagrams,
as indicated in \eqn{eq:2loopRemainder},
can be understood from the IR-blocking numerator properties,
which only allow vector propagators to produce poles in $\eps$.

As derived from the general principles in \sec{sec:IRfactorization},
the remainder~\eqref{eq:2loopRemainder} must obey
the factorization formula~\eqref{eq:Factorization2loopN4int}.
Since the amplitude has already been fully integrated in \rcite{Duhr:2019ywc},
for illustrative purposes we find it sufficient to perform
an integrand-level proof of the factorization
up to and including terms $\cO(\eps^{-2})$.
At this level, we can further simplify the formula
by noticing that the two-loop triangles~\eqref{eq:2Ltriangles} satisfy
\be
   \gTriBubB[scale=0.8,rotate=180,all=line,eA=doubleline,eLB=$i$,eLC=$j$]{}
 = 2 \gTriBubA[scale=0.8,rotate=180,all=line,eA=doubleline,eLB=$i$,eLC=$j$]{}
 + \cO(\eps^{-1})
 \label{eq:trianglerelation}
\ee
up to the first two orders in $\eps$.
It is therefore sufficient to prove
\be
   \cW_4^{(2)}
    = \left( \sum_{i<j}^4 s_{ij}
             \gTriA[scale=0.8,rotate=180,eLB=$i$,eLC=$j$,
                    eA=doubleline,eB=line,eC=line,iA=line,iB=line,iC=line]{}
             \bT_i\cdot\bT_j\!
      \right) \cW_4^{(1)}
    + \beta_0\!
      \left( \sum_{i<j}^4s_{ij}
             \gTriBubA[scale=0.8,rotate=180,all=line,
                       eA=doubleline,eLB=$i$,eLC=$j$]{}
            \bT_i\cdot\bT_j\!
      \right) \cM_4^{(0)}+ \cO(\eps^{-1})\, ,
\label{eq:2loopIR}
\ee
where expressions for $\cW_4^{(1)}$ and $\cM_4^{(0)}$ are given
in \eqns{eq:1loopRemainder}{eq:1loopUV}, respectively.

Below we explain how to interpret the divergent behavior of each of
the integrals appearing in the two-loop remainder $\cW_4^{(2)}$.
Once the integrals have been decomposed into their various divergent regions,
it is a simple (albeit cumbersome) task to assemble them
and reproduce \eqn{eq:2loopIR}.

\subsubsection{Planar and non-planar double boxes}
\label{sec:doubleboxes}

Up to relabelings of their loop momenta,
the following three planar and non-planar double-box numerators are equal
in any helicity configuration and are given by
\begin{subequations} \begin{align}\!
n\!\left(\gBoxBox[scale=.8,eLA=$1^-$,eLB=$2^-$,eLC=$3^+$,eLD=$4^+$,
iA=aquark,iB=aquark,iF=aquark,iE=aquark,iD=aquark,iC=aquark,
iLA=$\downarrow\!\ell_1$,iLD=$\ell_2\!\downarrow$]{}\right)\!=
n\!\left(\gBoxBox[scale=.8,eLA=$1^-$,eLB=$2^-$,eLC=$3^+$,eLD=$4^+$,
iA=aquark,iB=aquark,iF=aquark,iG=quark,
iLA=$\downarrow\!\ell_1$,iLG=$\!\!\downarrow\!\!\ell_2$]{}\right)\!=
n\!\left(\!\!\gBoxBoxNP[scale=.6,eLA=$1^-$,eLB=$2^-$,eLC=$4^+$,eLD=$3^{\;\!\!\!^+}$,
iA=aquark,iB=aquark,iC=aquark,iD=aquark,iE=aquark,
iLA=$\downarrow\!\ell_1$,iLD=$\ell_2\,$]{\draw[line width=0.45pt,-to]
(0.5,1.3) -- (0.15,0.95);}\!\right)\!
=&-\kappa_{12}\mu_{12}\,,\\\!
n\!\left(\gBoxBox[scale=.8,eLA=$1^-$,eLB=$2^+$,eLC=$3^-$,eLD=$4^+$,
iA=aquark,iB=aquark,iF=aquark,iE=aquark,iD=aquark,iC=aquark,
iLA=$\downarrow\!\ell_1$,iLD=$\ell_2\!\downarrow$]{}\right)\!=
n\!\left(\gBoxBox[scale=.8,eLA=$1^-$,eLB=$2^+$,eLC=$3^-$,eLD=$4^+$,
iA=aquark,iB=aquark,iF=aquark,iG=quark,
iLA=$\downarrow\!\ell_1$,iLG=$\!\!\downarrow\!\!\ell_2$]{}\right)\!=
n\!\left(\!\!\gBoxBoxNP[scale=.6,eLA=$1^-$,eLB=$2^+$,eLC=$4^+$,eLD=$3^{\;\!\!\!^-}$,
iA=aquark,iB=aquark,iC=aquark,iD=aquark,iE=aquark,
iLA=$\downarrow\!\ell_1$,iLD=$\ell_2\,$]{\draw[line width=0.45pt,-to]
(0.5,1.3) -- (0.15,0.95);}\!\right)\!
=&\,\frac{\kappa_{13}}{u^2}\trm(1\ell_124\ell_23)\,,\\\!
n\!\left(\gBoxBox[scale=.8,eLA=$1^-$,eLB=$2^+$,eLC=$3^+$,eLD=$4^-$,
iA=aquark,iB=aquark,iF=aquark,iE=aquark,iD=aquark,iC=aquark,
iLA=$\downarrow\!\ell_1$,iLD=$\ell_2\!\downarrow$]{}\right)\!=
n\!\left(\gBoxBox[scale=.8,eLA=$1^-$,eLB=$2^+$,eLC=$3^+$,eLD=$4^-$,
iA=aquark,iB=aquark,iF=aquark,iG=quark,
iLA=$\downarrow\!\ell_1$,iLG=$\!\!\downarrow\!\!\ell_2$]{}\right)\!=
n\!\left(\!\!\gBoxBoxNP[scale=.6,eLA=$1^-$,eLB=$2^+$,eLC=$4^-$,eLD=$3^{\;\!\!\!^+}$,
iA=aquark,iB=aquark,iC=aquark,iD=aquark,iE=aquark,
iLA=$\downarrow\!\ell_1$,iLD=$\ell_2\,$]{\draw[line width=0.45pt,-to]
(0.5,1.3) -- (0.15,0.95);}\!\right)\!
=&\,\frac{\kappa_{14}}{t^2}\trm(1\ell_123\ell_24)\,.
\end{align} \label{eq:doubleBoxes}%
\end{subequations}
The state configurations with flipped helicities, which are not shown,
can be obtained by an appropriate replacement
of the $\kappa$-prefactors ($\kappa_{12} \to \kappa_{34}$ and so on)
and switching the parity of the traces $\trm \to \trp$.

The integrals of all three numerators in the first column are entirely finite.
Indeed, setting the loop momentum of any internal edge carrying hypermultiplets
to zero forces the corresponding numerator to vanish;
similarly for any collinear region that might otherwise have produced a divergence.
The overall powers of $\ell_1$ and $\ell_2$ are also too low to produce UV divergences.
Finite expressions for the integrals involving six-term traces were
obtained by Caron-Huot and Larsen in \rcite{CaronHuot:2012ab};
the integral of $\mu_{12}$ can be done using a Schwinger parametrization,
as we shall explain in \sec{sec:integrated}.
For the present purpose of understanding divergences,
we can safely ignore these integrals,
which correspond to the first line of \eqn{eq:2loopRemainder}.

A similar examination of the non-planar numerators
in the third column of \eqn{eq:doubleBoxes}
reveals that the only permitted divergence
is in the collinear region $(\ell_{12}+p_1) || p_3$.
It naturally arises from the three-gluon vertex in the center of the diagram.
However, as it only gives rise to an $\cO(\eps^{-1})$ divergence
we do not require its precise form for our present analysis.

As for the three diagrams in the second column of \eqn{eq:doubleBoxes},
their integrals behave as $\cO(\eps^{-2})$
due to the possibility of simultaneous soft and collinear divergences
in the gluonic rung on the left.
An integrand analysis lets us effortlessly extract
the leading order in $\eps$ (see also \rcites{Badger:2015lda,Badger:2016ozq}):
\begin{subequations} \begin{align}
  \gBoxBox[yshift=-1,all=line,scale=0.6,eLA=$1$,eLB=$2$,eLC=$3$,eLD=$4$,iLE=$\uset{\leftarrow}{\ell_2}$,iLF=$\uset{\rightarrow}{\ell_1}$]{}
  [\mu_{13}]&=
  \gTriA[all=line,scale=0.8,eLB=$3$,eLC=$4$,eA=double,eB=line,eC=line,iA=line,iB=line,iC=line]{}
  \times
  \gBox[yshift=-1,all=line,scale=0.7,eLA=$1$,eLB=$2$,eLC=$3$,eLD=$4$,iLD=$\uset{\rightarrow}{\ell}$]{}
  [\mu^2]+\cO(\eps^{-1})\,,\\
  \gBoxBox[yshift=-1,all=line,scale=0.6,eLA=$1$,eLB=$2$,eLC=$3$,eLD=$4$,iLE=$\uset{\leftarrow}{\ell_2}$,iLF=$\uset{\rightarrow}{\ell_1}$]{}
  [\text{tr}_\pm(1\ell_124\ell_33)]&=
  \gTriA[all=line,scale=0.8,eLB=$3$,eLC=$4$,eA=double,eB=line,eC=line,iA=line,iB=line,iC=line]{}
  \times
  \gBox[yshift=-1,all=line,scale=0.7,eLA=$1$,eLB=$2$,eLC=$3$,eLD=$4$,iLD=$\uset{\rightarrow}{\ell}$]{}
  [\text{tr}_\pm(1\ell24\ell3)]+\cO(\eps^{-1})\,,\\
  \gBoxBox[yshift=-1,all=line,scale=0.6,eLA=$1$,eLB=$2$,eLC=$3$,eLD=$4$,iLE=$\uset{\leftarrow}{\ell_2}$,iLF=$\uset{\rightarrow}{\ell_1}$]{}
  [\text{tr}_\pm(1\ell_123\ell_34)]&=
  \gTriA[all=line,scale=0.8,eLB=$3$,eLC=$4$,eA=double,eB=line,eC=line,iA=line,iB=line,iC=line]{}
  \times
  \gBox[yshift=-1,all=line,scale=0.7,eLA=$1$,eLB=$2$,eLC=$3$,eLD=$4$,iLD=$\uset{\rightarrow}{\ell}$]{}
  [\text{tr}_\pm(1\ell23\ell4)]+\cO(\eps^{-1})\,.
\end{align} \end{subequations}
where we have relabelled the loop momenta and used $\ell_3=\ell_1+\ell_2$.
The six-term traces that emerge as the one-loop box numerators
on the right-hand side are familiar to us,
as they occur in the unitarity cuts of the one-loop amplitude
presented in \rcite{Kalin:2018thp}.
Together with the appearance of the triangle integral,
this implies that we have found terms belonging to $\cW_4^{(1)}$.

The other non-planar double boxes occurring in \eqn{eq:2loopRemainder} are
\be
n\!\left(\!\!\gBoxBoxNP[scale=.6,eLA=$1^-$,eLB=$2^-$,eLC=$4^+$,eLD=$3^{\!\!^+}$,
iC=aquark,iD=aquark,iF=aquark,iG=aquark,
iLE=$\uset{\rightarrow}{\ell_1}$,iLD=$\ell_2$]{%
\draw[line width=0.45pt,-to] (0.5,1.3) -- (0.15,0.95);
}\!\right)=
\frac{\kappa_{12}}{s}\trp(3\ell_3\ell_24)\,, \quad
n\!\left(\!\!\gBoxBoxNP[scale=.6,eLA=$1^-$,eLB=$2^+$,eLC=$4^-$,eLD=$3^{\!\!^+}$,
iC=aquark,iD=aquark,iF=aquark,iG=aquark,
iLE=$\uset{\rightarrow}{\ell_1}$,iLD=$\ell_2$]{%
\draw[line width=0.45pt,-to] (0.5,1.3) -- (0.15,0.95);
}\!\right)=
\frac{\kappa_{14}}{t^2}\big[\trm(1\ell_332\ell_24)-t^2\mu_{23}\big]\,,
\ee
where $\ell_3=\ell_1+\ell_2$.
These permit both soft and collinear divergences in the gluonic $\ell_1$ loop.
As before, we can extract the leading divergent behavior
at the integrand level:
\begin{subequations} \begin{align}
  \gBoxBoxNP[yshift=-.8,all=line,scale=0.6,eLA=$1$,eLB=$3$,eLC=$4$,eLD=$2$,
  iLE=$\uset{\rightarrow}{\ell_1}$,iLD=$\ell_2$]{%
    \draw[line width=0.45pt,-to] (0.6,1.25) -- (0.25,0.9);
    }[\trpm(2\ell_3\ell_24)]&=
  \gBox[yshift=-.8,all=line,scale=0.7,eLA=$1$,eLB=$2$,eLC=$3$,eLD=$4$,
  iLD=$\uset{\rightarrow}{\ell}$]{}[\trpm(2(\ell\!-\!p_1)\ell4)]\times
  \gTriA[rotate=180,all=line,scale=0.8,eLB=$1$,eLC=$3$,
  eA=double,eB=line,eC=line,iA=line,iB=line,iC=line]{}
  +\cO(\eps^{-1})\,,\\
  \gBoxBoxNP[yshift=-.8,all=line,scale=0.6,eLA=$1$,eLB=$3$,eLC=$4$,eLD=$2$,
  iLE=$\uset{\rightarrow}{\ell_1}$,iLD=$\ell_2$]{%
    \draw[line width=0.45pt,-to] (0.6,1.25) -- (0.25,0.9);
    }[\trpm(1\ell_323\ell_24)]&=
  \gBox[yshift=-.8,all=line,scale=0.7,eLA=$1$,eLB=$2$,eLC=$3$,eLD=$4$,
  iLD=$\uset{\rightarrow}{\ell}$]{}[\trpm(1\ell23\ell4)]\times
  \gTriA[rotate=180,all=line,scale=0.8,eLB=$1$,eLC=$3$,
  eA=double,eB=line,eC=line,iA=line,iB=line,iC=line]{}
  +\cO(\eps^{-1})\,,\\
  \gBoxBoxNP[yshift=-.8,all=line,scale=0.6,eLA=$1$,eLB=$3$,eLC=$4$,eLD=$2$,
  iLE=$\uset{\rightarrow}{\ell_1}$,iLD=$\ell_2$]{%
    \draw[line width=0.45pt,-to] (0.6,1.25) -- (0.25,0.9);
    }[\mu_{23}]&=
  \gBox[yshift=-.8,all=line,scale=0.7,eLA=$1$,eLB=$2$,eLC=$3$,eLD=$4$,
  iLD=$\uset{\rightarrow}{\ell}$]{}[\mu^2]\times
  \gTriA[rotate=180,all=line,scale=0.8,eLB=$1$,eLC=$3$,
  eA=double,eB=line,eC=line,iA=line,iB=line,iC=line]{}
  +\cO(\eps^{-1})\,.
\end{align} \end{subequations}
In the first case we observe that
$\trpm(2(\ell\!-\!p_1)\ell4)=\trmp(1\ell(\ell\!+\!p_4)3)$
is precisely the factor appearing in the one-loop box diagram~\eqref{eq:1lbox}.
All terms on the right-hand side are natural when assembling $\cW_4^{(1)}$,
and again the triangles belong to the dipole operator.

\subsubsection{UV-divergent topologies}
\label{sec:uvdivergent2loop}

We continue the analysis with numerators that carry the UV divergence
of the remainder in~\eqn{eq:2loopRemainder}, namely the pentagon-triangle and the box-bubble diagrams.
The numerator of the box-bubble diagram is simply
\be
n\Bigg(\gBoxBubA[scale=.8,yshift=-5,eLA=$1$,eLB=$2$,eLC=$3$,eLD=$4$,
iE=aquark,iF=aquark,
iLE=$\uset{\rightarrow}{\ell_1}$,
iLF={$\oset[-0.3ex]{\rightarrow}{\ell_2}$}]{}\Bigg)
=2\ell_1\cdot\ell_2\,\kappa_{ij}\,,
\ee
where $i$, $j$ are the legs carrying negative helicity.
This numerator is entirely reducible,
since it can be rewritten as $2\ell_1\cdot\ell_2=\ell_3^2-\ell_1^2-\ell_2^2$
in terms of three propagator denominators.
Two of the three resulting scalar integrals contain scaleless tadpoles
and therefore vanish.
The remaining scalar integral has four soft regions around the limits
$\ell_3 \to 0$, $\ell_3 \to p_1$, $\ell_3 \to p_{12}$, and $\ell_3 \to -p_4$.
Taking these limits at the integrand level,
we extract the finite denominators from the integrals and obtain
\be
   \gBoxBubC[scale=.8,all=line,eLA=$1$,eLB=$2$,eLC=$3$,eLD=$4$,
             iLD=$\uset{\rightarrow}{\ell_3}$]{}\!
 = \frac{1}{s}
   \gTriBubA[scale=.8,yscale=-1,rotate=-90,all=line,eA=doubleline,eLB=$1$,eLC=$4$]{}
 + \frac{1}{t}
   \gTriBubB[scale=.8,rotate=180,all=line,eA=doubleline,eLB=$1$,eLC=$2$]{}
 + \frac{1}{s}
   \gTriA[rotate=90,all=line,scale=0.8,eLB=$2$,eLC=$3$,
          eA=double,eB=line,eC=line,iA=line,iB=line,iC=line]{}
   \gBubA[eA=double,eB=double,iA=line,iB=line,
          eLA=$\substack{1\\2}$,eLB=$\substack{4\\3}$]{}
 + \frac{1}{t}
   \gTriBubB[scale=.8,yscale=-1,all=line,eA=doubleline,eLB=$4$,eLC=$3$]{}
 + \cO(\eps^{-1}) \,.
\label{eq:BoxBubbleExpansion}
\ee
The UV-divergent internal bubble cleanly decouples from the outer integral
only in the third region $\ell_3 \to p_{12}$.
The other regions naturally produce the two-loop triangle topologies
that appear in the IR factorization formula~\eqref{eq:Factorization2loopN4int}.
We have confirmed that the expansion~\eqref{eq:BoxBubbleExpansion}
holds for two orders in $\eps$ by numeric evaluation in FIESTA
\cite{Smirnov:2008py,Smirnov:2013eza,Smirnov:2015mct}.
Recall that the $\kappa$-prefactor contains an ordered tree amplitude
and we derive
\be
\bI\bigg[n\bigg(\gBoxBubA[scale=.6,eLA=$1$,eLB=$2$,eLC=$3$,eLD=$4$,
iE=aquark,iF=aquark]{}\bigg)\bigg]\!
=iM_4^{(0)}(1,2,3,4)\bigg\{\!
t\gTriBubA[scale=.8,rotate=-90,all=line,eA=doubleline,eLB=$4$,eLC=$1$]{}+
s\gTriBubB[scale=.8,rotate=180,all=line,eA=doubleline,eLB=$1$,eLC=$2$]{}+
t\gTriA[rotate=90,all=line,scale=0.8,eLB=$2$,eLC=$3$,
  eA=double,eB=line,eC=line,iA=line,iB=line,iC=line]{}
   \gBubA[eA=double,eB=double,iA=line,iB=line,
        eLA=$\substack{1\\2}$,eLB=$\substack{4\\3}$]{}+
s\gTriBubB[scale=.8,yscale=-1,all=line,eA=doubleline,eLB=$4$,eLC=$3$]{}\!\bigg\}
+\cO(\eps^{-1})\,.
\ee
This equation is true for all gluonic helicity configurations
and can be made to explicitly incorporate them all
using the symbolic amplitude expression
$iM_4^{(0)} = \frac{1}{st} \sum_{i<j} \kappa_{ij}$.

The pentagon-triangle contributions to \eqn{eq:2loopRemainder}
have numerators of the form
\begin{subequations} \begin{align}
n\!\!\left(\!\!\gTriPenta[scale=.6,eLA=$1^-$,eLB=$2^-$,eLC=$3^+$,eLD=$4^+$,
  iD=aquark,iG=aquark,iE=aquark,
  iLF=$\ell_1\,\,\,\,$,iLE=$\ell_2\,\,$]{%
  \draw[line width=0.45pt,-to] (-0.6,0.45) -- (-1.2,0.2);
  \draw[line width=0.45pt,-to] (-0.35,0.55) -- (0.05,0.7);
}\!\!\right)\!\!&=
-\frac{\kappa_{12}}{s}\big[\trp(3\ell_3\ell_24)-s\mu_{23}\big]\,,\\
n\!\!\left(\!\!\gTriPenta[scale=.6,eLA=$1^-$,eLB=$2^+$,eLC=$3^-$,eLD=$4^+$,
  iD=aquark,iG=aquark,iE=aquark,
  iLF=$\ell_1\,\,\,\,$,iLE=$\ell_2\,\,$]{%
  \draw[line width=0.45pt,-to] (-0.6,0.45) -- (-1.2,0.2);
  \draw[line width=0.45pt,-to] (-0.35,0.55) -- (0.05,0.7);
}\!\!\right)\!\!&=
-\frac{\kappa_{13}}{u}\big[\trm(2\ell_3\ell_24)-u\mu_{23}\big]\,,
\end{align} \label{eq:PentaTriangleNumerators}%
\end{subequations}%
where again $\ell_3=\ell_1+\ell_2$.
Since the chirally projected Dirac traces are taken strictly
in the four-dimensional sense,
the $\mu$-terms can be regarded as their extra-dimensional components.
Although soft and collinear divergences in the $\ell_2$ loop are blocked
by the numerators, UV poles do develop in the $\ell_2\to\infty$ limit.
Taken in combination with simultaneous soft and collinear singularities
in the $\ell_1$ loop,
this implies that leading poles occur at $\cO(\eps^{-3})$.

Considering the soft limits of a slightly more general integral, we find
\begin{subequations} \begin{align}
   \gTriPenta[scale=.6,all=line,eLA=$1$,eLB=$2$,eLC=$3$,eLD=$4$,
              iLF=$\ell_1\,\,\,\,$,iLE=$\ell_2\,\,$,iLG=$\ell_3\,\,$]{%
              \draw[line width=0.45pt,-to] (-0.6,0.45) -- (-1.2,0.2);
              \draw[line width=0.45pt,-to] (-0.35,0.55) -- (0.05,0.7);
              \draw[line width=0.45pt,-to] (-0.25,-0.2) -- (-0.25,0.2);}
   \big[\trpm(q\ell_3\ell_24)-2(p_4\!\cdot q)\mu_{23}\big] &
   \xrightarrow{~\ell_1 \to 0~\,}
   \frac{1}{s} \times
   \gTriBoxA[scale=.6,all=line,eB=double,
             eLA=$1$,eLB=$\substack{2\\3}$,eLC=$4$,iLD=$\ell_2$]{%
             \draw[line width=0.45pt,-to] (0.8,1.05) -- (0.4,0.75);}
   \big[ 2(p_4\!\cdot q) \ell_2^2 \big] \,, \\* &
   \xrightarrow{~\ell_1 \to p_1\;\!}
   \frac{1}{t} \times
   \gTriA[rotate=180,all=line,scale=0.8,eLA=$4$,eLB=$1$,eLC=$\substack{2\\3}$,
          iLA=$\ell_2$,eA=line,eB=line,eC=double,iA=line,iB=line,iC=line]{%
          \draw[line width=0.45pt,-to] (0.0,-0.35) -- (0.3,-0.15);}
   \big[ 2(p_4\!\cdot q) \ell_2^2 \big] \times
   \gTriA[rotate=180,all=line,scale=0.8,eLB=$1$,eLC=$2$,
          eA=double,eB=line,eC=line,iA=line,iB=line,iC=line]{} \!.
\end{align} \end{subequations}
For $\ell_1 \to 0$ we factor out $(\ell_1-p_{12})^2\to s$,
while the four-dimensional trace and $\mu_{23} \to \mu_{22}$ combine
into a $D$-dimensional $\ell_2^2$ in the numerator.
In the limit $\ell_1 \to p_1$
the propagators factorize into two one-loop topologies,
and we drop the numerator contribution $\trpm(q1\ell_24)$ that integrates to zero.
Analogous simplifications happen
for the regions $\ell_1 \to p_{12}$ and $\ell_1 \to -p_4$.
By integrand reduction we obtain
\begin{align}
\label{eq:PentaTriangleExpansion}
 & \gTriPenta[scale=.6,all=line,eLA=$1$,eLB=$2$,eLC=$3$,eLD=$4$,
              iLF=$\ell_1\,\,\,\,$,iLE=$\ell_2\,\,$,iLG=$\ell_3\,\,$]{%
              \draw[line width=0.45pt,-to] (-0.6,0.45) -- (-1.2,0.2);
              \draw[line width=0.45pt,-to] (-0.35,0.55) -- (0.05,0.7);
              \draw[line width=0.45pt,-to] (-0.25,-0.2) -- (-0.25,0.2);}
   \big[\trpm(q\ell_3\ell_24)-2(p_4\!\cdot q)\mu_{23}\big] \\* &
   \qquad
 = 2(p_4\!\cdot q) \bigg\{
   \frac{1}{s}
   \gTriBubB[scale=.8,all=line,eA=doubleline,eLB=$4$,eLC=$1$]{}
 + \frac{1}{t}
   \gBubA[eA=double,eB=double,iA=line,iB=line,
          eLA=$\substack{2\\3}$,eLB=$\substack{1\\4}$]{}
   \gTriA[rotate=180,all=line,scale=0.8,eLB=$1$,eLC=$2$,
          eA=double,eB=line,eC=line,iA=line,iB=line,iC=line]{}
 + \frac{1}{s}
   \gBubA[eA=double,eB=double,iA=line,iB=line,
          eLA=$\substack{1\\2}$,eLB=$\substack{4\\3}$]{}
   \gTriA[rotate=180,all=line,scale=0.8,eLB=$2$,eLC=$3$,
          eA=double,eB=line,eC=line,iA=line,iB=line,iC=line]{}
 + \frac{1}{t}
   \gTriBubB[scale=.8,all=line,eA=doubleline,eLB=$4$,eLC=$3$,yscale=-1]{}
   \bigg\} + \cO(\eps^{-1})  \nn
\end{align}
involving the two-loop triangles~\eqref{eq:2LtriangleB}.
Again, we have checked this expansion numerically using FIESTA
\cite{Smirnov:2008py,Smirnov:2013eza,Smirnov:2015mct}.

That our analysis of soft regions correctly predicts both the $1/\eps^3$ and $1/\eps^2$ terms
in \eqns{eq:BoxBubbleExpansion}{eq:PentaTriangleExpansion} is non-trivial.
Unlike \sec{sec:doubleboxes},
where the leading divergent behavior $\cO(\eps^{-2})$
was always associated with a particular soft region,
here one can imagine $1/\eps^2$ poles arising from a
collinear divergence in the $\ell_1$-loop and a UV divergence in the $\ell_2$-loop.
Such terms would not be detected merely by thinking about soft regions,
yet the above analysis leaves nothing out.
It seems to be a characteristic of these trace-based numerators
that they permit such behavior.
For the pentagon-triangle numerators~\eqref{eq:PentaTriangleNumerators},
the integral expansion~\eqref{eq:PentaTriangleExpansion} implies
\begin{align}
 & \bI\bigg[n\bigg(\!\!\gTriPenta[scale=.6,eLA=$1$,eLB=$2$,eLC=$3$,eLD=$4$,
                                  iD=aquark,iG=aquark,iE=aquark]{}\!\bigg)\bigg] \\*
 & = -iM_4^{(0)}(1,2,3,4) \bigg\{
     t \gTriBubB[scale=.8,all=line,eA=doubleline,eLB=$4$,eLC=$1$]{}
   + s \gTriBubB[scale=.8,all=line,eA=doubleline,eLB=$4$,eLC=$3$,yscale=-1]{}
   + t \gTriA[rotate=180,all=line,scale=0.8,eLB=$2$,eLC=$3$,
              eA=double,eB=line,eC=line,iA=line,iB=line,iC=line]{}
       \gBubA[eA=double,eB=double,iA=line,iB=line,eLA=$\substack{1\\2}$]{}
   + s \gTriA[rotate=180,all=line,scale=0.8,eLB=$1$,eLC=$2$,
              eA=double,eB=line,eC=line,iA=line,iB=line,iC=line]{}
       \gBubA[eA=double,eB=double,iA=line,iB=line,eLA=$\substack{2\\3}$]{}
     \bigg\} + \cO(\eps^{-1}) \,, \nn
\end{align}
in terms of the ordered tree-level amplitude with arbitrary external helicities.

This provides us with the last ingredient needed to prove
the IR factorization formula~\eqref{eq:2loopIR}.
At this order in $\eps$,
the second two-loop triangle topology~\eqref{eq:2LtriangleB}
cleanly cancels when the pentagon-triangle
and box-bubble contributions are combined.
Adding them to the leading-order divergences
of the planar and non-planar double boxes
we find perfect agreement with \eqn{eq:2loopIR}.

\subsection{External vectors + matter}

In the two-loop four-gluon amplitude we see a clear link between
the choice of IR subtraction scheme and the structure of the integrand.
Therefore, we naturally question whether a similarly revealing choice
of integrand basis might shed light on what IR-subtraction scheme best
enhances the transcendentality properties of amplitudes with external matter.
This question was already explored in \rcite{Kalin:2018thp} to a certain extent;
however, the requirement that all numerators be color-dual
limited our freedom to explore other possibilities.
As we have seen in previous examples,
such as the one-loop four-gluon amplitude in \sec{sec:1lvectors},
relaxing color-kinematics duality can be beneficial for exposing
the divergent structure of these amplitudes before integration.
So here we briefly investigate this possibility.

The following two double boxes were presented in \rcite{Kalin:2018thp}:
\begin{subequations}
\begin{align}
    n\!\left(\gBoxBox[scale=0.8,eA=quark,eB=aquark,eLA=$1$,eLB=$2$,eLC=$3^-$,eLD=$4^+$,
    iB=quark,iC=quark,iD=quark,iE=quark,iF=quark,iLD=$\ell_2$]{}\right)
    &= \frac{s\kappa_{(13)(23)}}{tu} \trp(13\ell_24)\,,\\
    n\!\left(\gBoxBox[scale=0.8,eA=quark,eB=aquark,eLA=$1$,eLB=$2$,eLC=$3^-$,eLD=$4^+$,
    iB=quark,iF=quark,iG=aquark,iLG=$\!\ell_3$]{}\right)
    &= \frac{s\kappa_{(13)(23)}}{tu} \trp(13\ell_34)\,,
\end{align}
\end{subequations}
where loop momenta follow the directions of the matter arrows.
These belong to the color-dual presentation of the four-point mixed amplitude
(which in \sec{sec:integrated} we will fully integrate).
Allowing ourselves to relax BCJ duality,
we find similarly compact expressions for the other six double boxes:
\begin{subequations}
\begin{align}
n\!\left(\gBoxBox[scale=.8,eLA=$1$,eLB=$2^-$,eLC=$3$,eLD=$4^+$,
eA=quark,eC=aquark,iF=quark,iG=aquark,iC=aquark,
iLC=$\ell_2$,iLF=$\ell_1$]{}\right)&=
\frac{\kappa_{(12)(23)}}{st}\trp(12\ell_1p_{12}\ell_24)\,,\\
n\!\left(\gBoxBox[scale=.8,eLA=$1$,eLB=$2^-$,eLC=$3^+$,eLD=$4$,
eA=quark,eD=aquark,iD=aquark,iC=aquark,iG=aquark,iF=quark,
iLD=$\ell_2$,iLF=$\ell_1$]{}\right)&=
\frac{\kappa_{(12)(24)}}{su}\trp(12\ell_1p_{12}\ell_23)\,,\\
n\!\left(\gBoxBox[scale=.8,eLA=$1$,eLB=$2^-$,eLC=$3$,eLD=$4^+$,
eA=quark,eC=aquark,iA=aquark,iB=aquark,iG=quark,iE=quark,iD=quark,
iLD=$\ell_2$,iLA=$\ell_1$]{}\right)&=
\frac{\kappa_{(12)(23)}}{st}\trp(12\ell_1p_{12}\ell_24)\,,\\
n\!\left(\gBoxBox[scale=.8,eLA=$1$,eLB=$2^-$,eLC=$3^+$,eLD=$4$,
eA=quark,eD=aquark,iE=quark,iF=quark,
iLE=$\ell_2$,iLF=$\ell_1$]{}\right)&=
\frac{\kappa_{(12)(24)}}{su}\trp(12\ell_1p_{12}\ell_23)\,,\\
n\!\left(\gBoxBox[scale=.8,eLA=$1$,eLB=$2^-$,eLC=$3^+$,eLD=$4$,
eA=quark,eD=aquark,iA=aquark,iB=aquark,iC=aquark,iD=aquark,
iLD=$\ell_2$,iLA=$\ell_1$]{}\right)&=
\frac{\kappa_{(12)(24)}}{su}\trp(12\ell_1p_{12}\ell_23)\,,\\
n\!\left(\gBoxBox[scale=.8,eLA=$1$,eLB=$2^-$,eLC=$3$,eLD=$4^+$,
eA=quark,eC=aquark,iA=aquark,iB=aquark,iC=aquark,
iLC=$\ell_2$,iLA=$\ell_1$]{}\right)&=
\frac{\kappa_{(12)(23)}}{st}\trp(12\ell_1p_{12}\ell_24)\,.
\end{align}
\end{subequations}
These numerators vanish whenever the loop momentum
carried by an internal matter edge vanishes.
Similar tricks work for the other planar numerators,
which suggests that a privileged integrand basis does exist.

Unfortunately, while relaxing the requirement of color-kinematics duality
gives us more freedom to write down desirable expressions in the planar sector,
it also makes it much harder to find suitable non-planars.
This is because, when enforcing the duality,
non-planar numerators are given to us automatically by
Jacobi or commutation relations in terms of the planars ---
writing down non-planar generalized unitarity cuts is therefore unnecessary.
While the iterated cut construction can be used to obtain non-planar cuts,
a generic mechanism for lifting them off-shell is still lacking.
This does not rule out the existence of such non-planar expressions in any way---
it simply complicates the task of finding them.

Another problem is that there is no natural
analogue of $\cM_4^{(2)[\cN=4]}$ for us to subtract,
so the IR divergences of all graphs must be analyzed.
As we shall see in \sec{sec:integrated},
this affects our ability to write down a suitable IR subtraction scheme.

\section{Integration \& transcendental weight}
\label{sec:integrated}

Having explored how the integrand structure reflects the
IR behavior of various different amplitudes,
we now proceed to study the analytic form of the three two-loop
$\cN=2$ SQCD amplitudes under consideration by direct integration.
The four-vector amplitude was already integrated in \rcite{Duhr:2019ywc},
so here we apply the same techniques to the other two amplitudes.
The algorithm is as follows:
\begin{enumerate}[label={(\theenumi)}]
\item All contributions are converted to scalar-type integrals using Schwinger pa\-ra\-me\-tri\-za\-tion~\cite{Anastasiou:1999bn,Anastasiou:2001}.
    Amongst the resulting terms are integrals in $(D+2n)$ dimensions (where $n \in \mathbb{Z}$) and with raised powers of propagators.
\item Higher-dimensional scalar integrals are shifted back to $D$ dimensions using dimensional recurrence relations
    \cite{Tarasov:1996br,Lee:2009dh}.
\item Using the \emph{Mathematica} package LiteRed~\cite{Lee:2012cn}, the resulting $D$-dimensional
    scalar integrals are reduced to a basis of masters using integration-by-parts relations (IBPs)~\cite{Chetyrkin:1981qh,Tkachov:1981wb}.
\item Known expressions for the four-point two-loop master integrals~\cite{Smirnov:1999wz,Tausk:1999vh,Anastasiou:2000mf,Gehrmann:2005pd,Henn:2013pwa} are inserted.
    Final manipulation of the results is performed using the \emph{Mathematica} package HPL~\cite{Maitre:2005uu}.
\end{enumerate}
The resulting $\eps$-pole structure in all three cases matches the one predicted in~\eqn{Factorization2loop2}.

We provide machine-readable ancillary text files for the complete expressions of the one- and two-loop four-point amplitudes as an attachment to the \texttt{arXiv} submission of this paper.
The expressions are written out for an $\textrm{SU}(N_c)$ gauge group,\footnote{The
	gluonic amplitude for a selection of different gauge groups can be found attached to \rcite{Duhr:2019ywc}.}
distinct flavors of external matter parton pairs,
and $N_f$ flavors for internal matter loops.
The equal-flavor amplitudes can be obtained by an appropriate summation over permutations of the distinct-flavor expressions.
For the four-gluon amplitude, we present all helicity configurations as in~\rcite{Duhr:2019ywc}. For the amplitudes involving external matter
we make a particular choice w.r.t. the helicities of the external particles, which can be readily observed from their $\kappa$-dependence as
shown in~\eqref{eq:kappaDef}.
The two-loop cusp and field anomalous dimensions
have already been advertised in \eqn{eq:N2AnomDimensions}.

Since we have the full analytic expressions at our disposal, we are able to study the range of transcendental weights appearing in the amplitudes.
In \rcite{Duhr:2019ywc}, where the analytic form of the four-gluon amplitude was studied,
a conspiracy between the IR subtraction scheme as defined by Catani~\cite{Catani:1998bh} (shown in \eqn{eq:IRfactorizationCatani}) and terms of lower transcendentality was observed at the conformal point of the theory.
More specifically, writing the gluonic remainder as~\cite{Duhr:2019ywc}
\be
\cW_n^{(L)} = \cR_n^{(L)} + (C_A-T_F N_f)\cS_n^{(L)} \,,
\ee
one sees $\cR_4^{(2)}$ contains terms of weight 2 through 4 at $\cO(\eps^0)$.
Upon IR subtraction via
\be
   \cR_4^{(2)}
    = \left( \sum_{i<j}^4 s_{ij}
             \gTriA[scale=0.8,rotate=180,eLB=$i$,eLC=$j$,
                    eA=doubleline,eB=line,eC=line,iA=line,iB=line,iC=line]{}
             \bT_i\cdot\bT_j\!
      \right) \cR_4^{(1)}+ \cR_4^{(2){\rm fin}} \,,
\label{eq:ConfFactorization2loopN4int}
\ee
it was found that $\cR_4^{(2){\rm fin}}$ is given by terms of weights 3 and 4 only at $\cO(\eps^0)$.
This intriguing cancellation was limited only to $\cR_4^{(2)}$,
as the Catani-style subtraction did not ameliorate the transcendental structure of the non-conformal part of the amplitude.

Bypassing this restriction, from the point of view
of transcendentality the scheme defined in \eqn{eq:Factorization2loopN4int}
seems to be a logical extension of \eqn{eq:ConfFactorization2loopN4int}.
Indeed, performing this subtraction we see that,
although $\cW^{(2)}_4$ contains terms of all
possible weights 0 through 4 at $\cO(\eps^0)$,
$\cW_4^{(2){\rm fin}}$ is described entirely by terms
of weights 3 and 4 at the same order in $\eps$.
As an illustration, we provide explicit expressions for two independent helicity
configurations of $\cW_4^{(2){\rm fin}}$ in \app{sec:2Lresults}.
Note that the precise form of \eqn{eq:Factorization2loopN4int}
is important in this regard.
For example, if we rewrite the non-conformal term
using only one two-loop triangle topology, as in \eqn{eq:trianglerelation},
this will reintroduce terms of weight 2 at $\cO(\eps^0)$.

Our discussion of the one-loop amplitudes in \sec{sec:oneloop}
provides us with some intuition for why this cancellation happens.
In the one-loop mixed amplitude (discussed in \sec{sec:1lmixed})
no bubbles appeared at any stage,
and the result had uniform weight regardless of our scheme choice,
as this property manifestly holds for the other topologies involved.
In the one-loop four-vector amplitude (discussed in \sec{sec:1lvectors})
the appearance of lower-weight terms in $\cH_4^{(1)}$ was attributed
to our inability to consistently remove bubble integrals.
In the two-loop gluonic amplitude,
while bubble integrals do appear and produce UV divergences,
they occur either as sub-integrals, and so can be incorporated into
the two-loop triangle topologies used to subtract IR divergences,
or they naturally arrange themselves into the one-loop remainder $\cW_4^{(1)}$.
While this does not constitute a full integrand-level understanding of the
weight grading of the two-loop result, it does motivate why one could
expect the scheme in~\eqn{eq:Factorization2loopN4int} to improve the transcendental
structure of the finite part of the amplitude.

Having identified this peculiar scheme for the four-gluon case,
it is natural to wonder whether such an enhanced cancellation of lower-weight
terms can also be achieved for the two-loop amplitudes
involving external matter fields.
In these cases, there is no analogous result in $\cN=4$ SYM which can be subtracted and a remainder as given by \eqn{N4diff} cannot be defined.
To this end, we can generalize
the scheme defined in \eqn{eq:Factorization2loopN4int} as follows:
\begin{align}
   \cM_n^{(2)}\!= &\,
      \bigg[
      \frac{1}{2} \mathbf{S}(\eps) \mathbf{S}(\eps)
    + \frac{1}{2} \big(\gamma_K^{(2)}-4\beta_0\big) \mathbf{S}(2\eps)
    + \frac{1}{\eps}\! \sum_{i=1}^n \gamma_i^{(2)}\!
      \bigg] \cM_n^{(0)}
    + \!\bigg(\! \sum_{i<j}^n s_{ij}
             \gTriA[scale=0.8,rotate=180,eLB=$i$,eLC=$j$,
                    eA=doubleline,eB=line,eC=line,iA=line,iB=line,iC=line]{}
             \bT_i\cdot\bT_j\!
      \bigg) \cH_n^{(1)} \nn \\* &
    + \beta_0
      \bigg( \sum_{i<j}^ns_{ij}
             \bigg[ \gTriBubB[scale=0.8,rotate=180,all=line,
                              eA=doubleline,eLB=$i$,eLC=$j$]{}\!\!
                  - \gTriBubA[scale=0.8,rotate=180,all=line,
                              eA=doubleline,eLB=$i$,eLC=$j$]{}
                  - \frac{1}{\eps}
                    \gTriA[scale=0.8,rotate=180,eLB=$i$,eLC=$j$,
      eA=doubleline,eB=line,eC=line,iA=line,iB=line,iC=line]{}
             \bigg] \bT_i\cdot\bT_j\!
      \bigg) \cM_n^{(0)} \nn \\* &
    + \frac{1}{2\eps}
      \bigg[ \frac{n}{2} \beta_0 + \sum_{i=1}^n \gamma_i^{(1)}
      \bigg]
      \big( \cM_n^{(1)} + \mathbf{S}(\eps) \cM_n^{(0)} + \cH_n^{(1)} \big) \nn \\* &
    + \frac{1}{\eps}
      \bigg[ \frac{n-2}{2} \beta_1
           + \frac{\beta_0}{2} \bigg( 1 - \frac{\zeta_2}{4} \bigg)
             \sum_{i=1}^n \bT_i^2
      \bigg] \cM_n^{(0)}
    + \cM_n^{(2){\rm fin}} \,,
\label{eq:TransSub2loop}
\end{align}
where $\cH^{(1)}_n$ was defined in \eqn{Factorization2loop1}.
Using this generalized scheme involving scalar triangle integrals, we study the transcendentality of $\cM^{(L){\rm fin}}_n$ for the various amplitudes computed in this paper.
Our findings are summarized in \Tab{tab:2Lweighttable}.

\begin{table}
\centering
\begin{tabular}{ c | c | c | c }
Weights at $\cO(\eps^0)$ & Vectors & Mixed & Matter  \\
  \hline
$\cH^{(2)}_4$ &  0,\,1,\,2,\,3,\,4  & 0,\,1,\,2,\,3,\,4  & 0,\,1,\,2,\,3,\,4 \\
$\cM^{(2){\rm fin}}_4$  & 3,\,4 & 2,\,3,\,4 & 0,\,1,\,2,\,3\,,4 \\
$\cH^{(2){\rm conf}}_4$ & 2,\,3,\,4  & 3,\,4  & 3,\,4 \\
$\cM^{(2){\rm fin,conf}}_4$  & 3,\,4 & 3,\,4 & 3,\,4 \\
  \hline
\end{tabular}
\caption{The transcendental weights appearing for various encodings of the finite part of the two-loop amplitudes under consideration.
  The first row shows the weights for the hard part of the various unrenormalized two-loop amplitudes as defined by minimal subtraction~\eqref{Factorization2loop2}; the second row shows the weights appearing after IR subtraction as defined by~\eqn{eq:TransSub2loop}.
  The third and fourth rows show the weights of the same terms evaluated at the conformal point $N_f = C_A/T_F$.}
\label{tab:2Lweighttable}
\end{table}

In the four-gluon case, \eqn{eq:TransSub2loop} is similar to \eqn{eq:Factorization2loopN4int} so the same cancellations occur.
For the amplitude involving both external vector and matter fields,
the scheme defined above also ameliorates the transcendental structure of the finite part.
We note that in this case, the uniform transcendentality of the
one-loop result simplifies the weight grading of \eqn{eq:TransSub2loop}.
Finally, for the amplitude with all external matter fields no cancellation is observed.
However, we should not rule out the possibility of a further generalization
of \eqn{eq:TransSub2loop} that could induce analogous behavior.
In particular, at weights 0 and 1 the color structures and transcendental objects which appear are sufficiently
restricted that the proportionality to the tree-level amplitude is remarkably simple.

\section{Summary and outlook}
\label{sec:conclusions}

In this paper we have explored a link between the degree of
uniform transcendentality violation and IR-subtraction schemes in $\cN=2$ SQCD.
Taking two-loop gluonic amplitudes as ``corrections'' to
uniformly transcendental $\cN=4$ SYM amplitudes,
with $\cW_n^{(L)}\equiv\cM_n^{(L)}-\cM_n^{(L)[\cN=4]}$,
we find that all two-loop divergences (including UV) may be
expressed in terms of scalar triangle integrals:
\be
   \cW_n^{(2)}\!
    =\! \left( \sum_{i<j}^n s_{ij}
             \gTriA[scale=0.8,rotate=180,eLB=$i$,eLC=$j$,
                    eA=doubleline,eB=line,eC=line,iA=line,iB=line,iC=line]{}
             \bT_i\cdot\bT_j\!
      \right) \cW_n^{(1)}
    + \beta_0\!
      \left( \sum_{i<j}^ns_{ij}
             \left[\!\gTriBubB[scale=0.8,rotate=180,all=line,
                              eA=doubleline,eLB=$i$,eLC=$j$]{}\!\!
                  - \gTriBubA[scale=0.8,rotate=180,all=line,
                              eA=doubleline,eLB=$i$,eLC=$j$]{}\right]
            \bT_i\cdot\bT_j\!
      \right) \cM_n^{(0)}+ \cW_n^{(2)\text{fin}} .
\label{eq:Factorization2loopN4intB}
\ee
The use of these integrals defines a specific
IR-subtraction scheme to all orders in $\eps$
and cancels all transcendental weight-0,1,2 terms
from $\cW_4^{(2)\text{fin}}$.
A similar scheme was used in \rcite{Duhr:2019ywc} to demonstrate the cancellation
of lower-weight terms from the finite part of the same
amplitude at the conformal point $\beta_0=2N_c-N_f=0$ ---
here we have generalized the result to the full
SQCD theory with arbitrary $N_c$ and $N_f$.

Our understanding of this scheme came from two opposite perspectives.
Firstly, in \sec{sec:IRfactorization} we provided a derivation
of generic formulae~\eqref{eq:IRfactorization2loop}
and~\eqref{Factorization2loop}
for the two-loop IR divergences ---
after and before UV renormalization, respectively.
These formulae are valid for any massless gauge theory in four dimensions
and improve on Catani's well-known formulae \eqref{eq:IRfactorizationCatani}
by excluding unnecessary color structures
of the form $\tf^{abc}\bT_i^a\bT_j^b\bT_k^c$.
In particular,
the formula~\eqref{Factorization2loop} for the divergences
of unrenormalized (bare) amplitudes $\cM_n^{(2)}$
places UV, soft, and collinear poles on an equal footing.
Even more cleanup happens for gluonic amplitudes
\eqref{eq:Factorization2loopN4diff}:
$\cW_n^{(L)}=\cM_n^{(L)}-\cM_n^{(L)[\cN=4]}$ diverges only as $1/\eps^{2L-1}$
since the leading $1/\eps^{2L}$ poles are absorbed into
the uniformly transcendental $\cN=4$ SYM amplitudes.
Finally, the IR scheme~\eqref{eq:Factorization2loopN4intB}
was obtained by specializing to $\cN=2$ SQCD
using our results~\eqref{eq:N2AnomDimensions} for
the two-loop anomalous dimensions.

A more intuitive picture came from studying the IR behavior of the loop integrands.
In \rcite{Kalin:2018thp} an iterated cut construction was used to write down
expressions for all two-loop four-point $\cN=2$ SQCD integrands.
Diagrams with internal matter lines were found to have a controlled IR behavior,
so that singular regions arising from  massless $i/p^2$ propagators are
``blocked'' by the appearance of local numerators that vanish in those regions.
This matches our physical intuition in QCD:
quarks obey Fermi-Dirac statistics,
so they should be distinguishable even in the high-energy (massless) limit ---
soft or collinear divergences should arise only from virtual gluon exchange.
The usual quark propagator $i\slashed{p}/p^2$ also ensures this property,
for instance, as $\slashed{p}$ vanishes in the soft limit $p^\mu\to0$.
Supersymmetric quarks and gluons follow the same qualitative behavior.

The controlled IR behavior made it possible to analyze
both the divergences and transcendentality structure
of multi-loop $\cN=2$ SQCD amplitudes before integration.
In the one-loop examples discussed in \sec{sec:oneloop}, this involved
expressing soft and collinear divergences in terms of unintegrated triangle integrals
and UV divergences in terms of bubbles.
The latter, we observed, source the unwanted lower-weight
terms in the finite parts of the amplitudes.
For the one- and two-loop gluonic amplitudes discussed in sections
\ref{sec:1lvectors} and \ref{sec:2Lextvectors}, respectively,
we used an off-shell supersymmetry decomposition to eliminate all purely-gluonic
diagrams from the remainder functions $\cW_4^{(L)}$;
the remaining diagrams all contain internal matter loops.
Analysis of the two-loop amplitude in \sec{sec:2Lextvectors}
gave rise to the two-loop triangles with UV-divergent bubbles embedded,
which fit naturally into the IR scheme presented in \eqn{eq:Factorization2loopN4intB}.

Finally, in this paper we also integrated the two-loop four-point
$\cN=2$ SQCD amplitudes with matter on external legs,
first presented in \rcite{Kalin:2018thp}.
Using a generalized scheme~\eqref{eq:TransSub2loop}
that allows for external matter,
we found a cancellation of weight-0,1 terms from the
four-point mixed amplitude but not the four-matter amplitude ---
results are presented in \Tab{tab:2Lweighttable}.
Integration of these amplitudes also provided us
with the quark anomalous dimension $\gamma_q^{(2)}$,
as well as a cross-check of the cusp and gluonic
anomalous dimensions $\gamma_K^{(2)}$ and $\gamma_g^{(2)}$,
all given in \eqn{eq:N2AnomDimensions}.
All three include an arbitrary dependence on the number of matter flavors $N_f$.
They are valid in the FDH regularization scheme~\cite{Bern:1991aq,Bern:2002zk},
and may in principle be converted to
the 't Hooft-Veltman scheme~\cite{tHooft:1972tcz}
or the conventional dimensional regularization scheme~\cite{Collins:1984xc}
using the dictionary of \rcite{Gnendiger:2014nxa}.

Let us now expand upon the features and consequences of our work.

\paragraph{Anomalous dimensions.}
Given our $\cN=2$ SQCD results~\eqref{eq:N2AnomDimensions}
for the two-loop anomalous dimensions,
one could wonder whether they contain hints of their lower-su\-per\-sym\-met\-ry counterparts.
Indeed, the result~\eqref{eq:N2AnomDimensionsCusp}
for the two-loop cusp anomalous dimension
is in perfect agreement with the general FDH formula
\be
   \gamma_K^{(2)}
    = \bigg(\frac{64}{9} - 2\zeta_2\bigg) C_A
    - \frac{10}{9} T_f n_f - \frac{4}{9} T_s n_s
\label{eq:QCDbeta}
\ee
for a gauge theory minimally coupled to $n_f$~Weyl fermions and $n_s$~real scalars.
It is well known to practitioners, as it can be inferred from interpolating between
the cases of QCD ($n_s=0$, $n_f=2N_f$, $T_f=T_F$ \cite{Bern:2002tk})
and $\cN=4$ SYM ($n_f=4$, $n_s=6$, $T_f=T_s=C_A$ \cite{Bern:2004kq})
based on its linearity with respect to the particle content.
It is difficult to write similarly generic formulae
for the two-loop gluonic and quark collinear anomalous dimensions,
as they are sensitive to how the adjoint scalars couple to
the fundamental fermions and scalars.
A supersymmetry interpolation is, however, possible for pure SYM theories,
obtained by setting $N_f=0$.
We find perfect consistency with the $0<\cN \leq 4$ supersymmetric results
of \rcites{Brandhuber:2018xzk,Brandhuber:2018kqb}
for the Catani terms which survive in the leading-color limit $N_c \to \infty$:
\be
   K_\text{FDH}^{[\text{SYM}]} =
      C_A \big[{-\zeta_2}+(4-\cN)\big] \,, \qquad \quad
   H_{g,\text{FDH}}^{[\text{SYM}]} =
      \frac{1}{4} C_A^2 \bigg[ 2\zeta_3 + \frac{4-\cN}{2} \zeta_2 \bigg] \,,
\label{eq:CataniTermsSYM}
\ee
where we have adjusted the overall prefactors
to match our conventions in \eqn{eq:CataniTermsConversion}.
For arbitrary $\cN$,
we can combine \eqns{eq:CataniTermsConversion}{eq:CataniTermsSYM} and easily find
\be
   \gamma_K^{(2)} = 2C_A\big[{-\zeta_2}+(4-\cN)\big] \,, \qquad \quad
   \gamma_g^{(2)} = \frac1{16}C_A^2\big[2\zeta_3+(4-\cN)\zeta_2-2(4-\cN)^2\big]\,,
\ee
having also used $\beta_0 = \frac12C_A\!\left(4-\cN\right)$.
Uniform transcendentality is manifest for $\cN=4$.

\paragraph{Local IR subtraction.}
Our analysis of the IR divergences before integration
was facilitated by the controlled IR behavior of the $\cN=2$ integrands,
in which the diagram numerators naturally ``blocked'' IR divergences
associated with certain edges.
These IR-blocking properties can be regarded
as a kind of local IR subtraction at the amplitude-integrand level,
which has been a subject of significant interest in QCD
\cite{Nagy:2003qn,Assadsolimani:2009cz,Anastasiou:2018rib}.
In our approach, such a (partial) subtraction is made possible
by tailoring the diagram numerators to generalized unitarity cuts exactly,
which is impossible without certain Levi-Civita terms.
Although such terms vanish upon integration,
they often participate in loop-dependent chiral Dirac traces $\trpm(\cdots)$, which we found to be natural building blocks
for well-behaved loop integrands.
It would be extremely interesting to see
if such guidance from unitarity cuts can also help
achieving local IR subtraction in QCD.

\paragraph{Transcendentality.}
Our $\cN=2$ results reveal an intriguing
interplay between IR physics and transcendentality,
which is well-studied in $\cN=4$ SYM.
The conjectured uniform transcendentality property
\cite{Kotikov:2001sc,Kotikov:2002ab}
is naturally implied for integrands
that can be written as so-called $d$log forms,
which is a consequence of those integrands having only
unit leading singularities (see \rcite{ArkaniHamed:2012nw} for a review).
Using loop-level recursion, all-loop $n$-point planar MHV integrands
\cite{ArkaniHamed:2010kv,ArkaniHamed:2010gh,ArkaniHamed:2012nw}
and two-loop planar N$^k$MHV integrands~\cite{Bourjaily:2015jna}
have been expressed in terms of only diagrams with unit leading singularities,
and these naturally involve manifestly IR-finite integrals.
Similar structures have now also been found beyond the planar limit
\cite{Bern:2015ple,Bourjaily:2019iqr,Bourjaily:2019gqu},
but a full proof of uniform transcendentality remains elusive.
A better understanding of how the property is violated in theories with
$\cN<4$ supersymmetries may shed light on this question.

It would therefore be desirable to confirm that the minimal violation
of uniform transcendentality for the finite amplitude,
as defined by the IR scheme \eqref{eq:Factorization2loopN4intB},
continues for $n>4$.
It will be particularly interesting to see how the two-loop IR-controlling
numerators generalize for more external legs.
At $n=5$ points integration of the full-color MHV amplitude
should be achievable using currently available technology,
given recent progress on integrated five-parton amplitudes in QCD
\cite{Badger:2018gip,Abreu:2018jgq,Abreu:2019odu,Badger:2019djh}.
Extensions to the next loop order or higher orders in $\eps$ are also within reach~\cite{Henn:2016jdu,Ahmed:2019qtg}.
Furthermore, the case of lower supersymmetry should also be explored.
In the high-energy limit, a connection between superconformal symmetry and
uniform transcendental weight was found for the BFKL ladder at next-to-leading
logarithmic accuracy~\cite{DelDuca:2017peo}. It would be interesting to see if $\cN=1$ SQCD
has a similar minimal departure from uniform transcendentality when tuned to
a conformal point and if a similar relation between infrared structure and transcendental weight
can be constructed in the $\cN=1$ case.

Finally, the precise form of \eqn{eq:Factorization2loopN4intB} suggests
a better interpretation may exist in the language of form factors.
Each of the triangle integrals has precisely one off-shell leg,
which might indicate an expectation value of some operator in $\cN=2$ SQCD.
Given that the same transcendentality properties of amplitudes
are expected to carry over to form factors,
as has been observed in $\cN=4$ SYM
\cite{vanNeerven:1985ja,Gehrmann:2011xn,Brandhuber:2012vm},
such an analysis may shed further light on transcendentality violations
in $\cN<4$ supersymmetric theories.

\begin{acknowledgments}

We thank Charalampos Anastasiou, Simon Badger, Lance Dixon, Claude Duhr,
Einan Gardi, Henrik Johansson, Ben Page, Alexander Penin, Oliver Schlotterer,
and Leonardo Vernazza for interesting and helpful discussions.
AO would also like to acknowledge
the hospitality of the Galileo Galilei Institute for Theoretical Physics
during the program ``Amplitudes in the LHC era.''
AO has received funding from the European Union's Horizon 2020 research and innovation programme under the Marie Sk{\l}odowska-Curie grant agreement 746138 and ERC grant \emph{PertQCD} (694712).
The research of GK and GM is supported by the Swedish Research Council
under grant 621-2014-5722,
the Knut and Alice Wallenberg Foundation under grants KAW 2013.0235, 2018.0116,
and the Ragnar S\"{o}derberg Foundation (Swedish Foundations' Starting Grant).
GK has also received funding from the Knut and Alice Wallenberg Foundation
under grant KAW 2018.0441, and is supported in part by the US Department of Energy
under contract DE--AC02--76SF00515.
BV is supported by the European Research Council under ERC grant \emph{UNISCAMP} (804286).

\end{acknowledgments}

\appendix

\section{Anomalous dimensions in \texorpdfstring{$\cN=4$}{N=4} SYM}
\label{sec:BDS}

Exact four-point (and five-point) amplitudes in planar $\cN=4$ SYM
(with $G={\rm SU}(N_c)$) are given by the ABDK/BDS ansatz~\cite{Anastasiou:2003kj,Bern:2005iz}
which exponentiates the one-loop amplitudes evaluated to all orders in $\eps$:
\begin{align}
   \cM_n^{[\cN=4]} &
    = \big(4\pi\alpha_{\rm s}\big)^{\frac{n-2}{2}}\!
      \sum_{\text{perms}}\!\frac{1}{n}
      \Tr[T^{a_1}\!\ldots T^{a_n}]
      A_n^{(0)}(1,\ldots,n) M_n(1,\ldots,n;\eps)
      \big(1 + \cO(1/N_c) \big) \,, \nn \\*
\label{BDS}
   M_n & \equiv 1 + \sum_{L=1}^{\infty}
      \bigg(\!\frac{\alpha_{\rm s} S_\eps}{4\pi}N_c\!\bigg)^{\!\!L}
      M_n^{(L)}(\eps) \\* &
    = \exp\bigg\{ \sum_{L=1}^{\infty}
                  \bigg(\!\frac{\alpha_{\rm s} S_\eps}{4\pi}N_c\!\bigg)^{\!\!L}
                  \Big[ f^{(L)}(\eps) M_n^{(1)}(L\eps)
                      + C^{(L)} + E_n^{(L)}(\eps) \Big]
          \bigg\}  \,, \qquad \quad n = 4,5 \,, \nn
\end{align}
where $f^{(L)}(\eps)$ and $C^{(L)}$ are independent of the external kinematics,
and at one loop $f^{(1)}(\eps)=1$, $C^{(1)}=E_n^{(1)}(\eps)=0$ by definition.
Recall that at four points, for instance,
the tree and one-loop color-ordered amplitudes are given by
\begin{equation}
   A_4^{(0)[\cN=4]} = -\frac{i\delta^8(Q)}{st}
      \frac{[12][34]}{\braket{12}\braket{34}} \,, \qquad \quad
   M_4^{(1)} = -st
   \gBox[scale=.8,yshift=-1,all=line,eLA=$1$,eLB=$2$,eLC=$3$,eLD=$4$]{}\,.
\end{equation}
The two-loop planar amplitude is expressed as~\cite{Anastasiou:2003kj}
\be
   M_n^{(2)}(\eps) = \frac{1}{2} \big[ M_n^{(1)}(\eps) \big]^2
    - 2(\zeta_2 + \zeta_3 \eps + \zeta_4 \eps^2) M_n^{(1)}(2\eps) + \cO(\eps^0) .
\label{ABDK}
\ee
Now let us compare that
with the $\cN=4$ factorization formulae~\eqref{Factorization2loopN4},
which take the following form for the color-ordered amplitudes:
\small
\begin{subequations} \begin{align}
\label{Factorization2loopN4planar1}
   M_n^{(1)} & = S(\eps) M_n^{(0)} + H_n^{(1)} \,, \qquad \quad
   S(\eps) = -\frac{1}{\eps^2} \sum_{i=1}^{n}
      \bigg[ 1 - \eps \log\!\bigg(\!\frac{-s_{i(i+1)}}{\mu^2}\!\bigg) \bigg]  \\
\label{Factorization2loopN4planar2}
   M_n^{(2)} & = S(\eps) M_n^{(1)}
    + \bigg[\!
   -\!\frac{1}{2} \big[ S(\eps) \big]^2
    + \frac{1}{2N_c} \gamma_K^{(2)[\cN=4]} S(2\eps)
    + \frac{n}{\eps N_c^2} \gamma_g^{(2)[\cN=4]}\!
      \bigg] M_n^{(0)}
    + H_n^{(2)} \,,
\end{align} \label{Factorization2loopN4planar}%
\end{subequations}
\normalsize
where the $N_c$ denominators are due to the explicit factors of $N_c$
in the expansion~\eqref{BDS}.
Note that $M_n^{(0)} \equiv 1$,
so we can equate \eqns{ABDK}{Factorization2loopN4planar2}:
\begin{align}
   M_n^{(2)} &
    = \frac{1}{2} \big[ S(\eps) \big]^2
    + S(\eps) H_n^{(1)}
    - 2(\zeta_2 + \zeta_3 \eps) S(2\eps)
    + \cO(\eps^0) \\* &
    = \frac{1}{2} \big[ S(\eps) \big]^2
    + S(\eps) H_n^{(1)}
    + \frac{1}{2N_c} \gamma_K^{(2)[\cN=4]} S(2\eps)
    + \frac{n}{\eps N_c^2} \gamma_g^{(2)[\cN=4]}
    + H_n^{(2)} \,. \nn
\end{align}
Therefore, consistent with \eqn{eq:N4AnomDimensions2}, we find
\be
   \gamma_K^{(2)[\cN=4]} = -4 \zeta_2 N_c \,, \qquad \quad
   \gamma_g^{(2)[\cN=4]} = \frac{1}{2} \zeta_3 N_c^2 \,,
\ee
where we have used $S(2\eps) = -n/(4\eps^2) + \cO(\eps^{-1})$ for the latter.

\section{Two-loop finite remainder in \texorpdfstring{$\cN=2$}{N=2} SQCD}
\label{sec:2Lresults}

Here we present the finite remainder function of the two-loop gluonic amplitude~$\cW^{(2){\rm fin}}_4$ as defined in~\eqref{eq:Factorization2loopN4int}.
We split the result into two parts
\be
\cW^{(2){\rm fin}}_4
= \cR^{(2){\rm fin}}_4 + (C_A-T_F N_f) \cS^{(2){\rm fin}}_4 \,,
\ee
where $\cR_4^{(2){\rm fin}}$ denotes the remainder for the conformal theory
and $\cS_4^{(2){\rm fin}}$ represents corrections thereof for the
generic theory (recall that $(C_A-T_F N_f)=\beta_0$).
We present the results in terms of color-ordered
building blocks in the trace basis of the gauge group ${\rm SU}(N_c)$;
we denote the kinematic coefficient of
$N_c^i M^{(0)}(1^-\!,2^-\!,3^+\!,4^+)\Tr(T^{a_1}T^{a_2}T^{a_3}T^{a_4})$
as $W^{(2)[i]{\rm fin}}_{(--++)}$
and that of the double trace
$N_c^i M^{(0)}(1^-\!,2^-\!,3^+\!,4^+) \Tr(T^{a_1}T^{a_2})\Tr(T^{a_3}T^{a_4})$
as $W^{(2)[i]{\rm fin}}_{(--)(++)}$.
Analogous notation is used
for the constituent $\cR^{(2){\rm fin}}_4$ and $\cS^{(2){\rm fin}}_4$.
The results for $\cR^{(2){\rm fin}}_4$ were already obtained in \rcite{Duhr:2019ywc}
and in the planar case in~\rcites{Dixon2008talk,Leoni:2015zxa},
but we list them here again for completeness.
Note that, as discussed in~\rcite{Duhr:2019ywc}, the components given below form a sufficient set to reconstruct the full-color answer.

We introduce the shorthand notation $\tau = -{t}/{s}$, $\upsilon = -{u}/{s}$
with their logarithms being written as $\cT = \log(\tau)$,
and $\cU = \log(\upsilon)$.
Furthermore, ${\rm Li}_n(z)$ are the classical polylogarithms~\cite{Goncharov:1998kja,
GoncharovMixedTate} and $S_{n,p}(z)$ are Nielsen generalized polylogarithms (see \eg \rcite{Kolbig:1983qt}).
We give results in the region $s>0$; $t,u<0$, so $\cT$ and $\cU$ are real.
\begin{subequations} \begin{align}
   R^{(2)[2]}_{(--++)}
   ~\:=~\frac{\tau}{6} &
      \Big[ 48 {\rm Li}_4(\tau) - 24 (\cT\!+\cU) {\rm Li}_3(\tau)
          - 24 \cT {\rm Li}_3(\upsilon)
          + 24 \cT \cU {\rm Li}_2(\tau) + 24 \cT \cU {\rm Li}_2(\upsilon) \nn \\* &
          - 24 S_{2,2}(\tau )
          + \cT^4 - 4 \cT^3 \cU + 18 \cT^2 \cU^2
          + 24 \zeta_2 {\rm Li}_2(\tau )
          - 12 \zeta_2 \cT^2 + 24 \zeta_2 \cT \cU \nn \\* &
          + 24 \zeta_3 \cU - 168 \zeta_4 \Big] \nn \\*
    - i \pi \frac{2\tau}{3} &
      \Big[ 6 {\rm Li}_3(\tau) + 6 {\rm Li}_3(\upsilon)
          - 6 \cU {\rm Li}_2(\tau) - 6 \cU {\rm Li}_2(\upsilon) \\* &
          - \cT^3 + 3 \cT^2 \cU - 6 \cT \cU^2
          + 6 \zeta_2 \cU - 6 \zeta_2 \cT \Big]
    + 12 \zeta _3
    + \cO(\eps) \,, \nn \\
   R^{(2)[1]{\rm fin}}_{(--)(++)}\!
    = \frac{2\tau}{3} &
      \Big[ 96 {\rm Li}_4(\tau) + 96 {\rm Li}_4(\upsilon)
          - 24 (3\cT\!-\cU) {\rm Li}_3(\tau)
          + 24 (\cT\!-3\cU) {\rm Li}_3(\upsilon) \nn \\* &
          + 24 \cT (\cT\!-\cU) {\rm Li}_2(\tau)
          - 24 \cU (\cT\!-\cU) {\rm Li}_2(\upsilon) \nn \\* &
          + (\cT\!+\cU)^4 - 24 \cT^2 \cU^2
          - 12 \zeta_2 (\cT\!- \cU)^2 - 654 \zeta_4 \Big] \nn \\*
    - i \pi \frac{8\tau}{3} &
      \Big[ 12 {\rm Li}_3(\tau) + 12 {\rm Li}_3(\upsilon)
          - 12 \cT {\rm Li}_2(\tau ) - 12 \cU {\rm Li}_2(\upsilon) \\* &
          - (\cT\!+\cU)^3 - 18 \zeta_2 (\cT\!+\cU) \Big]
    + \cO(\eps) \,. \nn
\end{align} \end{subequations}
\begin{subequations} \begin{align}
   S^{(2)[1]{\rm fin}}_{(--++)}
    = -\frac{\tau}{6} \,\,\,\,&\!\!\!\!
      \Big[ 48 {\rm Li}_4(\tau)
          - 24 (\cT\!+\cU) {\rm Li}_3(\tau) - 24 \cT {\rm Li}_3(\upsilon)
          + 24 \cT \cU {\rm Li}_2(\tau)
          + 24 \cT \cU {\rm Li}_2(\upsilon) \nn \\* \,\,\,\,&\!\!\!\!
          - 24 S_{2,2}(\tau)
          + \cT^4 -4 \cT^3 U + 18 \cT^2 \cU^2
          + 24 \zeta_2 {\rm Li}_2(\tau)
          - 12 \zeta_2 \cT^2 + 24 \zeta_2 \cT \cU \nn \\* \,\,\,\,&\!\!\!\!
          + 24 \zeta_3 \cU - 168 \zeta_4 \Big] \nn \\*
    + i \pi \frac{2\tau}{3}
      \Big[ 6 {\rm Li}_3(\tau) \,\,\,&\!\!\!
          + 6 {\rm Li}_3(\upsilon)
          - 6 \cU {\rm Li}_2(\tau) - 6 \cU {\rm Li}_2(\upsilon)
          - \cT^3\!+ 3 \cT^2 \cU - 6 \cT \cU^2
          - 6 \zeta_2 (\cT\!-\cU) \Big] \nn \\*
    - \frac{1}{3}
      \Big[ 6 {\rm Li}_3(\tau) \,\,\,&\!\!\!
          - 6 \cT {\rm Li}_2(\tau)
          - \cT^3 - 3 \cT^2 \cU + 24 \zeta_2 \cT + 12 \zeta_3 \Big] \nn \\*
    + i \pi \Big[ 2 {\rm Li}_2(\tau) \,\,\,&\!\!\!
          + \cT^2 + 2 \cT \cU - 4 \zeta_2 \Big]
    + \cO(\eps) \,, \\
   S^{(2)[0]{\rm fin}}_{(--)(++)}\!
   =\:\!\!-\frac{2}{3\upsilon} &
      \Big[ 12 \tau \upsilon {\rm Li}_3(\tau)
          + 12 \tau \upsilon {\rm Li}_3(\upsilon)
          - 12 \tau \upsilon \cT {\rm Li}_2(\tau)
          - 12 \tau \upsilon \cU {\rm Li}_2(\upsilon) \nn \\* &
          - 2 \tau^2 \cT^3 - 3 (1\!-\!2\tau^2) \cT^2 \cU
          + 3 (1\!-\!2\tau\!-\!2\tau \upsilon) \cT \cU^2
          - 2 \upsilon^2 \cU^3 \nn \\* &
          - 6 (\tau\!-\!3 \upsilon) \zeta_2 \cT
          - 6 (1\!-\!4\tau^2) \zeta_2 \cU \Big] \\*
     - i \pi\frac{2}{\upsilon} &
      \Big[ 2 \tau \upsilon (\cT\!-\cU)^2 - \cT^2 - \cU^2
          - 4 (1\!-\!3\tau \upsilon) \zeta_2 \Big]
    + \cO(\eps) \,, \nn \\
   S^{(2)[-1]{\rm fin}}_{(--++)}
    = -\frac{\tau}{12} &
      \Big[ 48 {\rm Li}_4(\tau)
          - 24 (\cT\!+\cU) {\rm Li}_3(\tau)
          - 24 \cT {\rm Li}_3(\upsilon) \nn \\* &
          + 24 \cT \cU {\rm Li}_2(\tau) + 24 \cT \cU {\rm Li}_2(\upsilon)
          - 24 S_{2,2}(\tau)
          + \cT^4 - 4 \cT^3 \cU + 18 \cT^2 \cU^2 \nn \\* &
          + 24 \zeta_2 {\rm Li}_2(\tau) - 12 \zeta_2 \cT^2
          + 24 \zeta_2 \cT \cU + 24 \zeta_3 \cU - 168 \zeta_4 \Big] \nn \\*
    + i \pi \frac{\tau}{3} &
      \Big[ 6 {\rm Li}_3(\tau) + 6 {\rm Li}_3(\upsilon)
          - 6 \cU {\rm Li}_2(\tau) - 6 \cU {\rm Li}_2(\upsilon) \\* &
          - \cT^3 + 3 \cT^2 \cU - 6 \cT \cU^2
          - 6 \zeta_2 (\cT\!-\cU) \Big]
    - 6 \zeta _3
    + \cO(\eps) \,. \nn
\end{align} \end{subequations}
\begin{subequations} \begin{align}
   R^{(2)[2]}_{(-+-+)}
    = \frac{\tau}{6 \upsilon^2} & \cT^2 \big( \cT^2\!- 24 \zeta_2 \big)
    + i\pi\frac{2\tau}{3 \upsilon^2} \cT^3 + 12 \zeta_3
    + \cO(\eps) \,, \\
   R^{(2)[1]{\rm fin}}_{(-+)(-+)}
    = \frac{2 \tau}{3 \upsilon^2} &
      \Big[ 48 {\rm Li}_4(\tau)
          - 24 \cT {\rm Li}_3(\tau) - 24 S_{2,2}(\tau) + \cT^4 \\* &
          + 24 \zeta_2 {\rm Li}_2(\tau)
          - 84 \zeta_2 \cT^2 + 24 \zeta_3 \cT - 102 \zeta_4 \Big]
    + i \pi \frac{16 \tau}{3 \upsilon^2} \cT
      \big[ \cT^2 - 3 \zeta_2 \big] \nn \\*
    - \frac{16 \tau}{\upsilon^2} &
      \Big[ \tau {\rm Li}_3(\tau) + \upsilon {\rm Li}_3(\upsilon)
          - \tau \cT {\rm Li}_2(\tau) - \upsilon \cU {\rm Li}_2(\upsilon)
          - 5 \tau \zeta_2 \cT - 5 \upsilon \zeta_2 \cU - \zeta_3 \Big] \nn \\*
    + i \pi \frac{8 \tau}{\upsilon^2} &
      \Big[ 2 (\tau\!-\!\upsilon) {\rm Li}_2(\tau)
          - \tau \cT^2 - 2 \upsilon \cT \cU + \upsilon \cU^2
          - 2 \tau \zeta_2 \Big]
    + \cO(\eps) \,. \nn
\end{align} \end{subequations}
\begin{subequations} \begin{align}
   S^{(2)[1]{\rm fin}}_{(-+-+)}\!\!\!&\,\,\,
    = \frac{\tau}{4 \upsilon^2} \cT^2
      \big[ \cT^2 - 32 \zeta_2 \big]
    + i \pi \frac{\tau }{\upsilon^2} \cT \big[ \cT^2 - 4 \zeta_2 \big] \\*
    -\:\,\frac{1}{3 \upsilon} &
      \Big[ 6 \upsilon {\rm Li}_3(\tau)
          - 6 \upsilon \cT {\rm Li}_2(\tau)
          - (1\!+\!2\tau) \cT^3 - 3 (1\!-\!\tau) \cT^2 \cU
          + 24 (1\!+\!\tau) \zeta_2 \cT
          + 12 \upsilon \zeta_3 \Big] \nn \\*
    + i \pi \frac{1}{\upsilon} &
      \big[ 2 \upsilon {\rm Li}_2(\tau)
          + (1\!+\!2\tau) \cT^2 + 2 \upsilon \cT \cU - 4 \zeta _2 \Big]
    + \cO(\eps) \,, \nn \\
   S^{(2)[0]{\rm fin}}_{(-+)(-+)}\!\!\!\!&\,\,\,
    = \frac{\tau}{3 \upsilon^2}
      \Big[ 72 {\rm Li}_4(\tau)
          - 24 (2\cT\!+\cU) {\rm Li}_3(\tau)
          - 48 \cT {\rm Li}_3(\upsilon) \\* & \qquad~\:\quad
          + 12 \cT (\cT\!+2\cU) {\rm Li}_2(\tau)
          + 48 \cT \cU {\rm Li}_2(\upsilon)
          - 60 S_{2,2}(\tau) + 4 \cT^3 \cU + 24 \cT^2 \cU^2 \nn \\* & \qquad~\:\quad
          + 84 \zeta_2 {\rm Li}_2(\tau) - 54 \zeta_2 \cT^2
          + 72 \zeta_2 \cT \cU
          + 24 \zeta_3 (2\cT\!+\!\cU) - 267 \zeta_4 \Big] \nn \\*
    - i \pi \frac{4 \tau}{3 \upsilon^2} &
      \Big[ 3 {\rm Li}_3(\tau)\!+\!12 {\rm Li}_3(\upsilon)
          - 6 \cU {\rm Li}_2(\tau)\!-\!12 \cU {\rm Li}_2(\upsilon)
          - \cT^3\!- 6 \cT \cU^2\!- 3 \zeta _2 (\cT\!-2\cU)
          - 3 \zeta_3 \Big] \nn \\*
    + \frac{2}{3 \upsilon^2} &
      \Big[ 6 \tau (1\!+\!\upsilon) {\rm Li}_3(\tau)
          - 6 \tau \upsilon {\rm Li}_3(\upsilon)
          - 6 \tau (1\!+\!\upsilon) \cT {\rm Li}_2(\tau)
          + 6 \tau \upsilon \cU {\rm Li}_2(\upsilon)
          - 2 \tau^2 \cT^3 \nn \\* &
          + 3 \upsilon^2 \cT \cU (\cT\!-\cU)
          + 2 \upsilon^2 \cU^3
          - 6(3\!+\!2\tau\!-\!4\tau^2) \zeta_2 \cT
           + 6 (1\!+\!3\tau\!-\!4\tau^2)\zeta_2 \cU
          - 6 \tau \zeta_3 \Big] \nn \\*
    - \frac{2i \pi}{\upsilon^2} &
      \Big[ 2 \tau (3\!-\!2\tau) {\rm Li}_2(\tau)
          - (1\!-\!2\tau^2) \cT^2 + 2 \tau \upsilon \cT \cU
          + \upsilon (\tau\!-\!\upsilon) \cU^2
          - 2 (2\!-\!\tau^2)\zeta_2 \Big]
    + \cO(\eps) \,, \nn \\
   S^{(2)[-1]{\rm fin}}_{(-+-+)}\!\!\!&\,\,\,
    = -\frac{\tau}{12 \upsilon^2} \cT^2
       \big[ \cT^2 - 24 \zeta_2 \big]
    - i \pi \frac{\tau}{3 \upsilon^2} \cT^3 - 6 \zeta_3
    + \cO(\eps) \,.
\end{align} \end{subequations}

\bibliographystyle{JHEP}
\bibliography{references}
\end{document}